\documentclass[longauth]{aa}  
\usepackage{graphicx}
\usepackage{txfonts}
\usepackage[]{hyperref}
\hypersetup{colorlinks=false}
\usepackage[utf8]{inputenc}
\usepackage[english]{babel}
\usepackage{url}
\usepackage{amssymb}
\usepackage{color}
\usepackage{multirow}
\usepackage{todonotes}
\usepackage{soul}
\usepackage{ulem}

\newcommand{\emm}[1]{\ensuremath{#1}}
\newcommand{\emr}[1]{\emm{\mathrm{#1}}}

\newcommand{\paren}[1]  {\emm{\left(  #1 \right) }} 
\newcommand{\cbrace}[1] {\emm{\left\{ #1 \right\}}} 
\newcommand{\bracket}[1]{\emm{\left[  #1 \right] }} 

\newcommand{\unit}[1]{\emr{\,#1}}
\newcommand{\kms}{\unit{km\,s^{-1}}}
\newcommand{\pscm}{\unit{cm^{-2}}}
\newcommand{\pccm}{\unit{cm^{-3}}}

\newcommand{\HII}{\ion{H}{ii}}

  \newcommand{\TabData}{%
  \setlength{\tabcolsep}{5pt}
  \begin{table}
    \centering %
    \caption{Properties of the 30m datasets.}
   \resizebox{\hsize}{!}{   
    \begin{tabular}{ccccc}
  \hline \hline
  Line & Rest frequency & Resolution & Beam      & Noise \\
       & (GHz)            & (km.s$^{-1}$)     & ($\arcsec$) & (mK)    \\
  \hline
  C$^{18}$O (1$-$0)  & 109.782173  & 0.5 & 23.6 & 162 \\
  $^{13}$CO (1$-$0)  & 110.201354  & 0.5 & 23.5 & 164 \\
  \hline
  \end{tabular}
  }  
    \label{tab:details-resulting-maps}
  \end{table}}

\newcommand{\TabHIIregions}{%
  \setlength{\tabcolsep}{5pt}
  \begin{table*}
    \centering %
    \caption{Properties of the \HII{} regions ionizing stars.}
\begin{tabular}{lcccccc}
\hline \hline
\multicolumn{1}{c}{\hfill}  & \multicolumn{2}{c}{\hfill}   \\
Star     & Spectral type & R.A., DEC (J2000) & Offset & Distance & HII region & Radius$^{\star}$    \\
 & &  (h:m:s, $^{\circ}$:$\arcmin$:$\arcsec$) & ($\arcmin$, $\arcmin$) & (pc) &  & ($\arcmin$) \\
\hfill  & \hfill  & \hfill    \\
\hline
$\sigma$Ori & O9.5V B  & 05 38 44.779, -02 36 00.12  &  (-33.35, 00.07) &  387.5 $\pm$ 1.3\tablefootmark{(1)} & IC 434 & 42\\
HD 38087   &  B5V D &  05 43 00.573 , -02 18 45.38 &  (32.87, 01.33) &   169 $\pm$ 37\tablefootmark{(2)} & IC 435 & 3\\
HD 37903    & B1.5V C  & 05 41 38.388, -02 15 32.48  & (13.72, 09.42)  &   362 $\pm$ 35\tablefootmark{(2)} & NGC 2023 & 2.5\\
Alnitak & O9.7Ib+B0III C  & 05 40 45.527, -01 56 33.26  &  (05.49, 31.04) & 294 $\pm$ 21\tablefootmark{(4)} &  & 11\\
IRS2b & O8V-B2V  & 05 41 45.50, -01 54 28.7  &  (20.54, 29.43) &  415\tablefootmark{(3)} & NGC 2024 & 12\\
V$^{\star}$ V901 Ori & B2V C  & 05 40 56.370, -01 30 25.857  & (14.44, 55.99)  &  437 $\pm$ 11\tablefootmark{(5)} &  & 3.5\\
HD 37674  &  B5V(n) C & 05 40 13.539, -01 27 45.25  &  (4.69, 55.99)  &  425 $\pm$ 15\tablefootmark{(5)} &  & 5\\
\hline
\end{tabular}
    \tablefoot{\tablefoottext{$\star$}{Radius of the circles drawn in
        Fig. \ref{fig:zoom-regions-NH-maps:layers-cloud}. They are
        determined according to the size of the \HII{} region emission either
        in H$\alpha$ or in radio continuum (see
        Sect. \ref{sec:detection-filaments}).}} %
    \tablebib{(1) \cite{Schaefer16}; (2) \cite{Gaia16}; (3)
      \cite{Anthony-Twarog82} ; (4) \cite{Hummel13}; (5) \cite{Gaia18}.}
    \label{tab:regions-HII}
  \end{table*}}

\newcommand{\FigRGB}{%
  \begin{figure*}
    \centering %
    \includegraphics[width=0.85\linewidth]{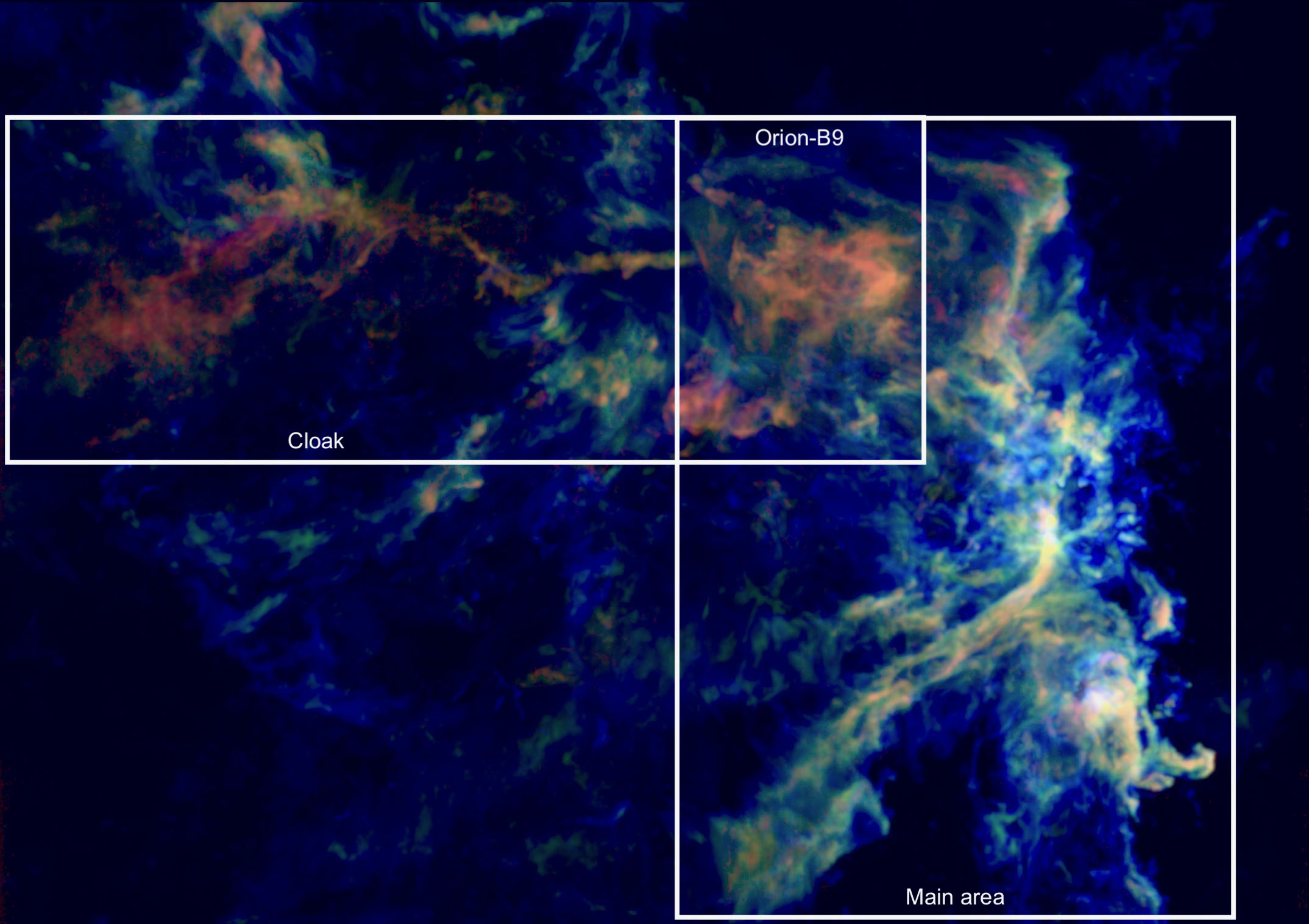}
    \caption{Carbon monoxide emission in the Orion B cloud. The image
      combines the peak temperatures of $^{12}$CO (1$-$0), $^{13}$CO
      (1$-$0) and C$^{18}$O (1$-$0) coded in blue, green and red,
      respectively.}
    \label{fig:regions-orionb-cloud}
  \end{figure*}}

\newcommand{\FigData}{%
  \begin{figure*}
    \centering %
    \includegraphics[width=\linewidth]{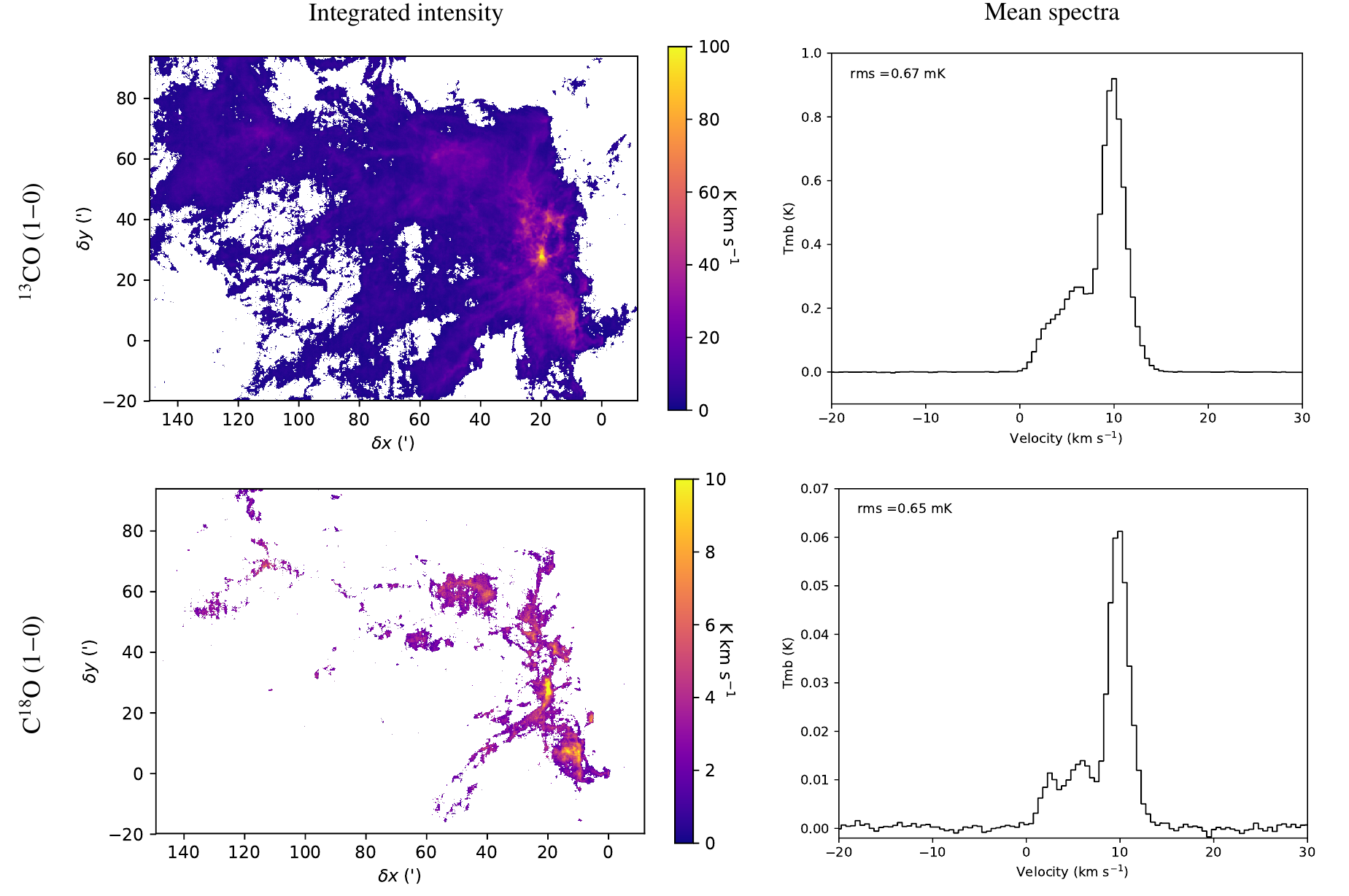}
    \caption{Zero-order moment maps (integrated intensity) of the total
      emission of the Orion B cloud (left) and mean spectra (right) of the
      $^{13}$CO (1$-$0) (top) and C$^{18}$O (1$-$0) (bottom) molecular
      lines from the 30m data sets. For the zero-order moment map, the
      integration was done on a velocity range between -5 and 20
      \kms{}. Isolated pixels and pixels with a signal-to-noise ratio
      lower than 3 are blanked.}
    \label{fig:moment0-maps:initial-datasets}
  \end{figure*}}

\newcommand{\FigLayerSpectra}{%
  \begin{figure}
    \centering %
    \includegraphics[width=1.03\linewidth]{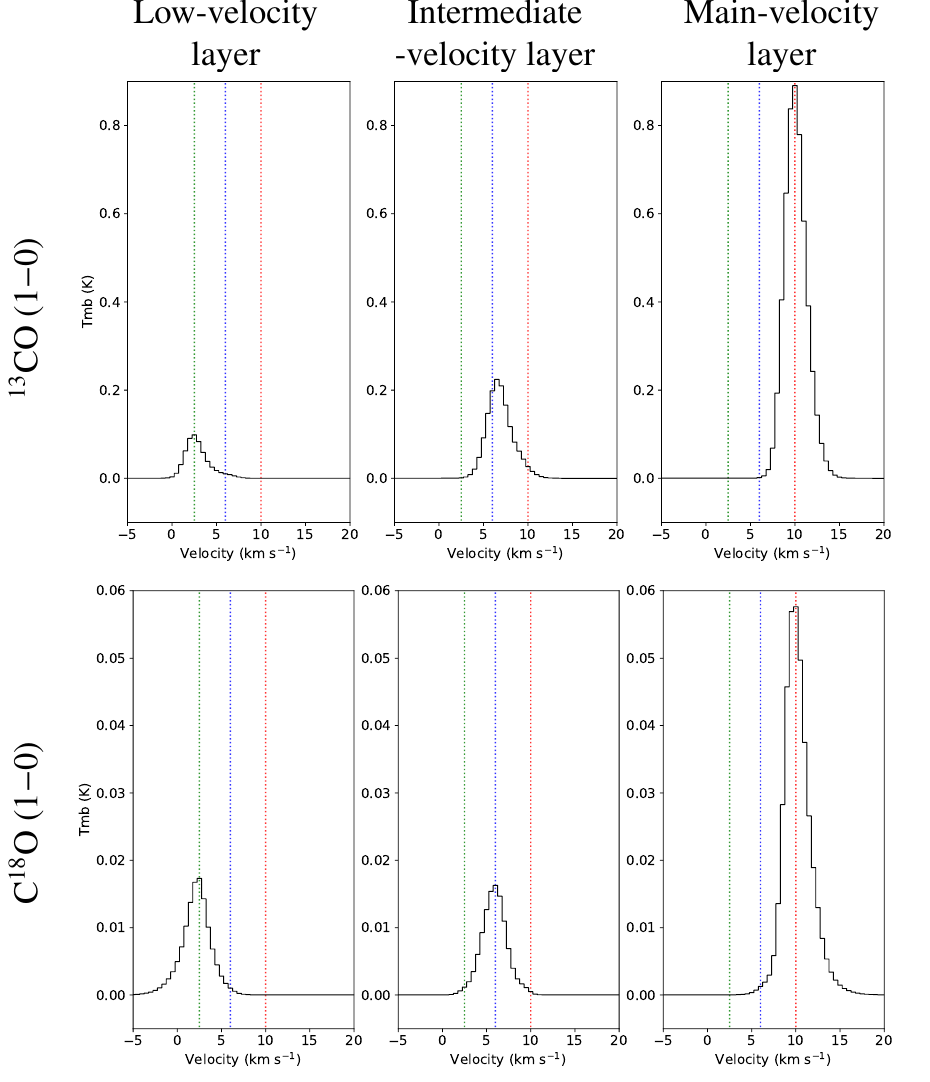}
    \caption{Mean spectra of the three layers of the Orion B cloud, namely
      the low-velocity layer (left), the intermediate-velocity layer
      (middle), and the main-velocity layer (right), from the $^{13}$CO
      (1$-$0) (top) and C$^{18}$O (1$-$0) (bottom) emission rebuilt by
      ROHSA. The color dotted lines show the systemic LSR velocity of the
      three cloud layers.}
    \label{fig:mean-spectrum-ROHSA-cloud-layers}
  \end{figure}}

\newcommand{\FigLayerMoments}{%
  \begin{figure*}
    \centering %
    \includegraphics[width=0.965\linewidth]{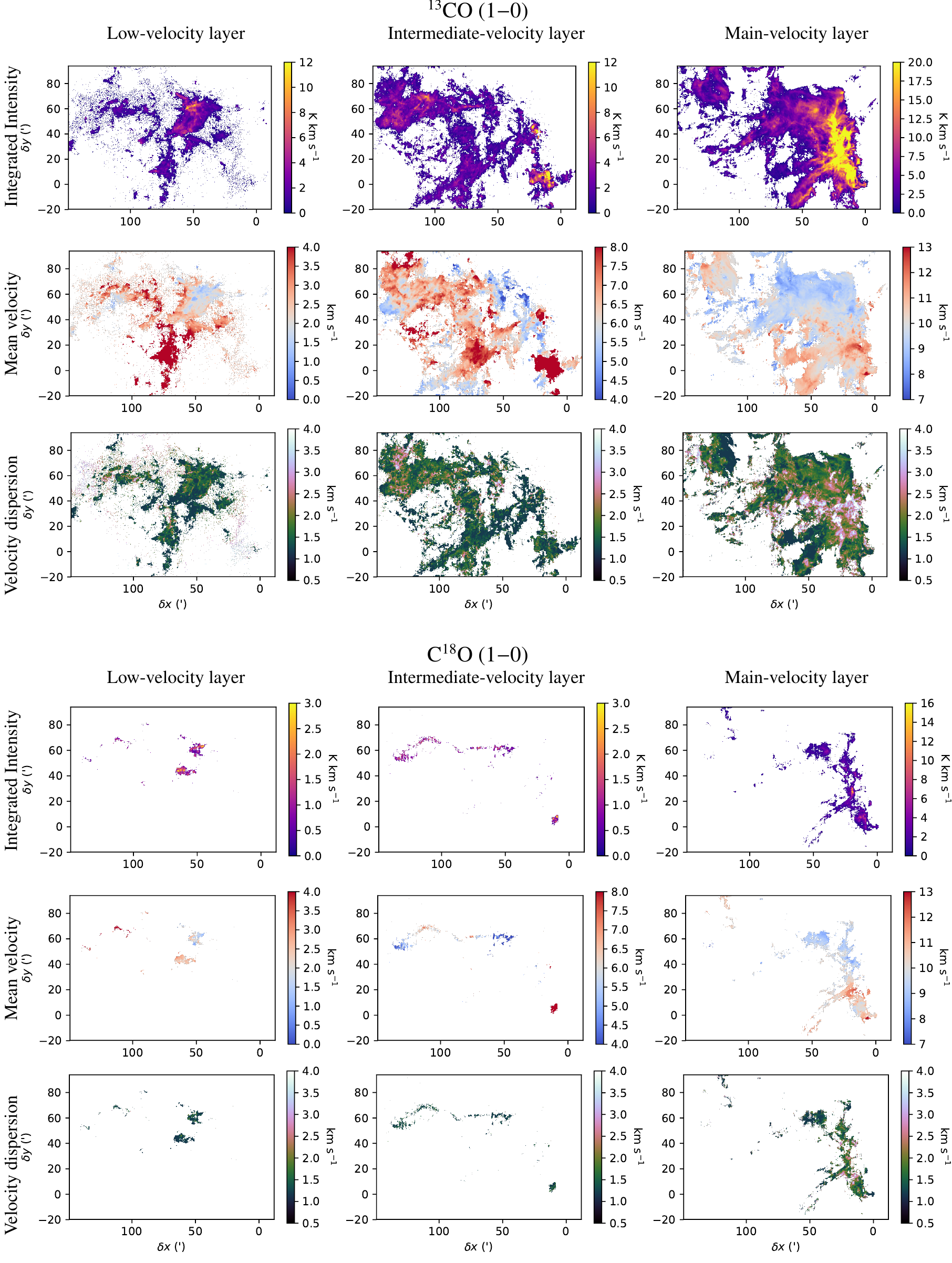}
    \caption{Zero-order (integrated intensity, top), first-order (mean
      velocity, middle), and second-order (velocity dispersion, bottom)
      moment maps of each of the identified layer of the Orion B cloud,
      namely the low-velocity layer (left), the intermediate-velocity layer
      (middle), and the main-velocity layer (right), from the $^{13}$CO
      (1$-$0) (top) and C$^{18}$O (1$-$0) (bottom) emission reconstructed
      by ROHSA. Isolated pixels, pixels with a signal-to-noise ratio lower
      than 3, and pixels with a mean velocity error higher than 0.25 km
      s$^{-1}$ are blanked. }
    \label{fig:moment-maps:ROHSA-layers-cloud}
  \end{figure*}}

\newcommand{\FigLayerNH}{%
  \begin{figure*}
    \centering %
    \includegraphics[width=\linewidth]{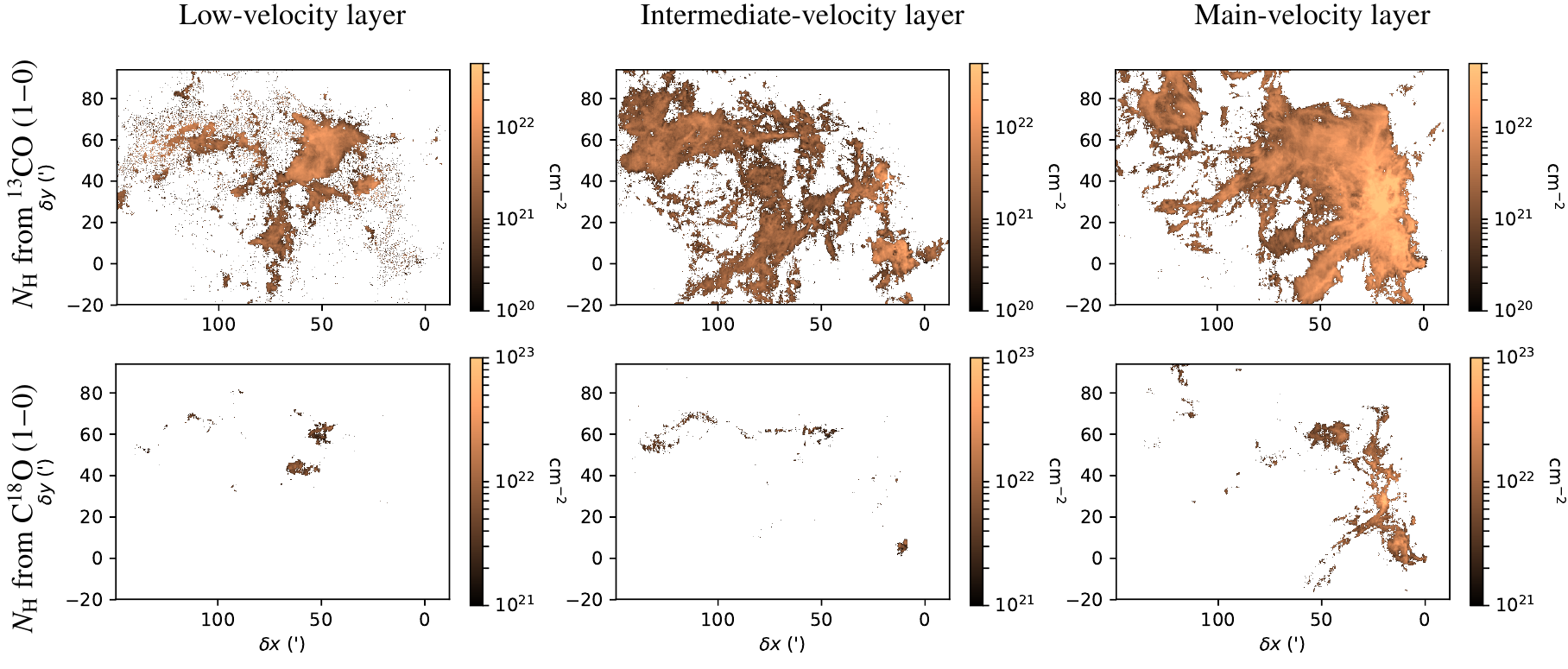}
    \caption{$N_\mathrm{H}$ maps of each of the identified layer of the
      Orion B cloud, namely the low-velocity layer (left), the
      intermediate-velocity layer (middle), and the main-velocity layer
      (right), from the $^{13}$CO (1$-$0) (top) and C$^{18}$O (1$-$0)
      (bottom) emission reconstructed by ROHSA. Isolated pixels, pixels
      with a signal-to-noise ratio lower than 3, and pixels with a mean
      velocity error higher than 0.25 km s$^{-1}$ are blanked.}
    \label{fig:NH-maps:ROHSA-layers-cloud}
  \end{figure*}}

\newcommand{\FigSubRegions}{%
  \begin{figure*}
    \includegraphics[width=\linewidth]{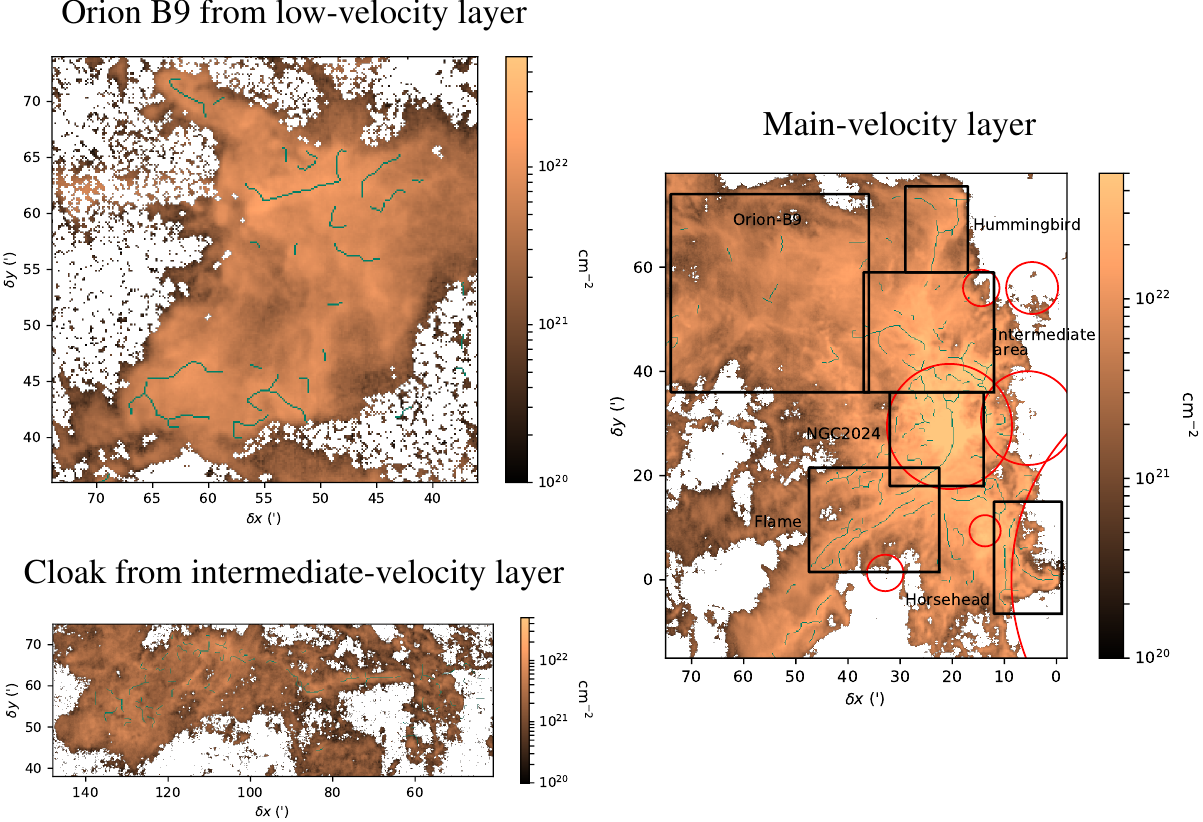}
    \caption{Subregions of interest overlaid on the $N_\mathrm{H}^{13}$
      maps of each of the identified layer of the Orion B cloud, namely the
      low-velocity layer (top left), the intermediate-velocity layer
      (bottom left), and the main-velocity layer (right), from the
      $^{13}$CO (1$-$0) emission reconstructed by ROHSA. The green lines
      show the filamentary structures extracted from the C$^{18}$O (1$-$0)
      emission. The red circles show the typical extensions of the \HII{}
      regions associated to massive stars (see Table
      \ref{tab:regions-HII}).  }
    \label{fig:zoom-regions-NH-maps:layers-cloud}
  \end{figure*}}

\newcommand{\FigAngGradientsExample}{%
  \begin{figure*}
    \centering %
    \includegraphics[width=0.655\linewidth]{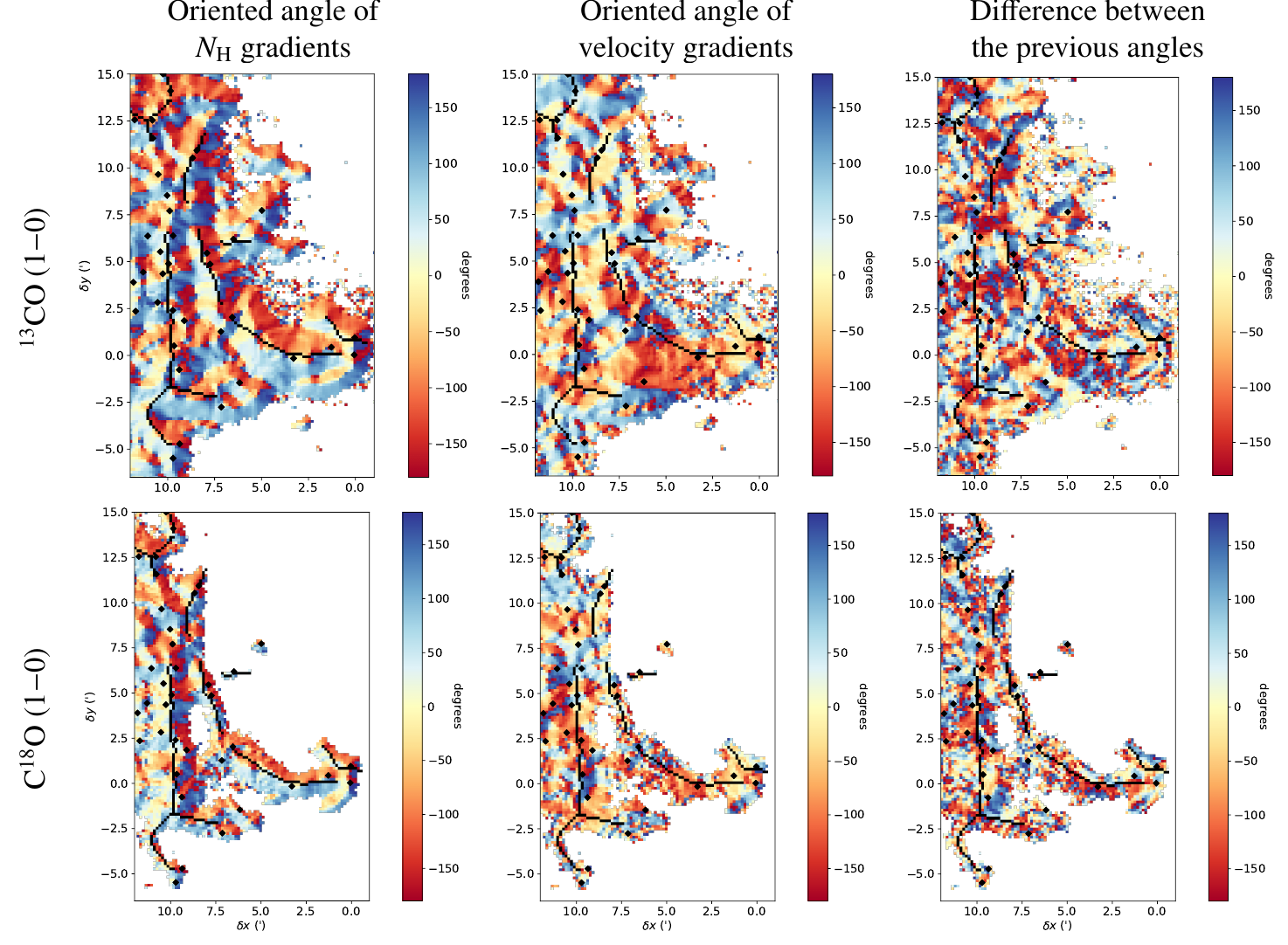}
    \caption{Maps of the column density gradient oriented angles (left) and
      the velocity gradient oriented angles (middle) in the Horsehead
      nebula from the main-velocity layer from the $^{13}$CO (1$-$0) (top)
      and the C$^{18}$O (1$-$0) (bottom) emission. Right: maps of the
      relative orientation between the oriented angles of the column
      density and the velocity gradients. Oriented angles are defined north
      from east. The black lines show the filamentary structure and the
      diamond dots show the dense cores identified by the Herschel Gould
      Belt Survey \citep{Konyves2020}.}
    \label{fig:oriented-angle-maps:example}
  \end{figure*}}

\newcommand{\FigDivMapsExample}{%
  \begin{figure*}
    \centering %
    \includegraphics[width=0.655\linewidth]{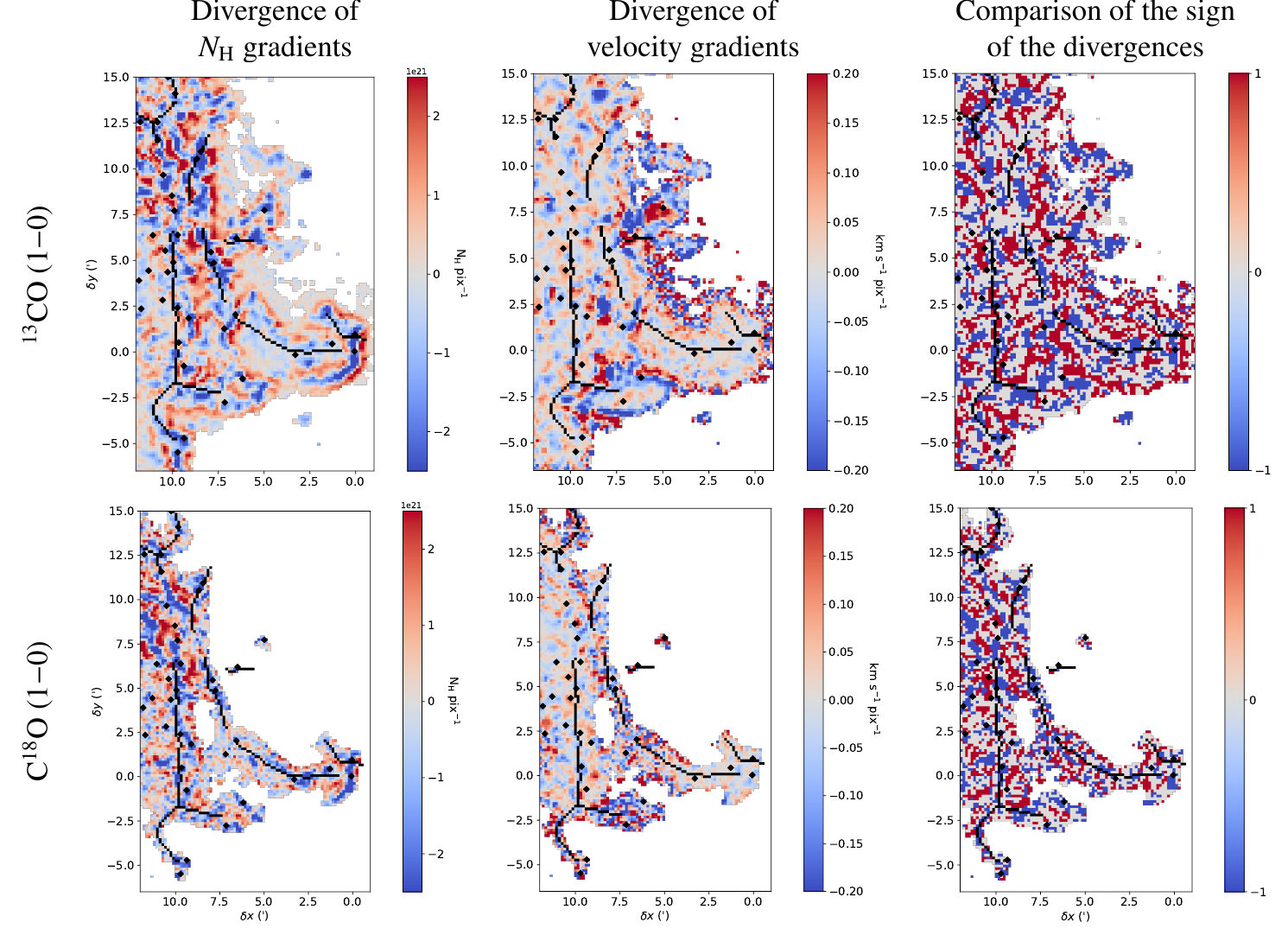}
    \caption{Divergence maps of the column density gradients in
      N$_\mathrm{H}$ pix$^{-1}$ (left) and the velocity gradients in km
      s$^{-1}$ pix$^{-1}$ (middle) in the Horsehead nebula from the
      main-velocity layer from the $^{13}$CO (1$-$0) (top) and the
      C$^{18}$O (1$-$0) (bottom) emission. Right: comparison maps of the
      sign between the velocity gradient divergence and the column density
      gradient divergence. The values -1 (blue)  or 1 (red) show the pixels
      where velocity and column density gradients both
      converge or both diverge. The value 0 (gray) indicates regions where the sign
      of at least one gradient is ill-defined because its modulus is close
      to zero. The black lines show the filamentary structure and the
      diamond dots show the dense cores identified by the Herschel Gould
      Belt Survey \citep{Konyves2020}.}
    \label{fig:divergence-maps:example}
  \end{figure*}}

\newcommand{\FigCircularHistoOne}{%
  \begin{figure*}
    \centering %
    \includegraphics[width=0.75\linewidth]{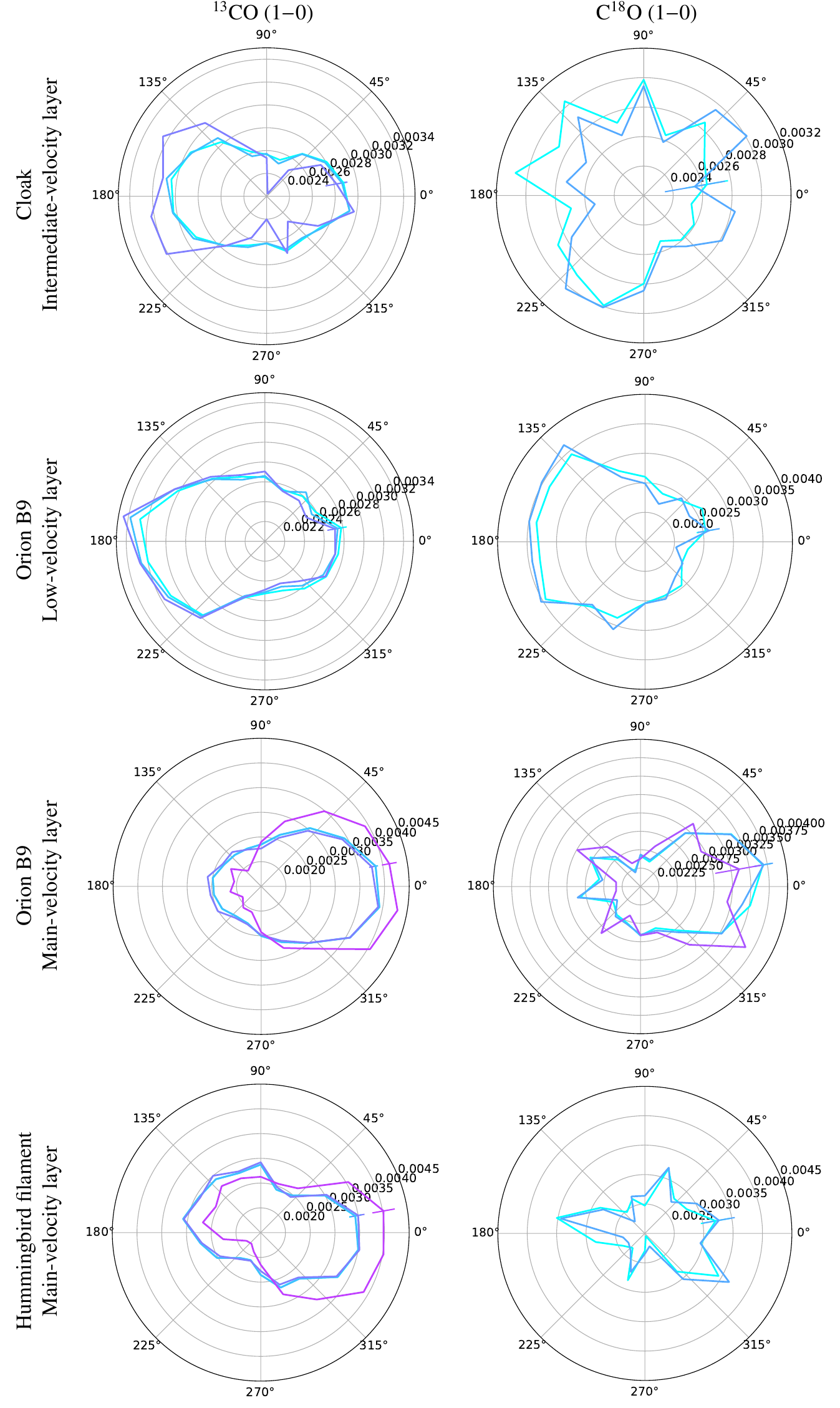}
    \caption{Polar histogram of the relative orientation between the
      oriented angles of the column density and the velocity gradients from
      the $^{13}$CO (1$-$0) (left) and the C$^{18}$O (1$-$0) (right)
      emission for the different sub-regions of interest within the Orion B
      cloud. The different colors correspond to the column density
      thresholds.  For $^{13}$CO (1$-$0): data with
      $\log(N_\mathrm{H}^{13}) > 20.00$ cm$^{-2}$ are shown in cyan,
      $\log(N_\mathrm{H}^{13}) > 21.00$ cm$^{-2}$ in blue,
      $\log(N_\mathrm{H}^{13}) > 21.55$ cm$^{-2}$ in blue-violet,
      $\log(N_\mathrm{H}^{13}) > 22.00$ cm$^{-2}$ in purple, and
      $\log(N_\mathrm{H}^{13}) > 22.40$ cm$^{-2}$ in fuchsia.  For
      C$^{18}$O (1$-$0): data with $\log(N_\mathrm{H}^{18}) > 21.00$
      cm$^{-2}$ are shown in cyan, $\log(N_\mathrm{H}^{18}) > 21.55$
      cm$^{-2}$ in blue, $\log(N_\mathrm{H}^{18}) > 22.00$ cm$^{-2}$ in
      purple, and $\log(N_\mathrm{H}^{18}) > 22.40$ cm$^{-2}$ in fuchsia.
      The colored bars at 10$^{\circ}$ show the error bars of the
      histograms.}
    \label{fig:histogram-difference-oriented-angle-gradients}
  \end{figure*}}

\newcommand{\FigCircularHistoTwo}{%
  \begin{figure*}
    \addtocounter{figure}{-1} \centering %
    \includegraphics[width=0.75\linewidth]{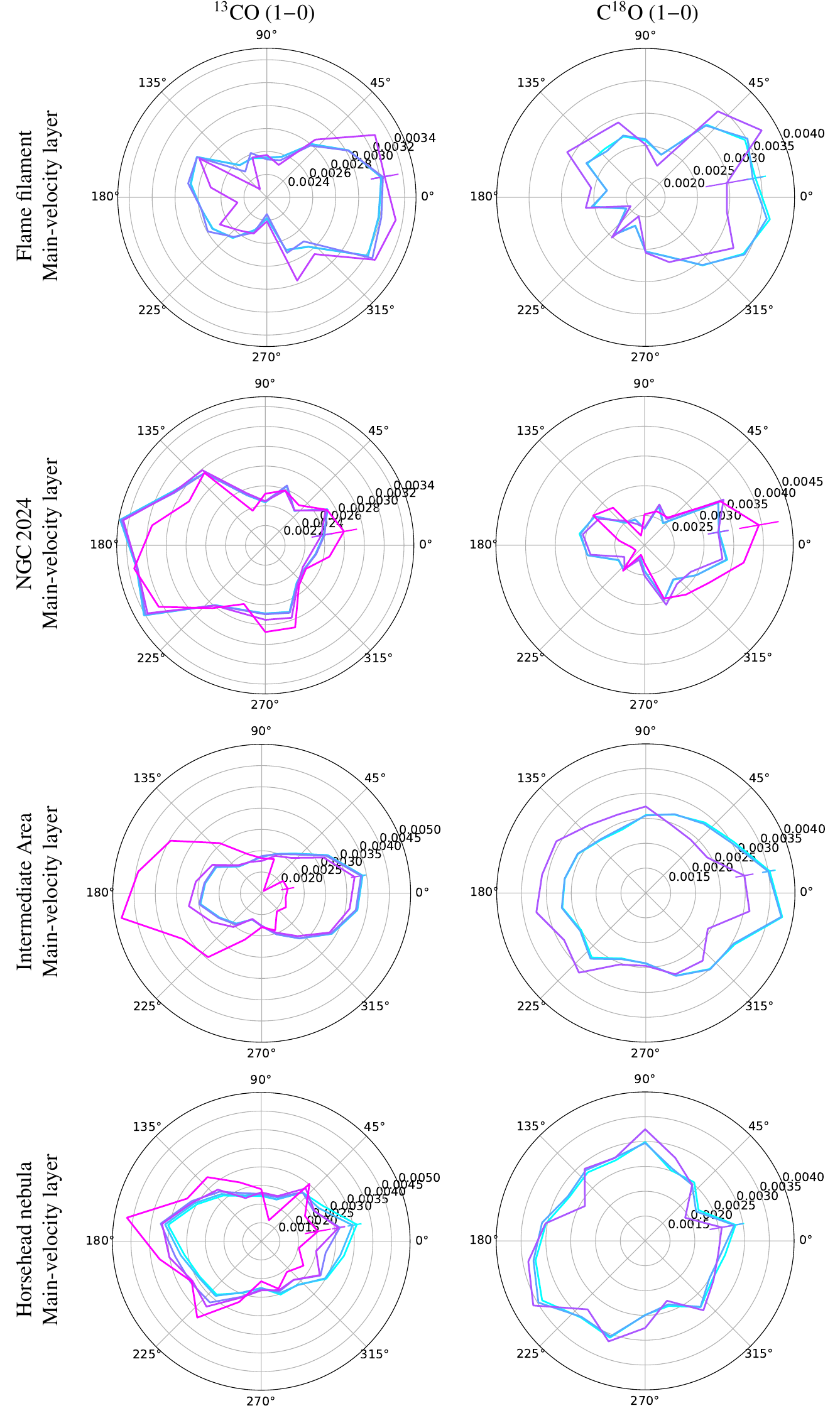}
    \caption{Continued.}
  \end{figure*}}

\newcommand{\FigCircularHistoSim}{%
  \begin{figure*}
    \centering %
    \includegraphics[width=1.05\linewidth]{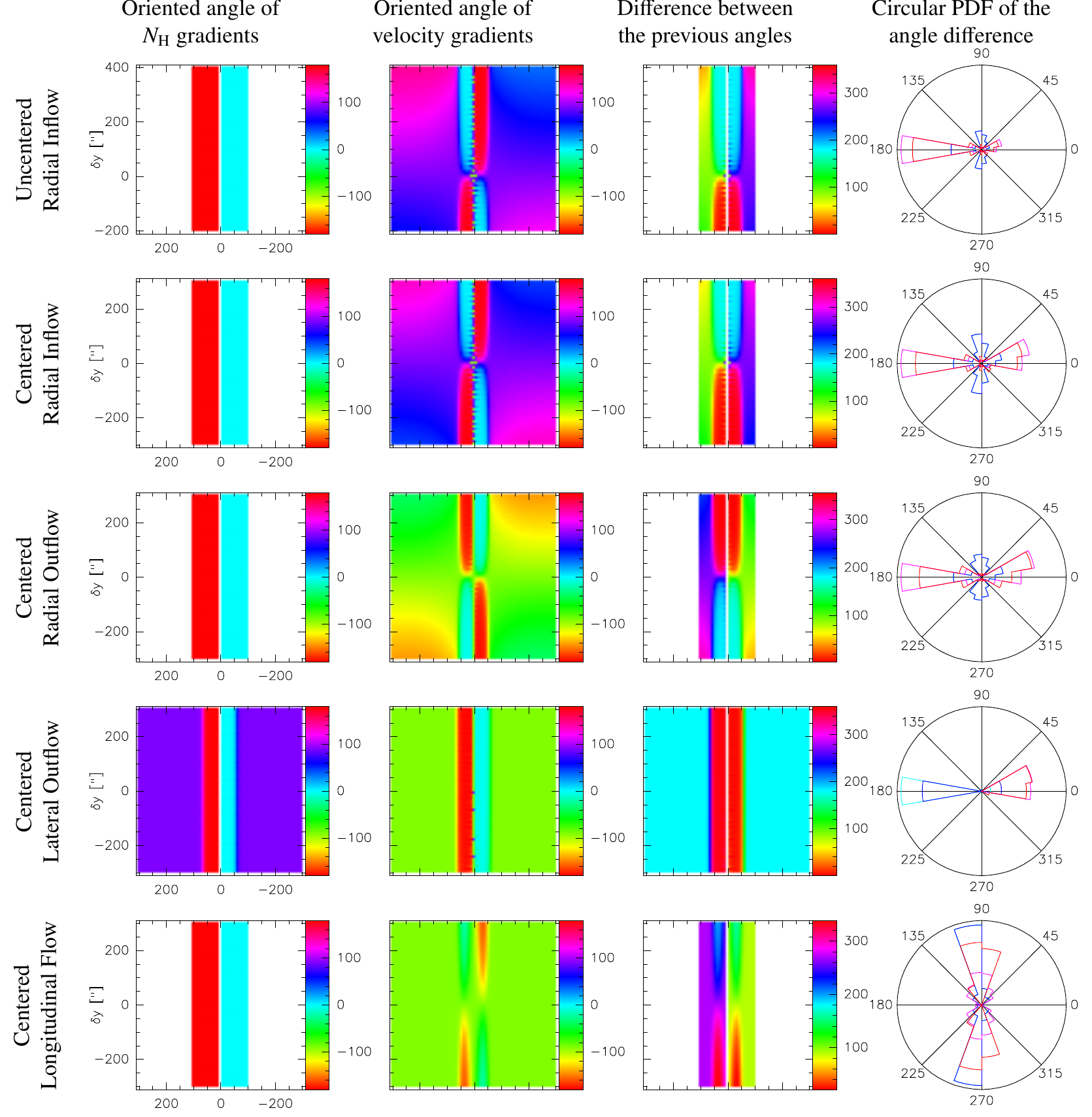}
    \caption{Oriented angles of the gradient of column density and centroid
      velocity, and their differences for five different toy models of the
      density and velocity fields. All angles are displayed in degree. The
      white areas on the images show regions where the gradient is too
      small to reliably compute its orientation. The different colors of
      the circular probability distribution function (PDF) correspond to the column density thresholds which
      are used in the Orion B data set analysis:
      $\log(N_\mathrm{H}) > 21.00$ cm$^{-2}$ are shown in cyan,
      $\log(N_\mathrm{H}) > 21.55$ cm$^{-2}$ in blue,
      $\log(N_\mathrm{H}) > 22.00$ cm$^{-2}$ in pink, and
      $\log(N_\mathrm{H}) > 22.40$ cm$^{-2}$ in red.}
    \label{fig:histogram-difference-oriented-angle-gradients-modeles}
  \end{figure*}}

\newcommand{\Figchidist}{%
  \begin{figure*}
    \centering %
    \includegraphics[width=\linewidth]{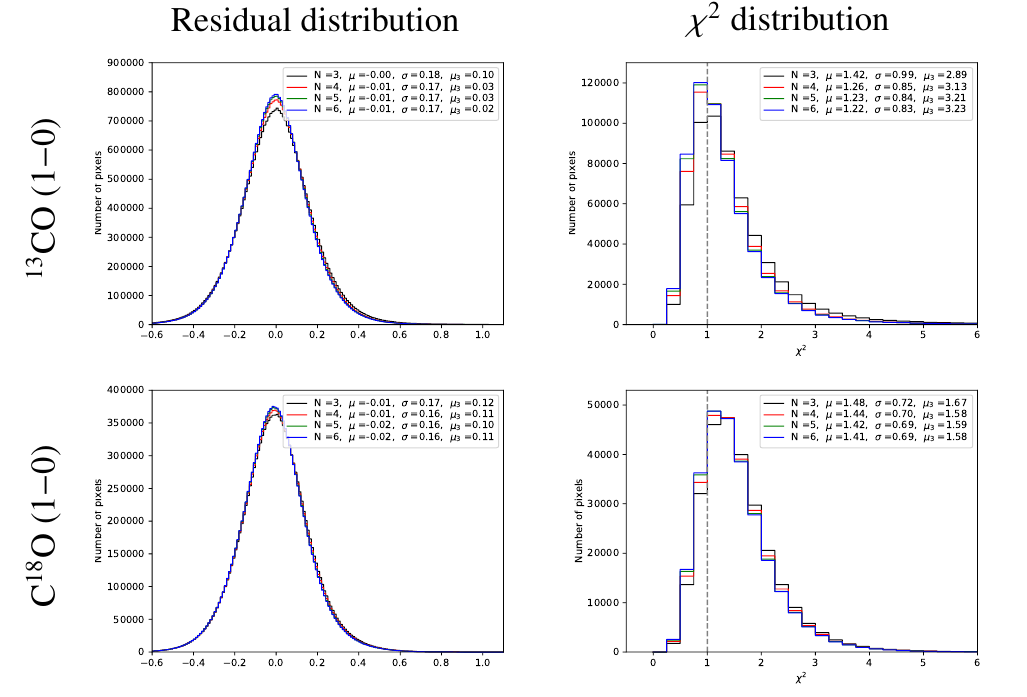}
    \caption{Residual (left) and $\chi^2$ (right) distributions between the
      original and the ROHSA reconstructed cubes for different numbers $N$
      of Gaussian components and $\lambda_\mathrm{i}=100$ from the
      $^{13}$CO (1$-$0) (top) and C$^{18}$O (1$-$0) (bottom) emission,
      respectively. The mean, variance and skewness of each distribution
      are shown in legend.}
    \label{fig:chi2-residuals-distribution-ngauss-ROHSA}
  \end{figure*}}

\newcommand{\FigDistCompOne}{%
  \begin{figure*}
    \centering %
    \includegraphics[width=0.94\linewidth]{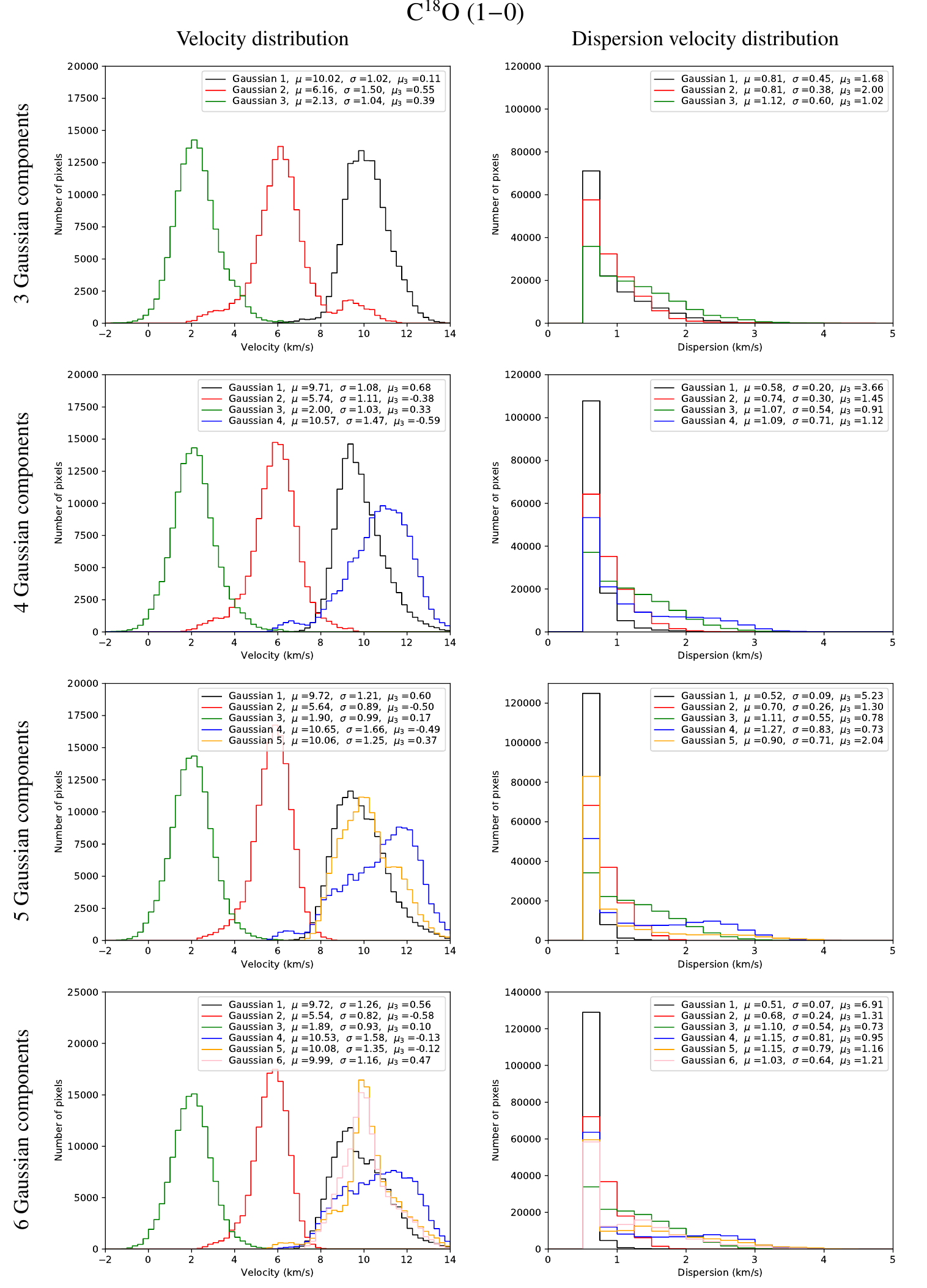}
    \caption{Distributions in velocity (left) and velocity dispersion
      (right) of each Gaussian component from the different decompositions
      using different numbers $N$ of Gaussian components. The mean,
      variance and skewness of each distribution are shown in legend.}
    \label{fig:distribution-position-sigma-ngauss-c18o-ROHSA}
  \end{figure*}}

\newcommand{\FigDistCompTwo}{%
  \begin{figure*}
    \centering %
    \includegraphics[width=0.94\linewidth]{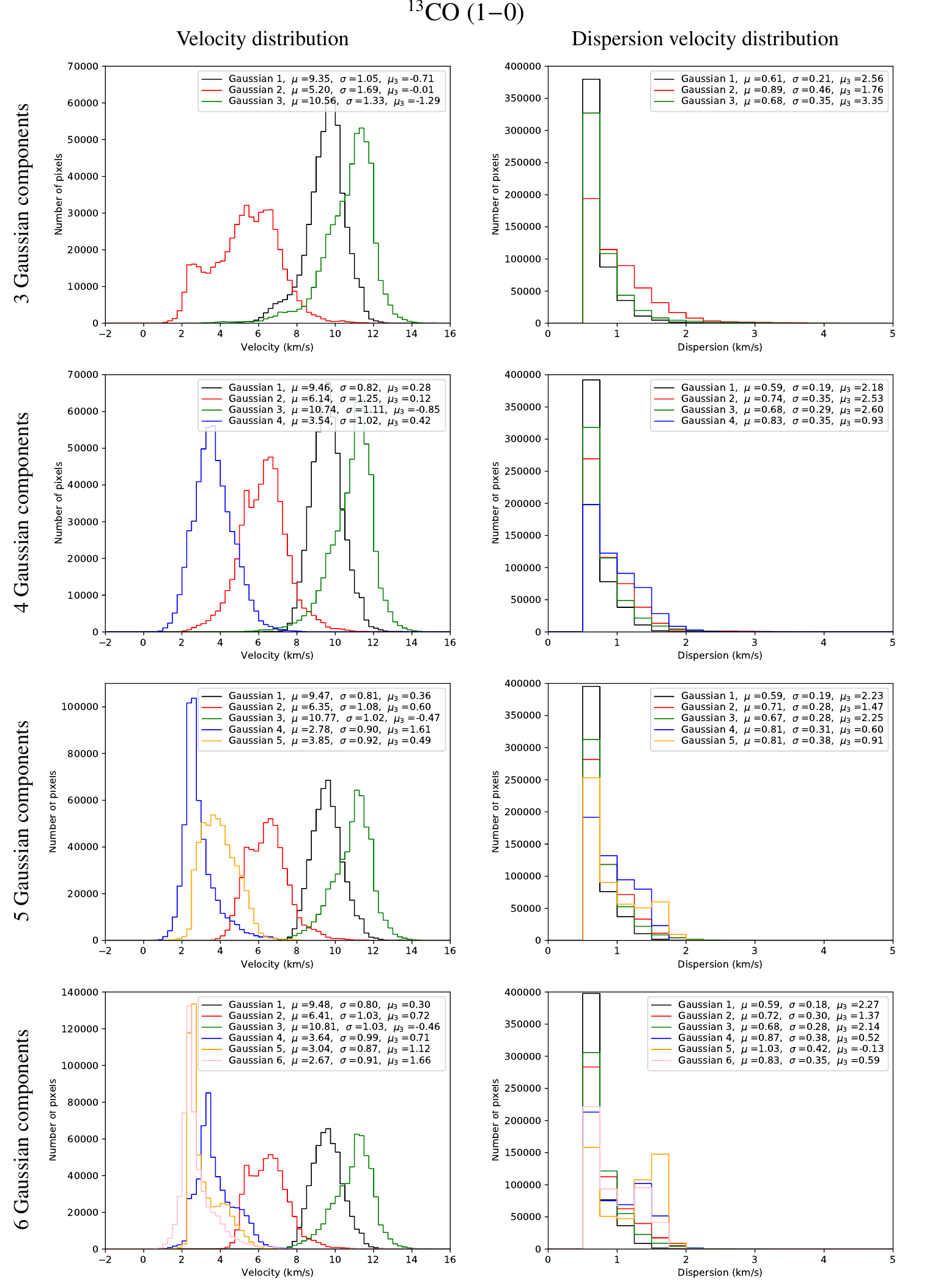}
    \caption{Distributions in velocity (left) and velocity dispersion
      (right) of each Gaussian component from the different decompositions
      using different numbers $N$ of Gaussian components. The mean,
      variance and skewness of each distribution are shown in legend.}
    \label{fig:distribution-position-sigma-ngauss-13co-ROHSA}
  \end{figure*}}

\newcommand{\FigchidistTwo}{%
  \begin{figure*}
    \centering %
    \includegraphics[width=\linewidth]{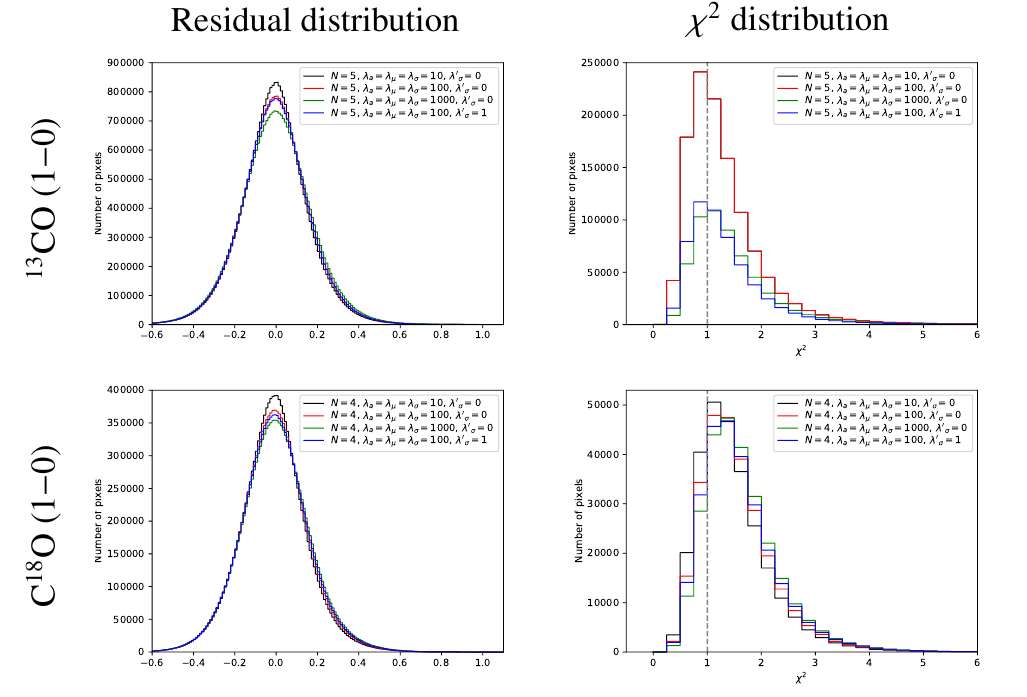}
    \caption{Residual (left) and $\chi^2$ (right) distributions between the
      original and the ROHSA reconstructed cubes for a given number of
      Gaussian components (four for C$^{18}$O and five for $^{13}$CO) and
      different hyper-parameters $\lambda_\mathrm{i}$ from the $^{13}$CO
      (1$-$0) (top) and C$^{18}$O (1$-$0) (bottom) emission, respectively.}
    \label{fig:chi2-residuals-distribution-param-ROHSA}
  \end{figure*}}

\newcommand{\FigGaussSpectra}{%
  \begin{figure}
    \centering %
    \includegraphics[width=\linewidth]{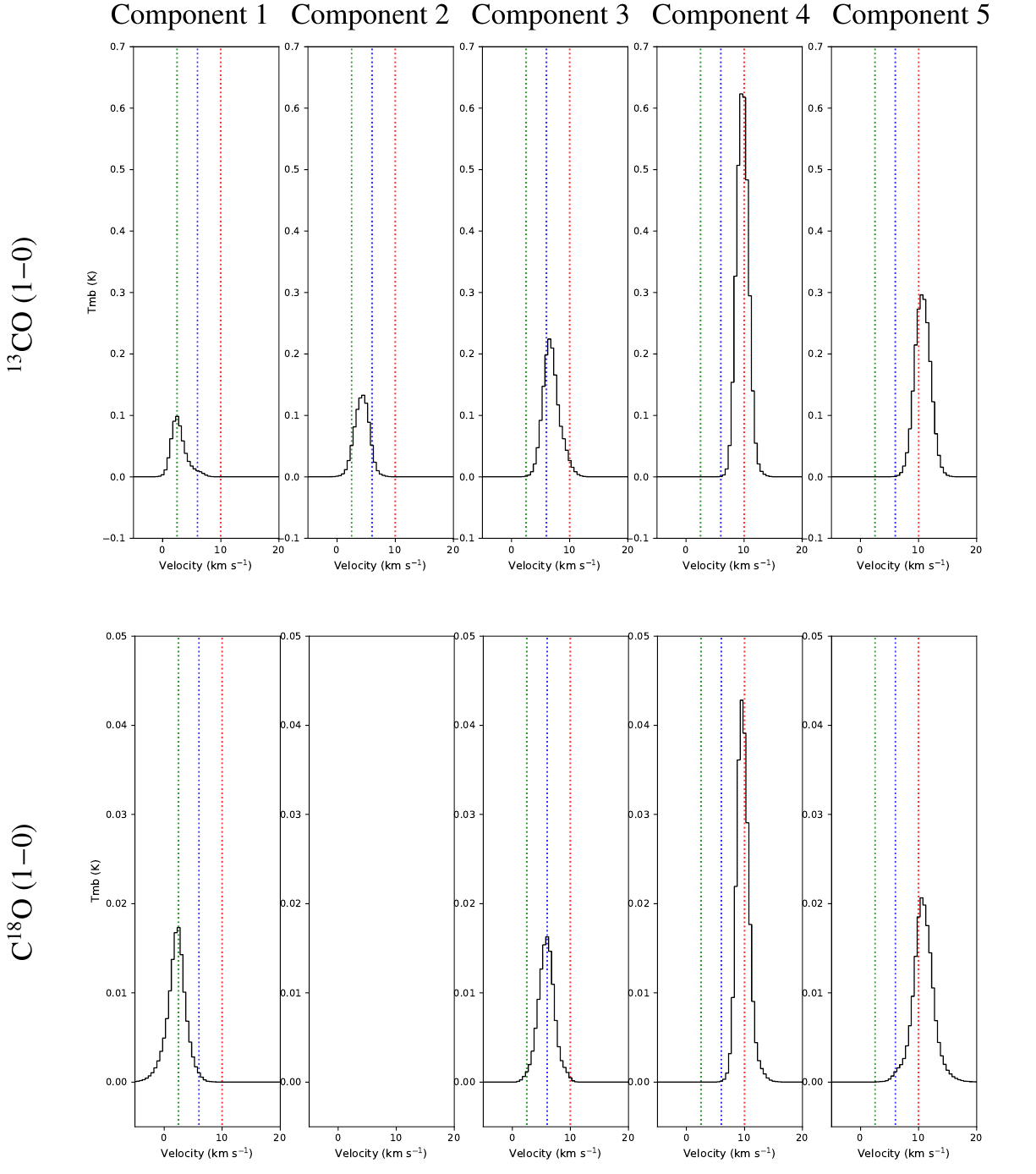}
    \caption{Mean spectrum of each Gaussian component fit by ROHSA for the
      $^{13}$CO (1$-$0) (top) and C$^{18}$O (bottom) (1$-$0) line. The
      color dotted lines show the systemic velocity of the three cloud
      layers.}
    \label{fig:mean-spectrum-ROHSA-gaussian-fit}
  \end{figure}}

\newcommand{\FigGaussMomentsCeO}{%
  \begin{figure*}
    \centering %
    \includegraphics[width=0.75\linewidth]{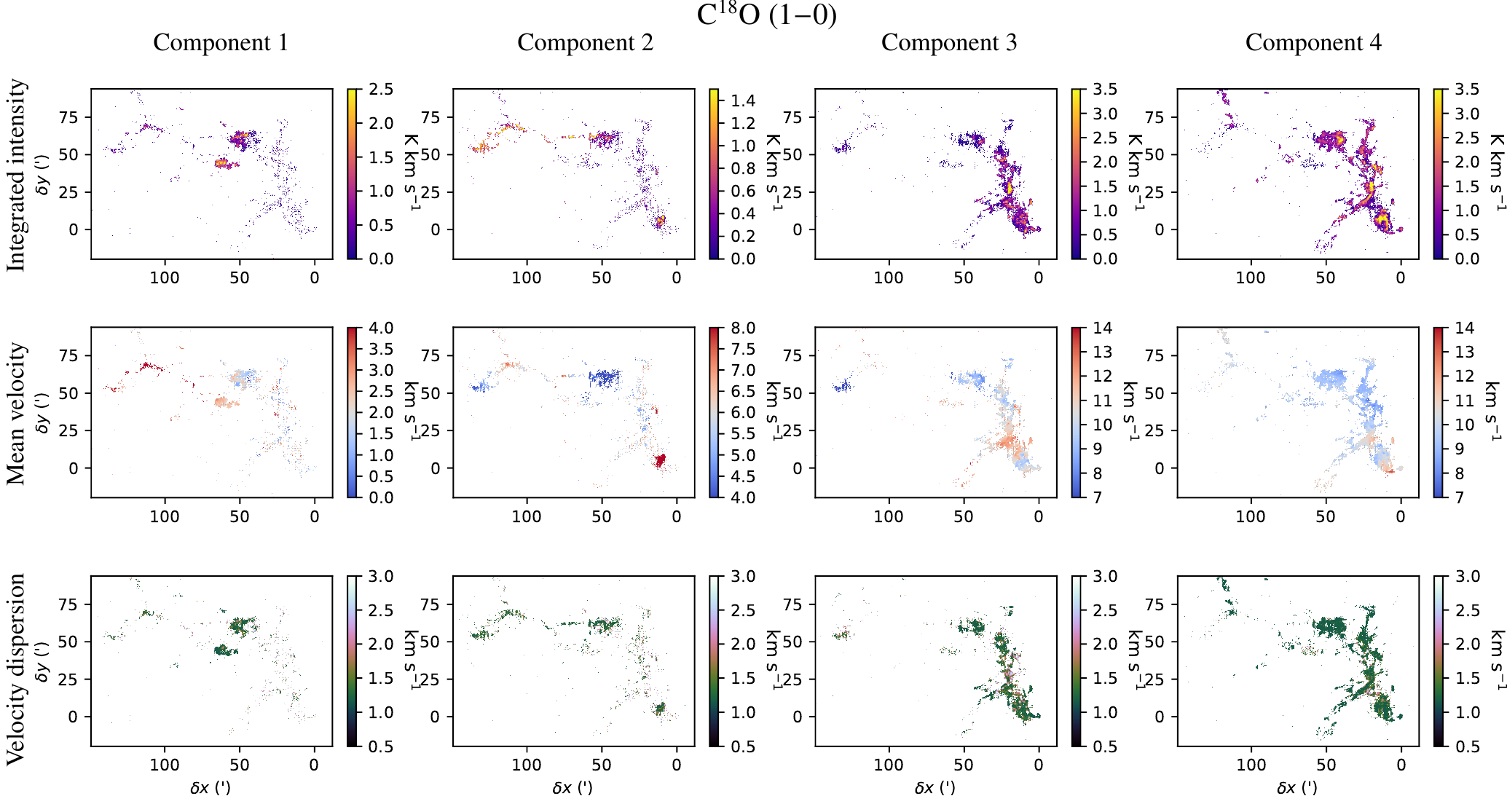}
    \caption{Zero-order (integrated intensity, top), first-order (mean
      velocity, middle), and second-order (velocity dispersion, bottom)
      moment maps of each of the four Gaussian components fit by ROHSA for the
      C$^{18}$O (1$-$0) line. Isolated pixels and pixels with a
      signal-to-noise ratio lower than 3 are blanked.}
    \label{fig:moment-maps:ROHSA-gaussian-fit-c18o}
  \end{figure*}}

\newcommand{\FigGaussMomentstCO}{%
  \begin{figure*}
    \centering %
    \includegraphics[width=0.56\linewidth]{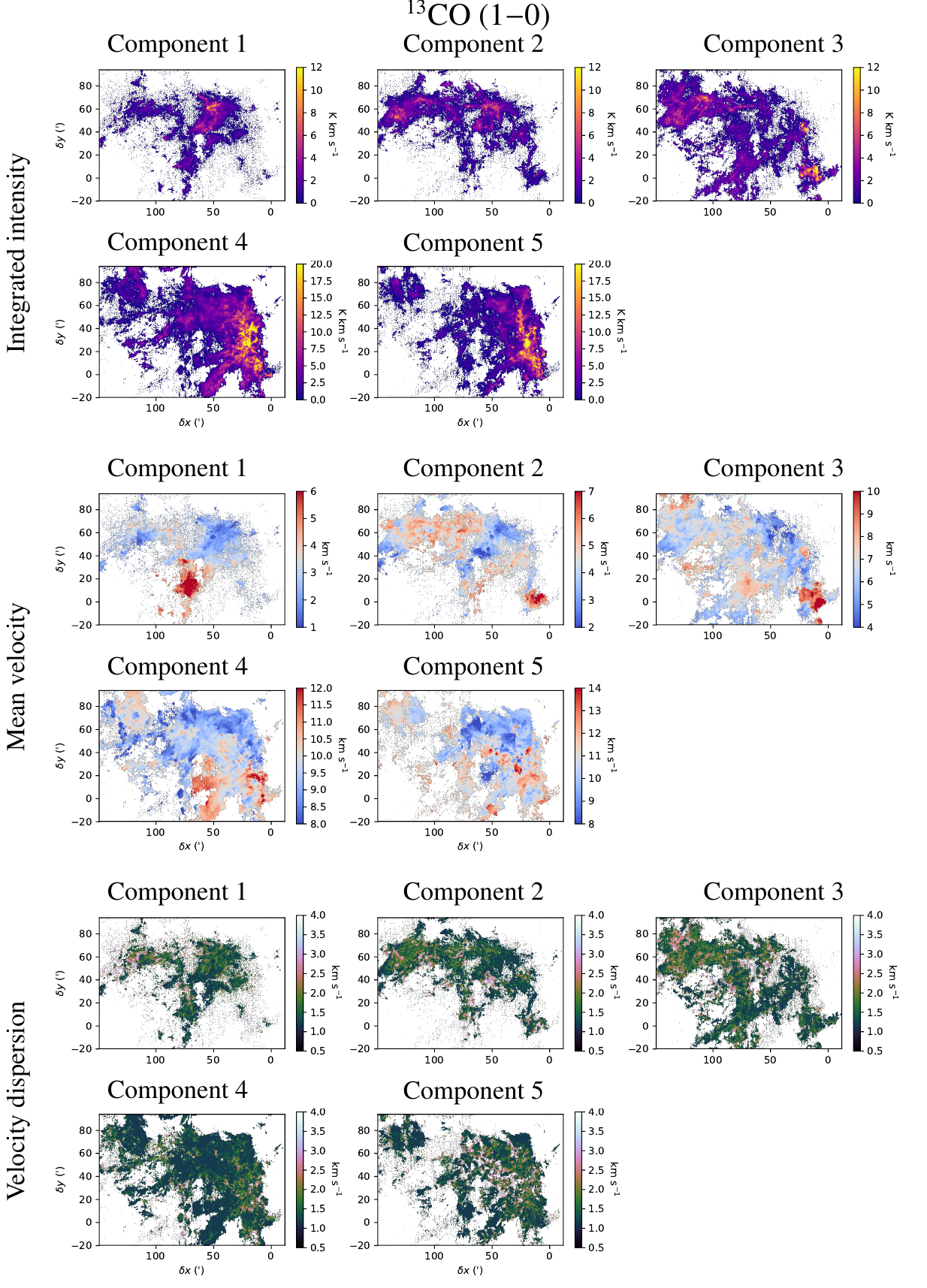}
    \caption{Zero-order (integrated intensity, top), first-order (mean
      velocity, middle), and second-order (velocity dispersion, bottom)
      moment maps of each of the five Gaussian components fit by ROHSA for the
      $^{13}$CO (1$-$0) line. Isolated pixels and pixels with a
      signal-to-noise ratio lower than 3 are blanked.}
    \label{fig:moment-maps:ROHSA-gaussian-fit-13co}
  \end{figure*}}

\newcommand{\FigMonteCarloNH}{%
  \setlength{\tabcolsep}{0pt}
  \begin{figure*}
    \centering %
    \includegraphics[width=\linewidth]{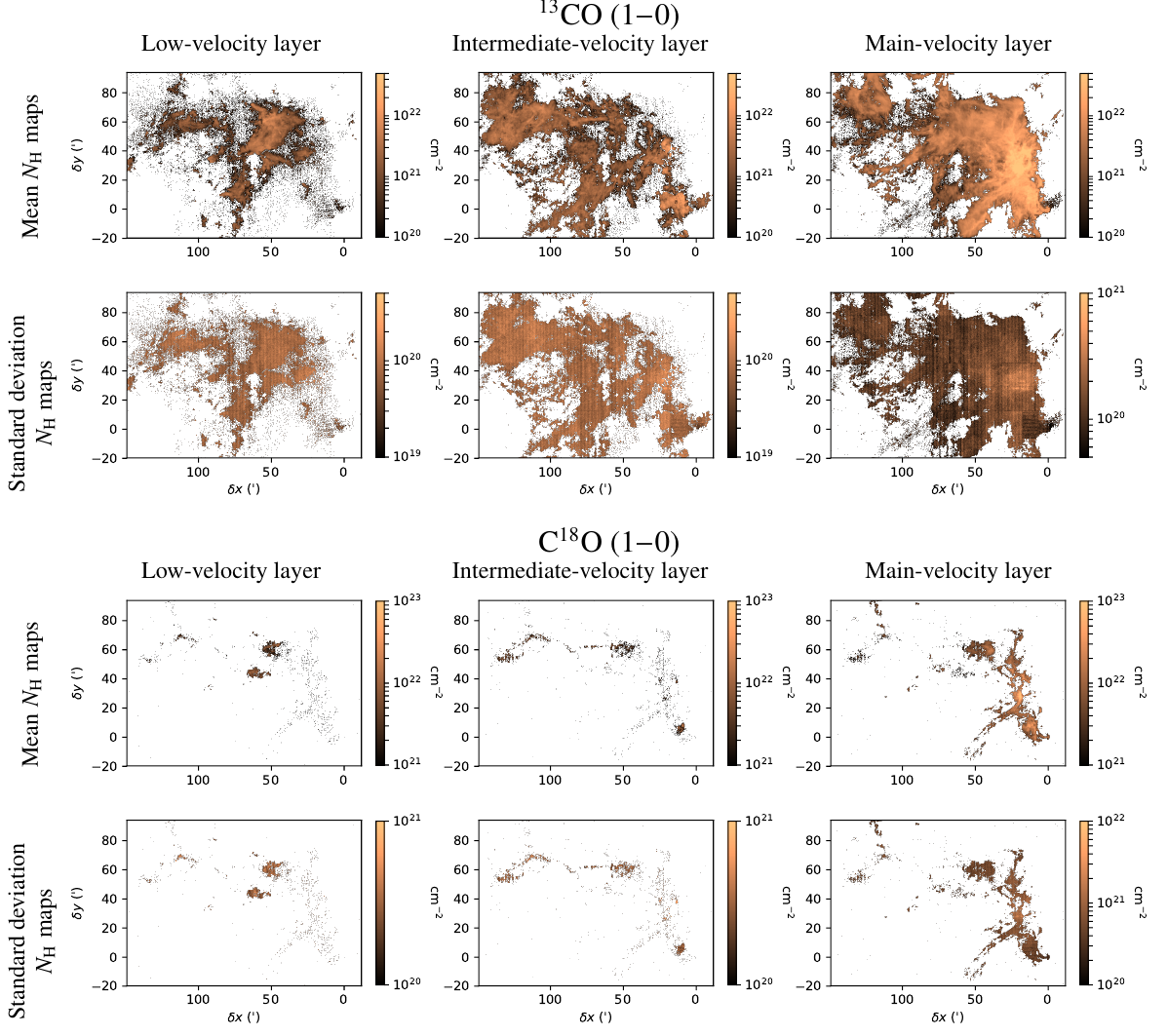}
    \caption{Mean and standard deviation on the $N_\mathrm{H}$ values from
      $^{13}$CO (1$-$0) (top) and C$^{18}$O (1$-$0) (bottom) implemented
      during the Monte-Carlo process for the three layers of the Orion B
      cloud, namely the low-velocity layer (left), the
      intermediate-velocity layer (middle), and the main-velocity layer
      (right) rebuilt by ROHSA. Isolated pixels and pixels with a
      signal-to-noise ratio lower than 3 are blanked.}
    \label{fig:NH-maps:monte-carlo}
  \end{figure*}}

\newcommand{\FigMonteCarloVelo}{%
  \setlength{\tabcolsep}{0pt}
  \begin{figure*}
    \centering %
    \includegraphics[width=\linewidth]{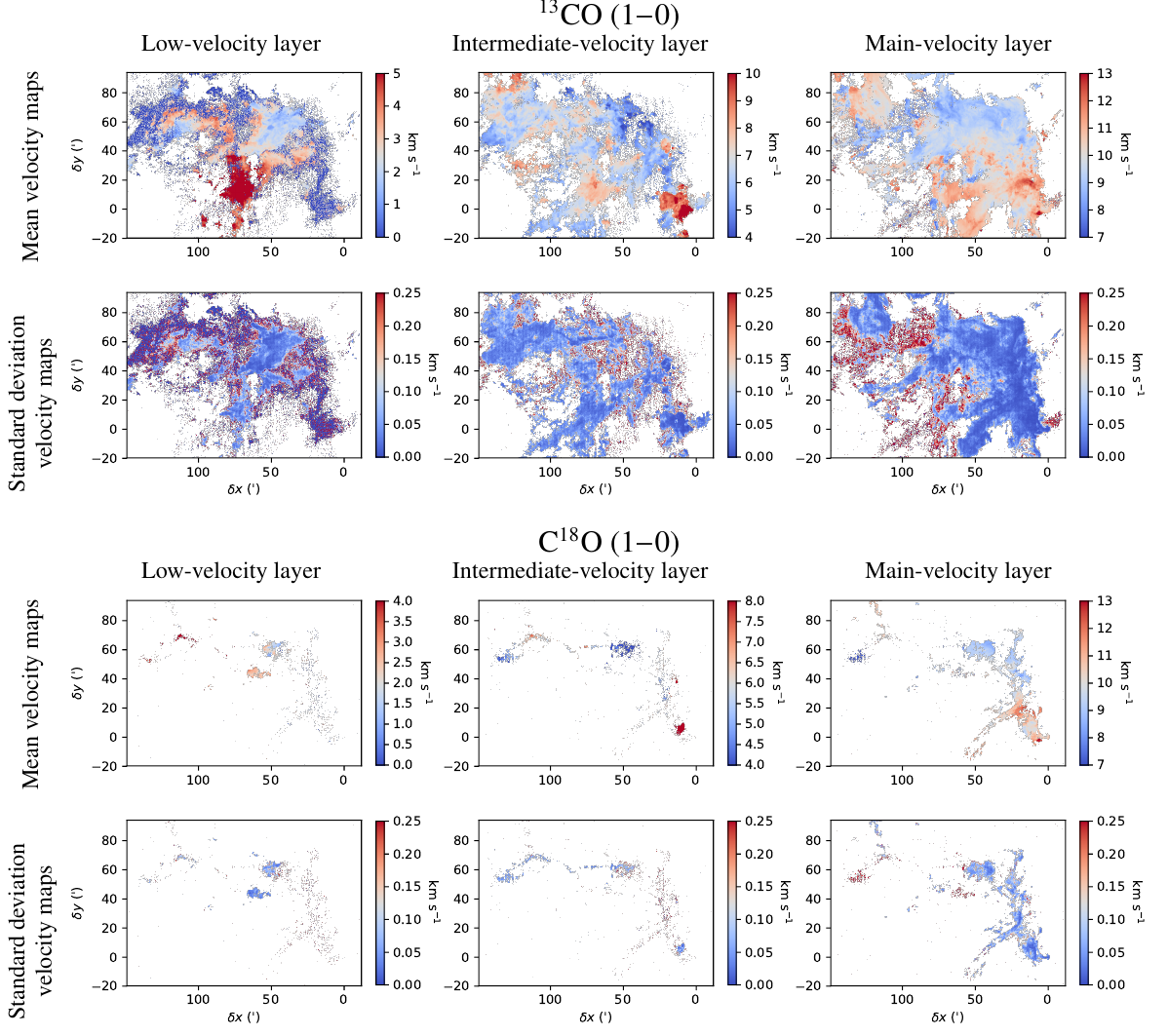}
    \caption{Mean and standard deviation on the centroid velocity values
      from $^{13}$CO (1$-$0) (top) and C$^{18}$O (1$-$0) (bottom)
      implemented during the Monte-Carlo process for the three layers of
      the Orion B cloud, namely the low-velocity layer (left), the
      intermediate-velocity layer (middle), and the main-velocity layer
      (right) rebuilt by ROHSA. Isolated pixels and pixels with a
      signal-to-noise ratio lower than 3 are blanked.}
    \label{fig:moment1-maps:monte-carlo}
  \end{figure*}}

\newcommand{\FigMomentsCloak}{%
  \begin{figure*}
    \centering %
    \includegraphics[width=\linewidth]{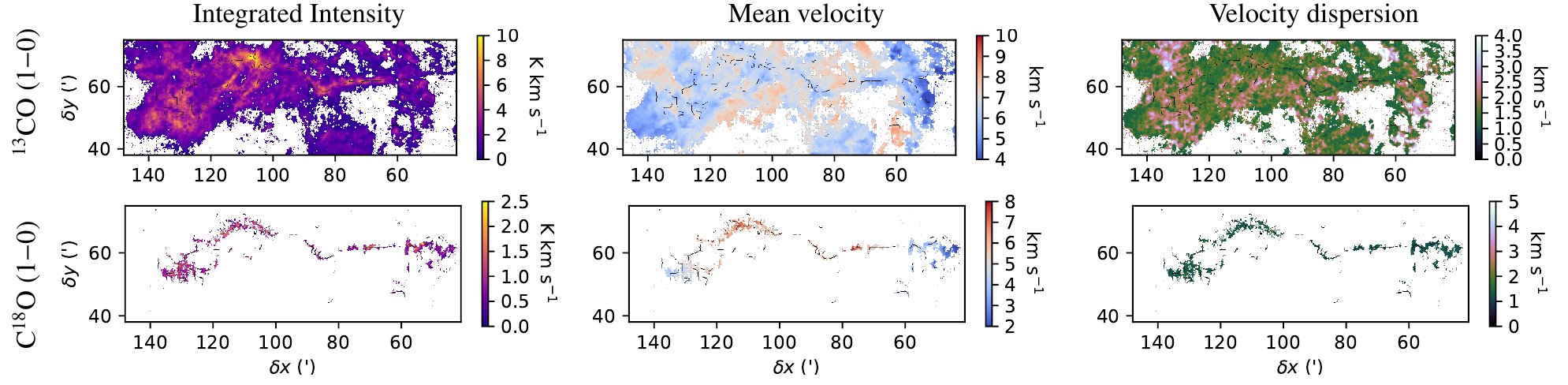}
    \caption{Zero-order (integrated intensity, left), first-order (mean
      velocity, middle), and second-order (velocity dispersion, right)
      moment maps in the Cloak area from the intermediate-velocity layer,
      from the $^{13}$CO (1$-$0) (top) and C$^{18}$O (1$-$0) (bottom)
      emission reconstructed by ROHSA. The black lines show the filamentary
      structures extracted from C$^{18}$O (1$-$0).}
    \label{fig:moment-maps:cloak}
  \end{figure*}}

\newcommand{\FigAngGradientsCloak}{%
  \begin{figure*}
    \centering %
    \includegraphics[width=\linewidth]{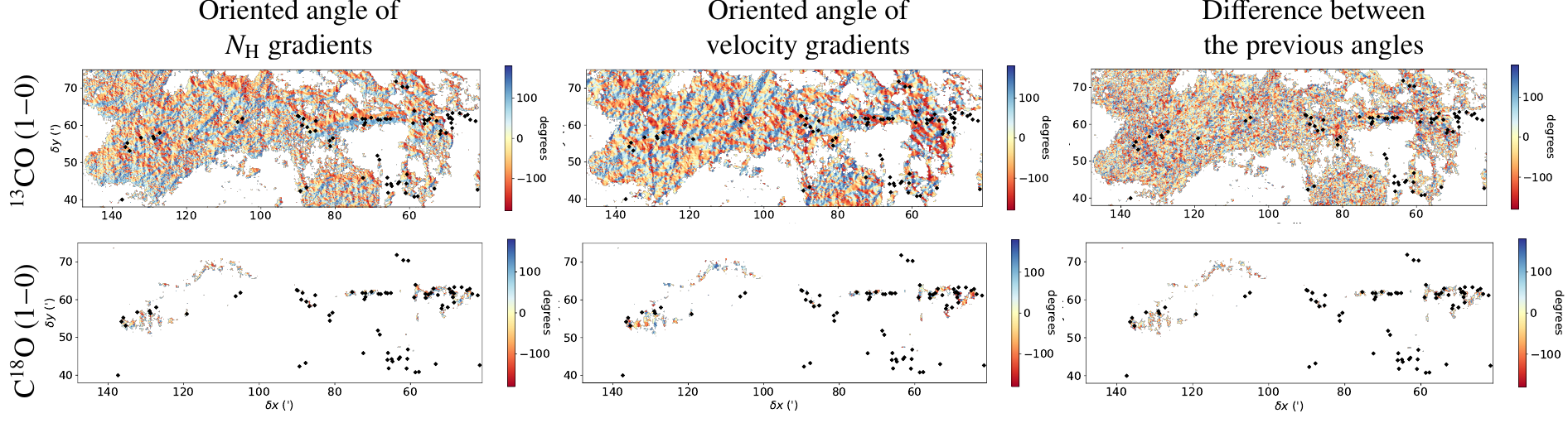}
    \caption{Maps of the column density gradient oriented angles (top) and
      the velocity gradient oriented angles (middle) in the Cloak area from
      the intermediate-velocity from the $^{13}$CO (1$-$0) (left) and the
      C$^{18}$O (1$-$0) (right) emission. Bottom: maps of the relative
      orientation between the oriented angles of the column density and the
      velocity gradients. Oriented angles are defined north from east. The
      black lines show the filamentary structure and the diamond dots show
      the dense cores.}
    \label{fig:oriented-angle-maps:cloak}
  \end{figure*}}
  
\newcommand{\FigDivMapsCloak}{%
  \begin{figure*}
    \centering %
    \includegraphics[width=\linewidth]{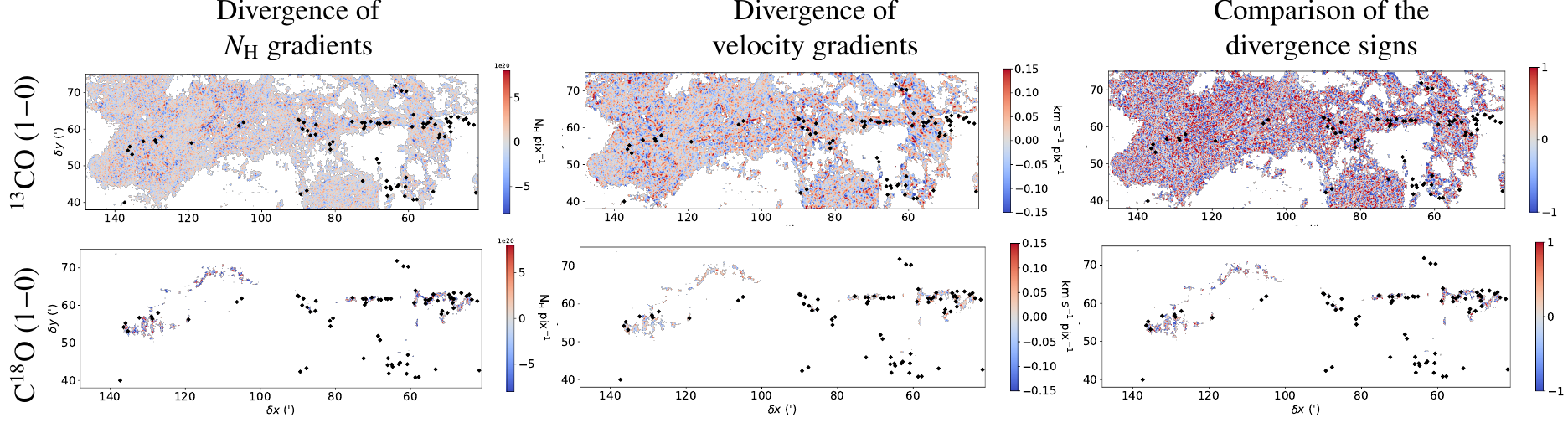}
    \caption{Divergence maps of the column density gradients in
      N$_\mathrm{H}$ pix$^{-1}$ (left) and the velocity gradients in km
      s$^{-1}$ pix$^{-1}$ (middle) for the intermediate-velocity layer of
      the Cloak area from the $^{13}$CO (1$-$0) (top) and the C$^{18}$O
      (1$-$0) (bottom) emission. Right: comparison maps of the sign between
      the velocity gradient divergence and the column density gradient
      divergence. The values -1 (blue) or 1 (red) show the pixels where
      velocity and column density gradients both converge or both diverge. The
      value 0 (gray) indicates regions where the sign of at least one
      gradient is ill-defined because its modulus is close to zero. The
      black lines show the filamentary structure and the diamond dots show
      the dense cores identified by the Herschel Gould Belt Survey
      \citep{Konyves2020}.}
    \label{fig:divergence-maps:cloak}
  \end{figure*}}

\newcommand{\FigMomentsBnineLow}{%
  \begin{figure}
    \centering %
    \includegraphics[width=\linewidth]{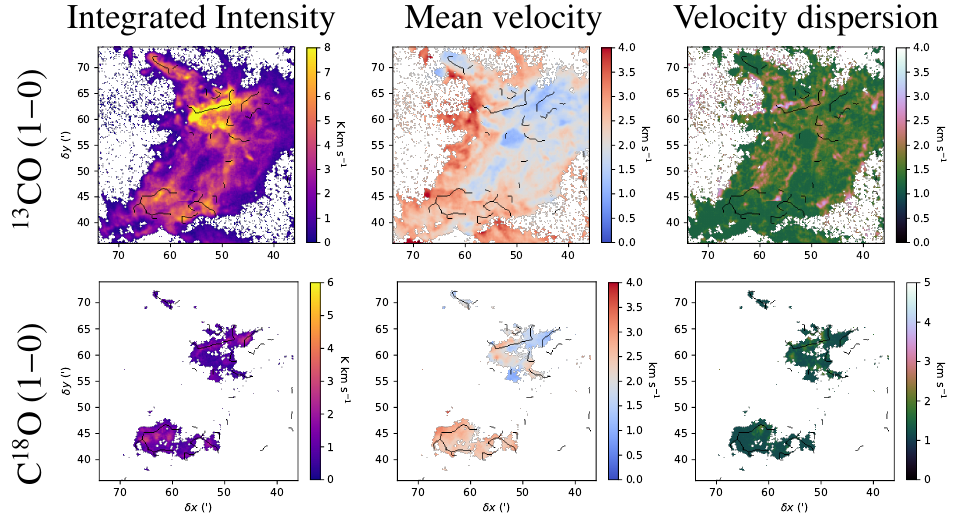}
    \caption{Moment maps for the low-velocity layer of the Orion B9 area.
      The figure layout is identical to Fig.~\ref{fig:moment-maps:cloak}.}
    \label{fig:moment-maps:orion-b9-low}
  \end{figure}}

\newcommand{\FigAngGradientsBnineLow}{%
  \begin{figure}
    \centering %
    \includegraphics[width=\linewidth]{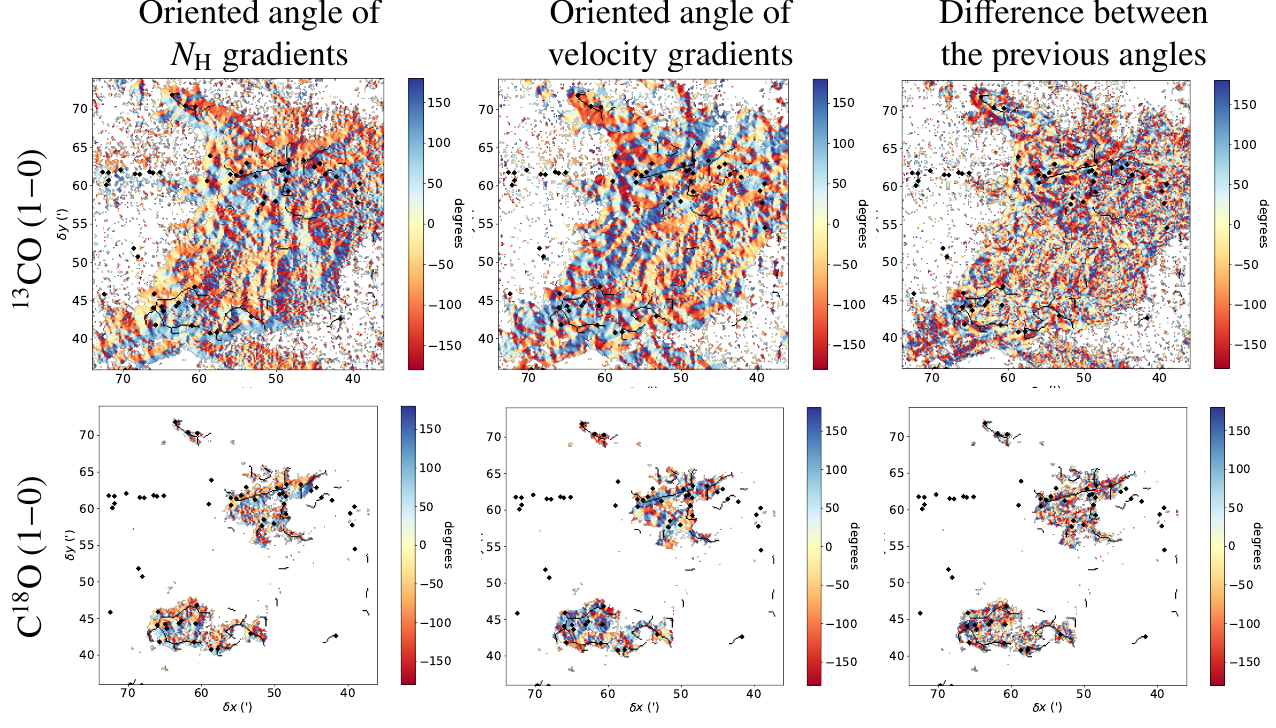}
    \caption{Maps of oriented angles of the gradients of the column density
      and centroid velocity gradients for the low-velocity layer of the
      Orion B9 area.  The figure layout is identical to
      Fig.~\ref{fig:oriented-angle-maps:cloak}}
    \label{fig:oriented-angle-maps:orion-b9-low}
  \end{figure}}

\newcommand{\FigDivMapsBnineLow}{%
  \begin{figure}
    \centering %
    \includegraphics[width=\linewidth]{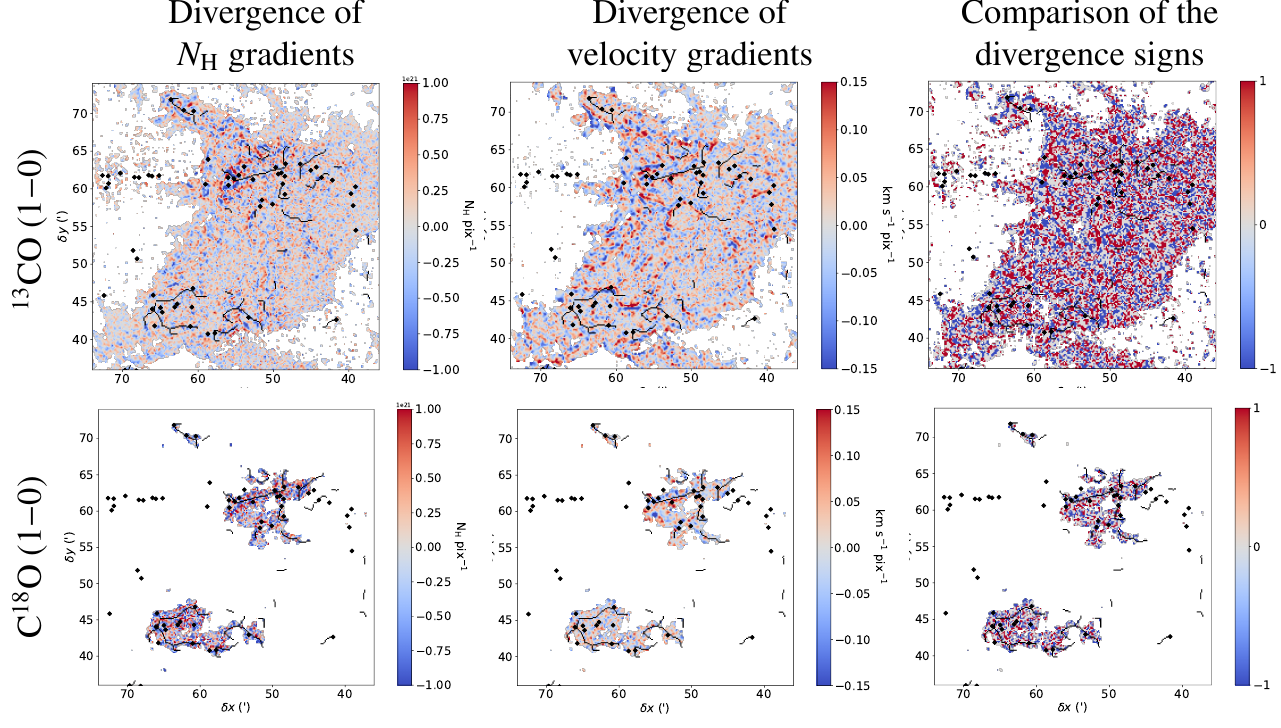}
    \caption{Divergence maps for the low-velocity layer of the Orion B9
      area.  The figure layout is identical to
      Fig.~\ref{fig:oriented-angle-maps:cloak}}
    \label{fig:divergence-maps:orion-b9-low}
  \end{figure}}

\newcommand{\FigMomentsBnineMain}{%
  \begin{figure}
    \centering %
    \includegraphics[width=\linewidth]{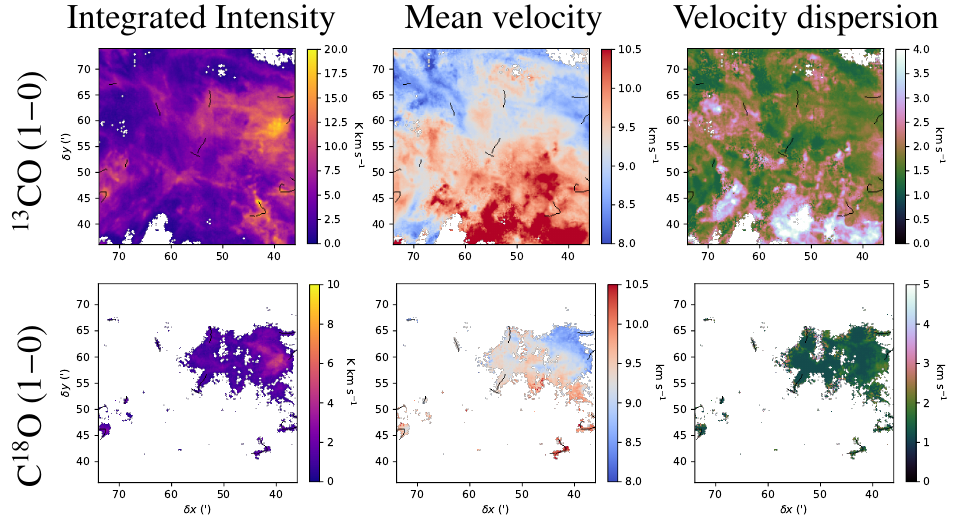}
    \caption{Moment maps for the main-velocity layer of the Orion B9
      area. The figure layout is identical to
      Fig.~\ref{fig:moment-maps:cloak}.}
    \label{fig:moment-maps:orion-b9-main}
  \end{figure}}

\newcommand{\FigAngGradientsBnineMain}{%
  \begin{figure}
    \centering %
    \includegraphics[width=\linewidth]{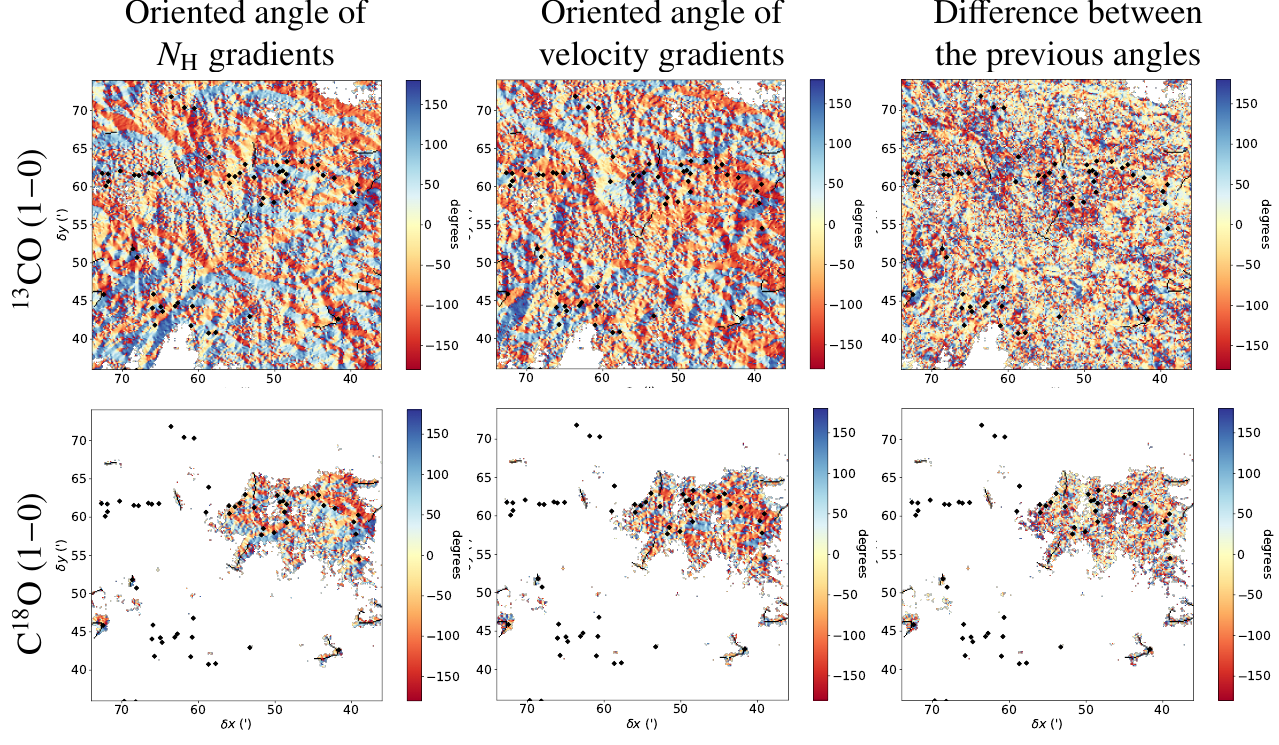}
    \caption{Maps of oriented angles of the gradients of the column density
      and centroid velocity gradients for the main-velocity layer of the
      Orion B9 area.  The figure layout is identical to
      Fig.~\ref{fig:oriented-angle-maps:cloak}}
    \label{fig:oriented-angle-maps:orion-b9-main}
  \end{figure}}

\newcommand{\FigDivMapsBnineMain}{%
  \begin{figure}
    \centering %
    \includegraphics[width=\linewidth]{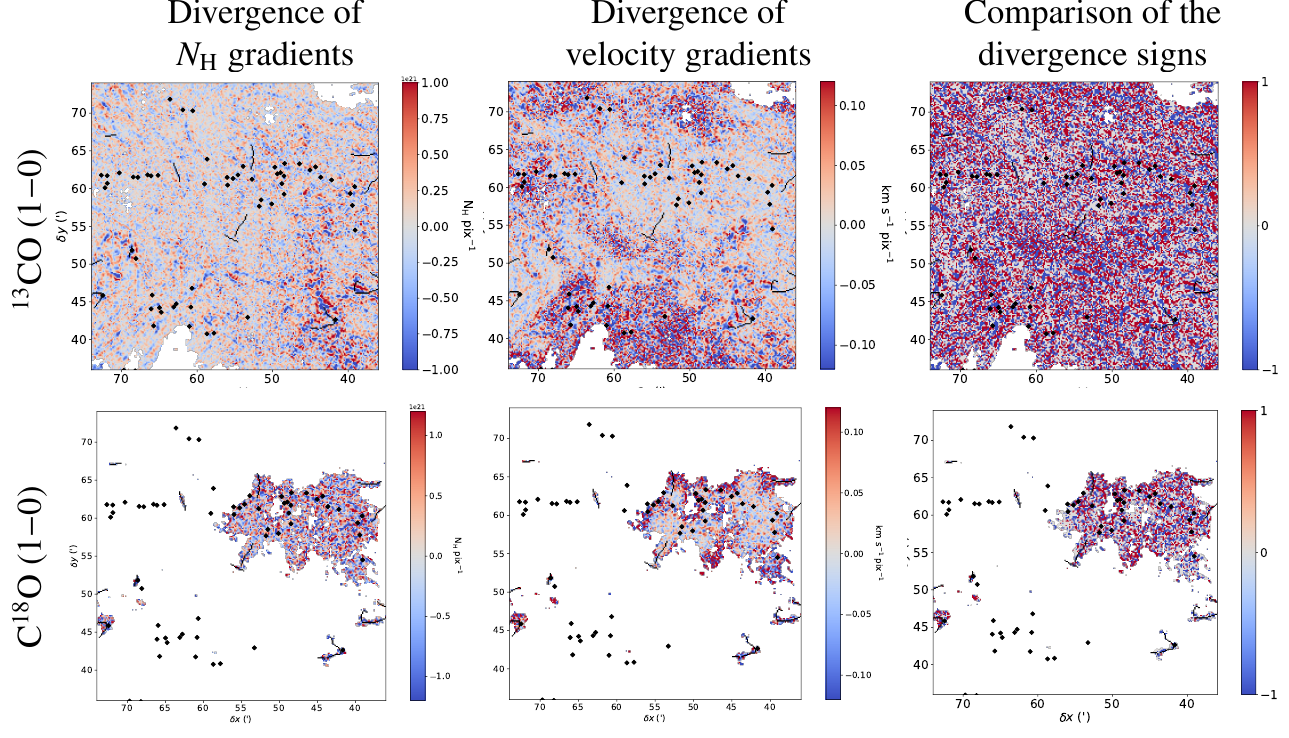}
    \caption{Divergence maps for the main-velocity layer of the Orion B9
      area.  The figure layout is identical to
      Fig.~\ref{fig:oriented-angle-maps:cloak}}
    \label{fig:divergence-maps:orion-b9-main}
  \end{figure}}

\newcommand{\FigMomentsHummingbird}{%
  \begin{figure}
    \centering %
    \includegraphics[width=\linewidth]{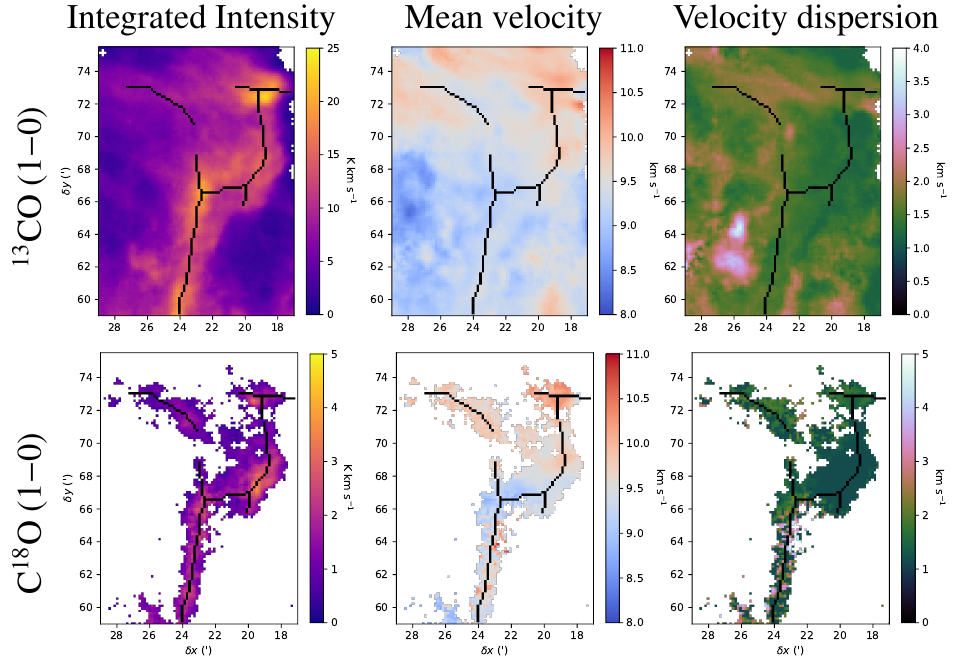}
    \caption{Moment maps for the main-velocity layer of the Hummingbird
      area. The figure layout is identical to
      Fig.~\ref{fig:moment-maps:cloak}.}
    \label{fig:moment-maps:hummingbird}
  \end{figure}}

\newcommand{\FigAngGradientsHummingbird}{%
  \begin{figure}
    \centering %
    \includegraphics[width=\linewidth]{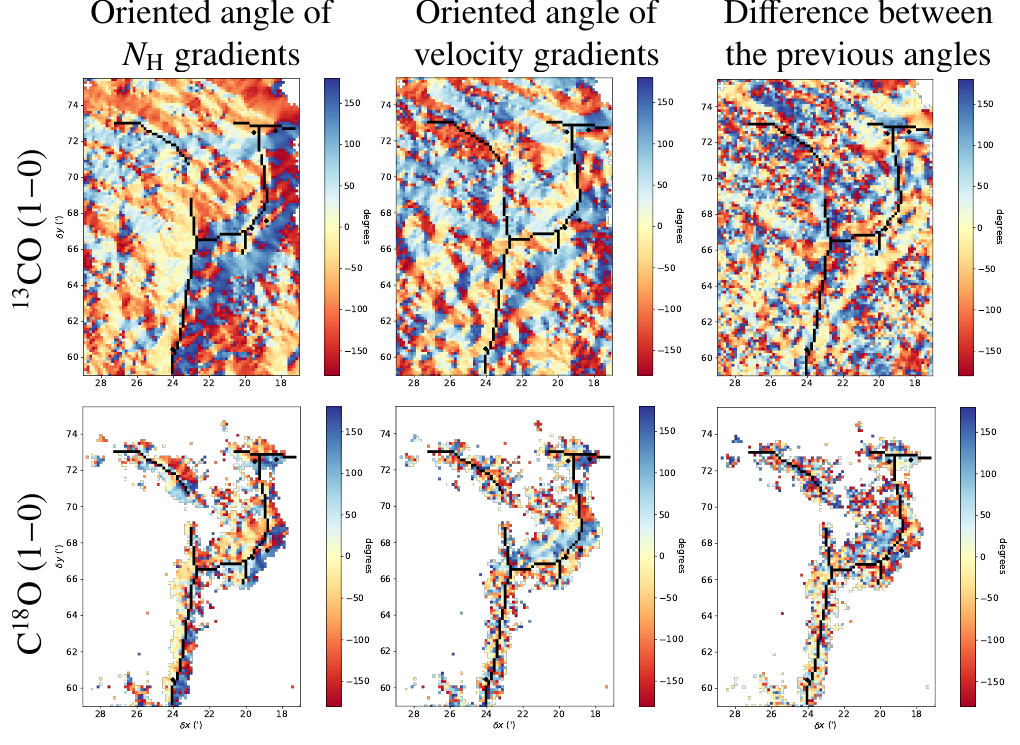}
    \caption{Maps of oriented angles of the gradients of the column density
      and centroid velocity gradients for the main-velocity layer of the
      Hummingbird area.  The figure layout is identical to
      Fig.~\ref{fig:oriented-angle-maps:cloak}}
    \label{fig:oriented-angle-maps:hummingbird}
  \end{figure}}

\newcommand{\FigDivMapsHummingbird}{%
  \begin{figure}
    \centering %
    \includegraphics[width=\linewidth]{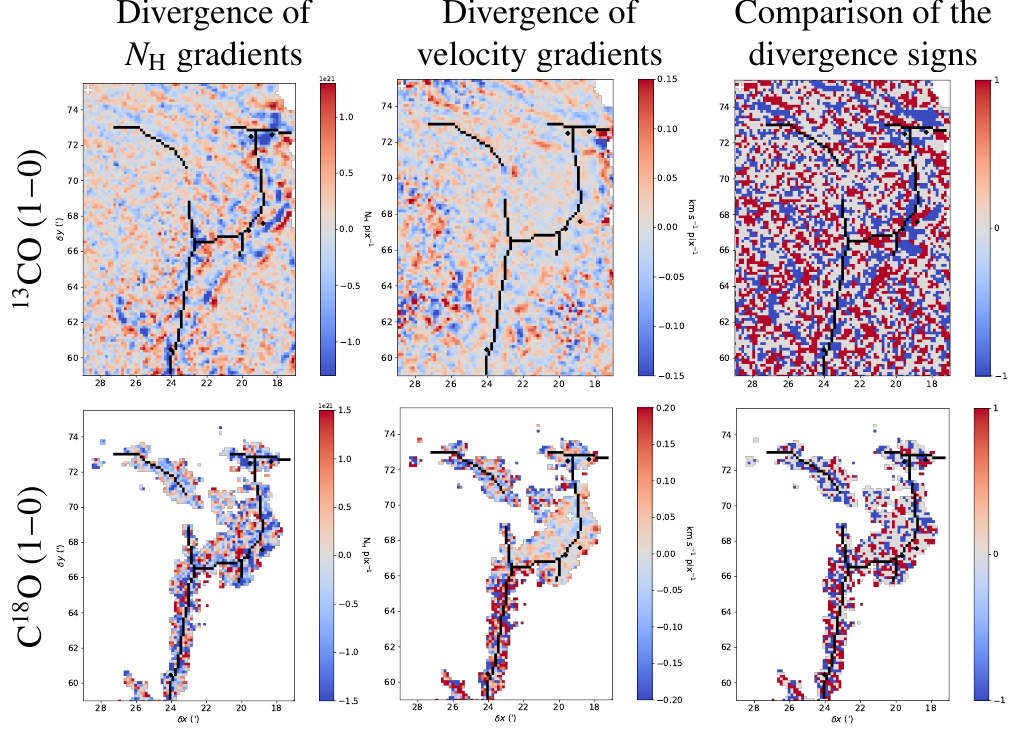}
    \caption{Divergence maps for the main-velocity layer of the Hummingbird area.  The figure layout is identical to
      Fig.~\ref{fig:oriented-angle-maps:cloak}}
    \label{fig:divergence-maps:hummingbird}
  \end{figure}}

\newcommand{\FigMomentsNGC}{%
  \setlength{\tabcolsep}{0pt}
  \begin{figure}
    \centering %
    \includegraphics[width=\linewidth]{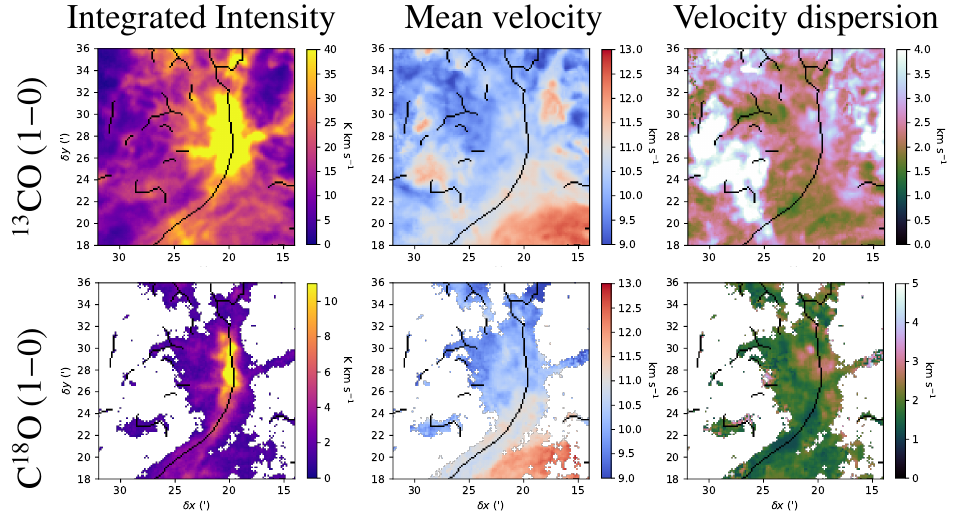}
    \caption{Moment maps for the main-velocity layer of the NGC\,2024
      area. The figure layout is identical to
      Fig.~\ref{fig:moment-maps:cloak}.}
    \label{fig:moment-maps:ngc2024}
  \end{figure}}
  
\newcommand{\FigAngGradientsNGC}{%
  \begin{figure}
    \centering %
    \includegraphics[width=\linewidth]{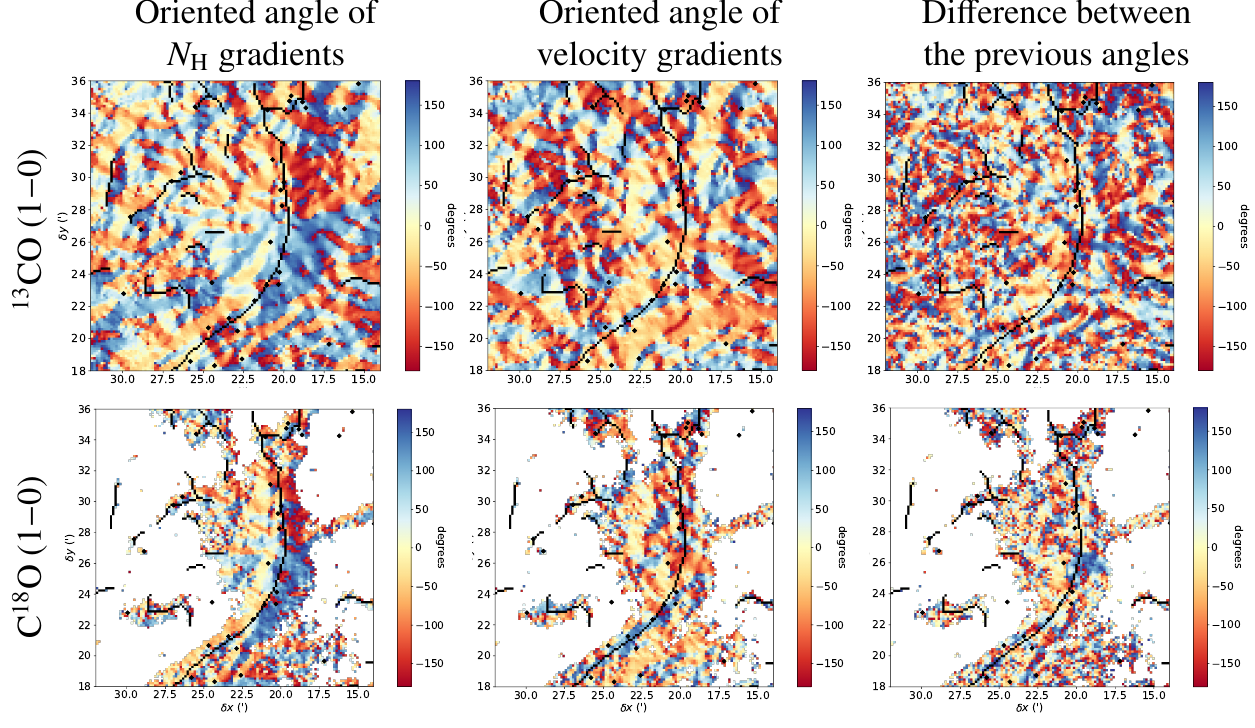}
    \caption{Maps of oriented angles of the gradients of the column density
      and centroid velocity gradients for the main-velocity layer of the
      NGC\,2024 area.  The figure layout is identical to
      Fig.~\ref{fig:oriented-angle-maps:cloak}}
    \label{fig:oriented-angle-maps:ngc2024}
  \end{figure}}

\newcommand{\FigDivMapsNGC}{%
  \begin{figure}
    \centering %
    \includegraphics[width=\linewidth]{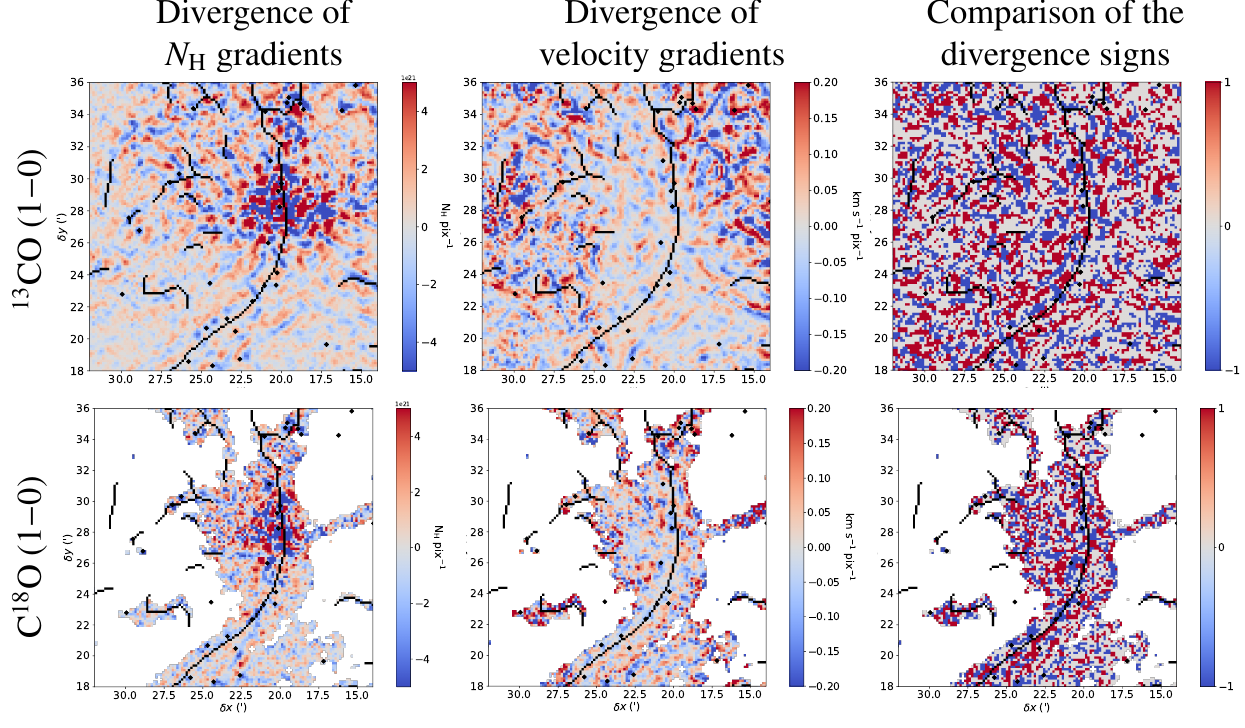}
    \caption{Divergence maps for the main-velocity layer of the NGC\,2024
      area.  The figure layout is identical to
      Fig.~\ref{fig:oriented-angle-maps:cloak}}
    \label{fig:divergence-maps:ngc2024}
  \end{figure}}

\newcommand{\FigMomentsFlame}{%
  \begin{figure}
    \centering %
    \includegraphics[width=\linewidth]{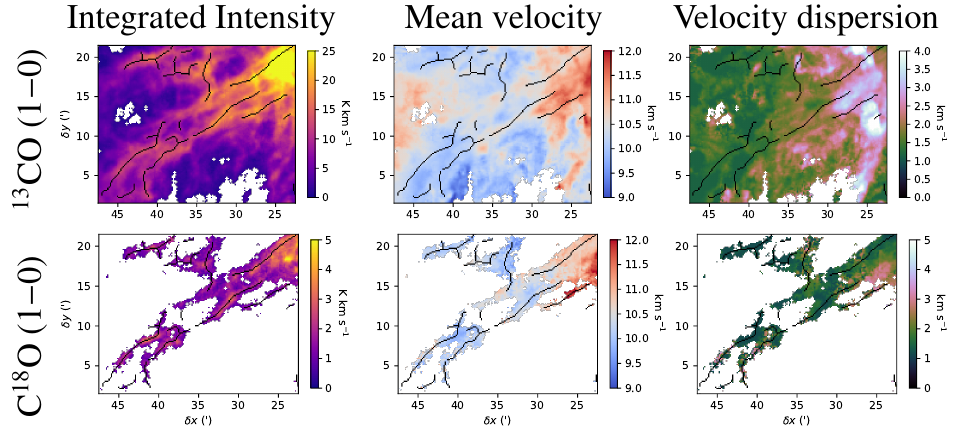}
    \caption{Moment maps for the main-velocity layer of the Flame Filament
      area. The figure layout is identical to
      Fig.~\ref{fig:moment-maps:cloak}.}
    \label{fig:moment-maps:flame}
  \end{figure}}

\newcommand{\FigAngGradientsFlame}{%
  \begin{figure}
    \centering %
    \includegraphics[width=\linewidth]{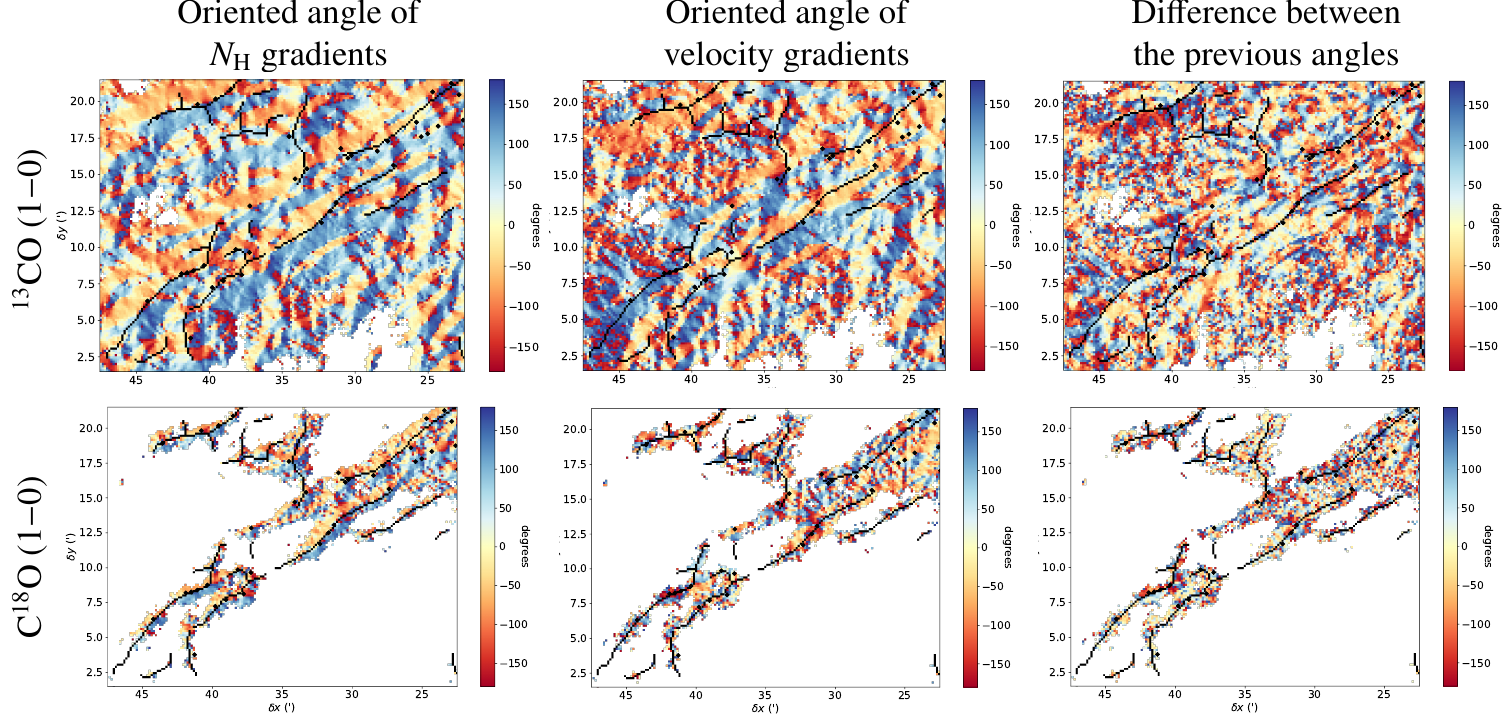}
    \caption{Maps of oriented angles of the gradients of the column density
      and centroid velocity gradients for the main-velocity layer of the
      Flame Filament area.  The figure layout is identical to
      Fig.~\ref{fig:oriented-angle-maps:cloak}}
    \label{fig:oriented-angle-maps:flame}
  \end{figure}}

\newcommand{\FigDivMapsFlame}{%
  \begin{figure}
    \centering %
    \includegraphics[width=\linewidth]{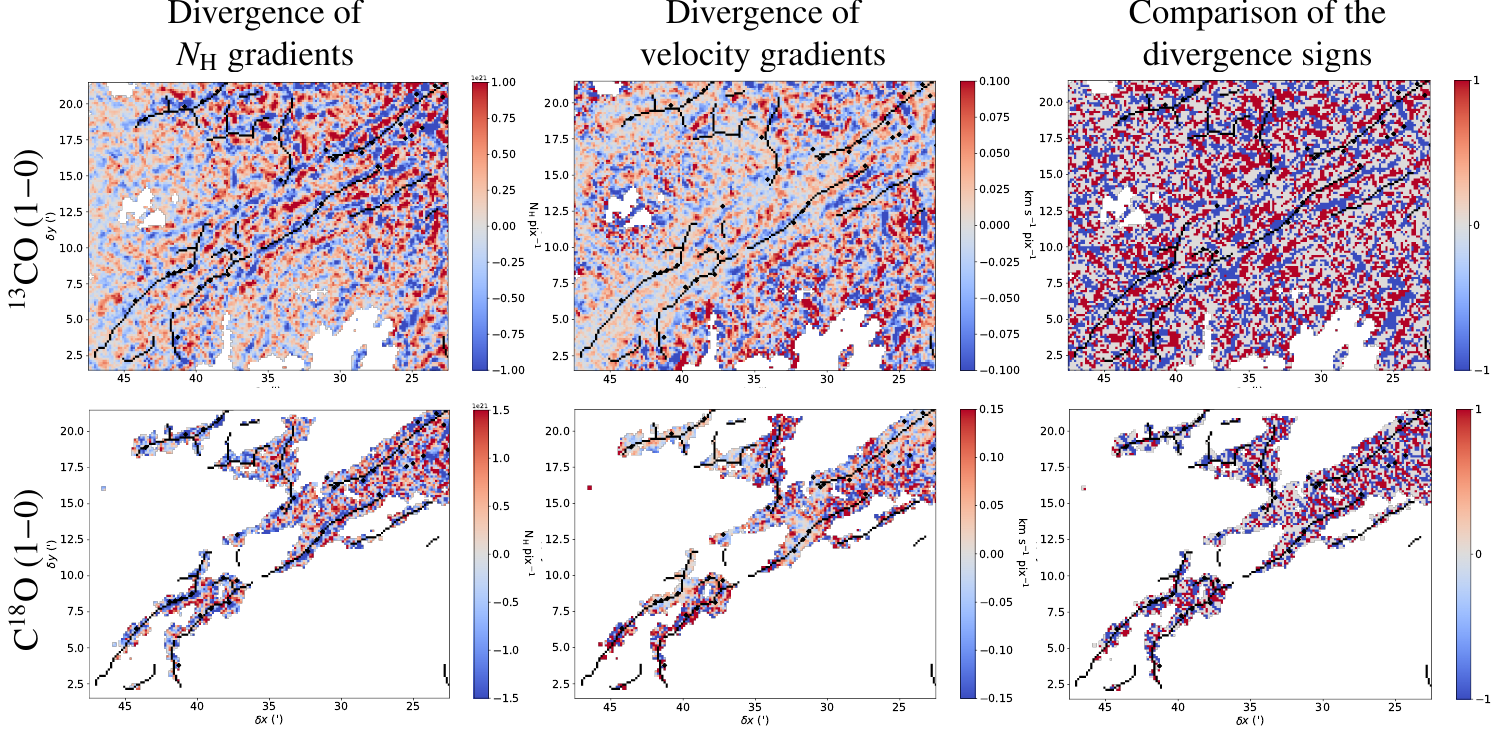}
    \caption{Divergence maps for the main-velocity layer of the Flame
      Filament area.  The figure layout is identical to
      Fig.~\ref{fig:oriented-angle-maps:cloak}}
    \label{fig:divergence-maps:flame}
  \end{figure}}

\newcommand{\FigMomentsIntermediate}{%
  \setlength{\tabcolsep}{0pt}
  \begin{figure}
    \centering %
    \includegraphics[width=\linewidth]{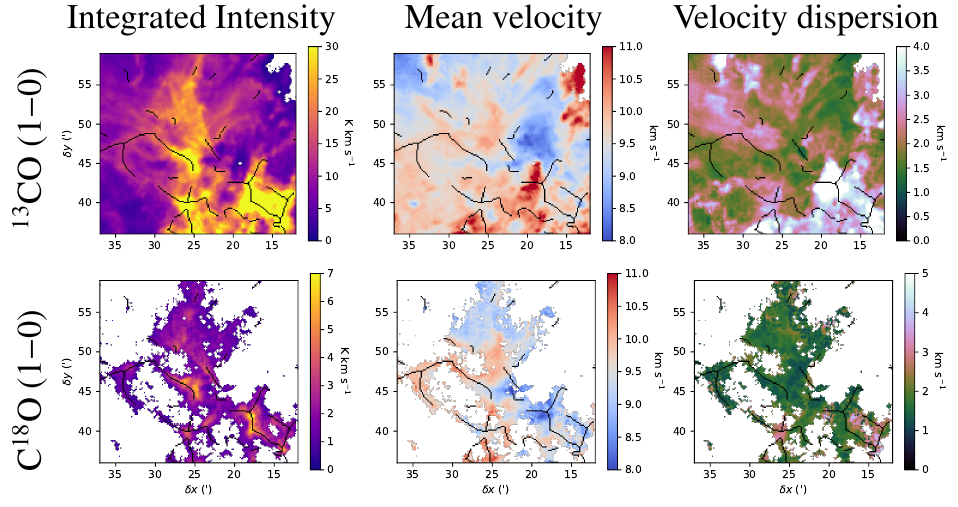}
    \caption{Moment maps for the main-velocity layer of the Intermediate
      area. The figure layout is identical to
      Fig.~\ref{fig:moment-maps:cloak}.}
    \label{fig:moment-maps:intermediate-area}
  \end{figure}}
  
\newcommand{\FigAngGradientsIntermediate}{%
  \begin{figure}
    \centering %
    \includegraphics[width=\linewidth]{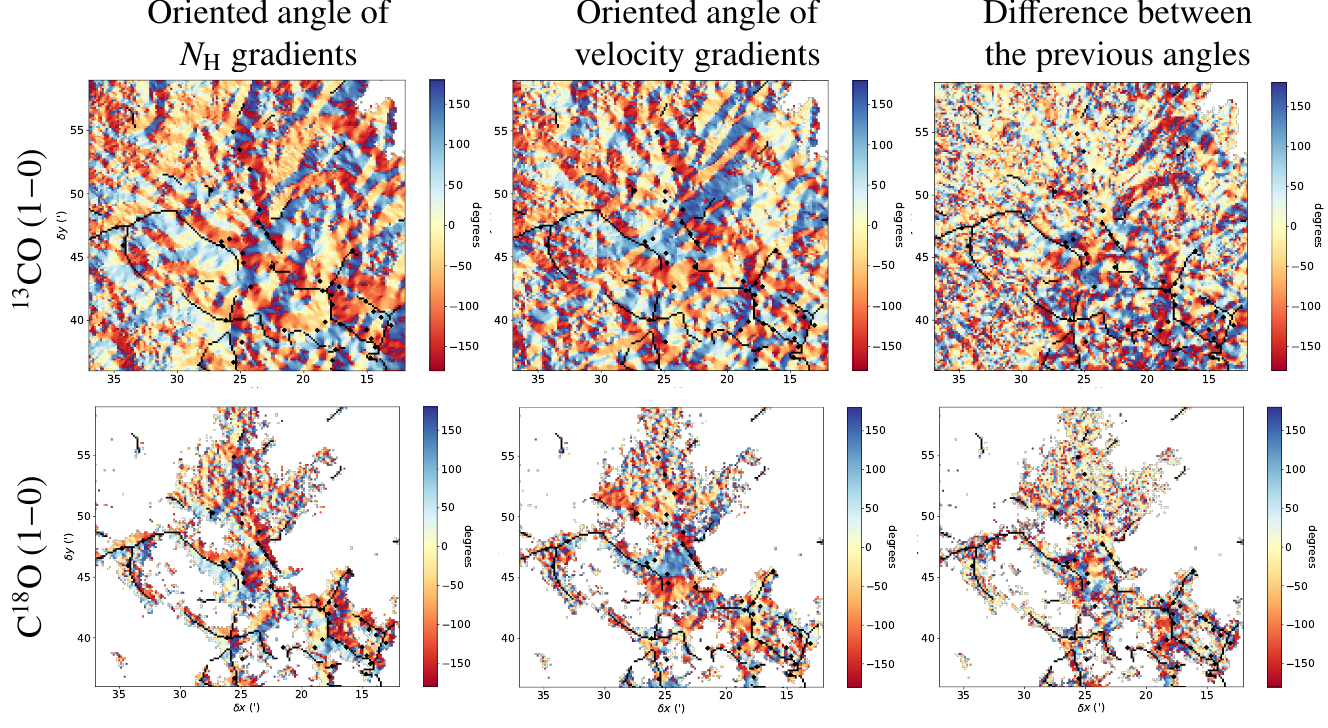}
    \caption{Maps of oriented angles of the gradients of the column density
      and centroid velocity gradients for the main-velocity layer of the
      Intermediate area.  The figure layout is identical to
      Fig.~\ref{fig:oriented-angle-maps:cloak}}
    \label{fig:oriented-angle-maps:intermediate}
  \end{figure}}

\newcommand{\FigDivMapsIntermediate}{%
  \begin{figure}
    \centering %
    \includegraphics[width=\linewidth]{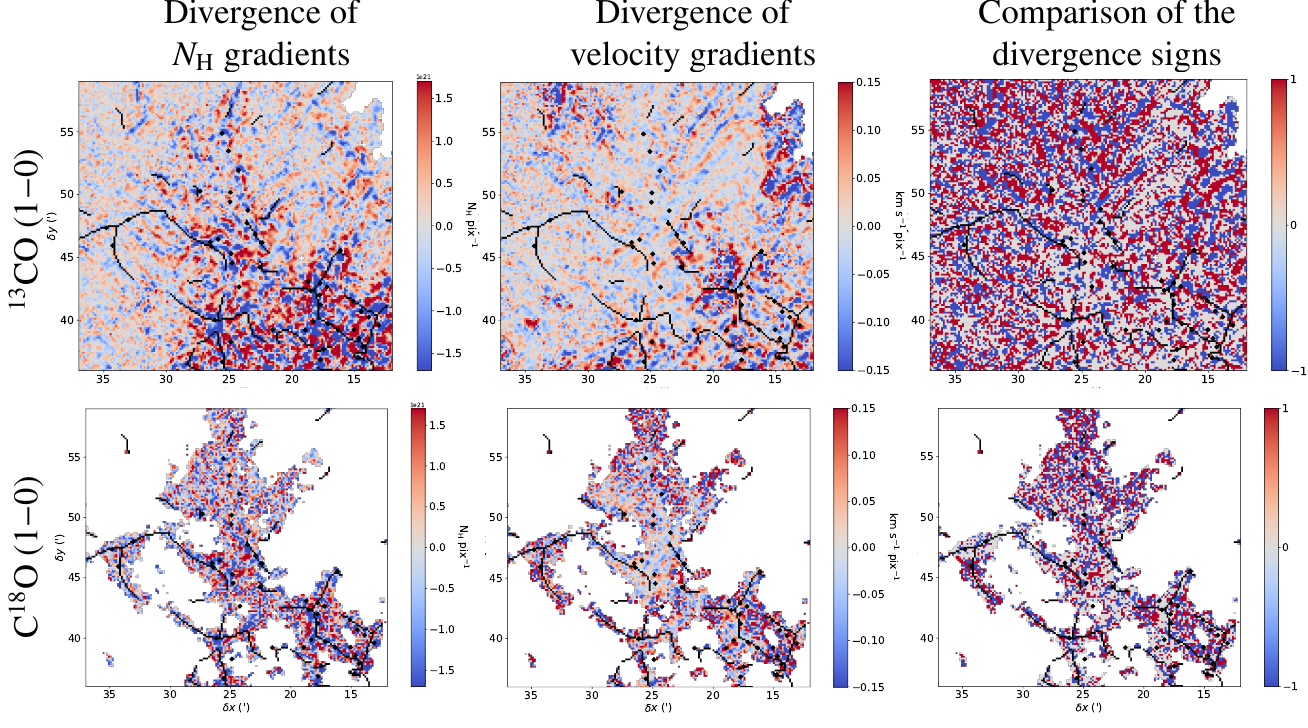}
    \caption{Divergence maps for the main-velocity layer of the
      Intermediate area.  The figure layout is identical to
      Fig.~\ref{fig:oriented-angle-maps:cloak}}
    \label{fig:divergence-maps:intermediate}
  \end{figure}}

\newcommand{\FigMomentsHorsehead}{%
  \begin{figure}
    \centering %
    \includegraphics[width=\linewidth]{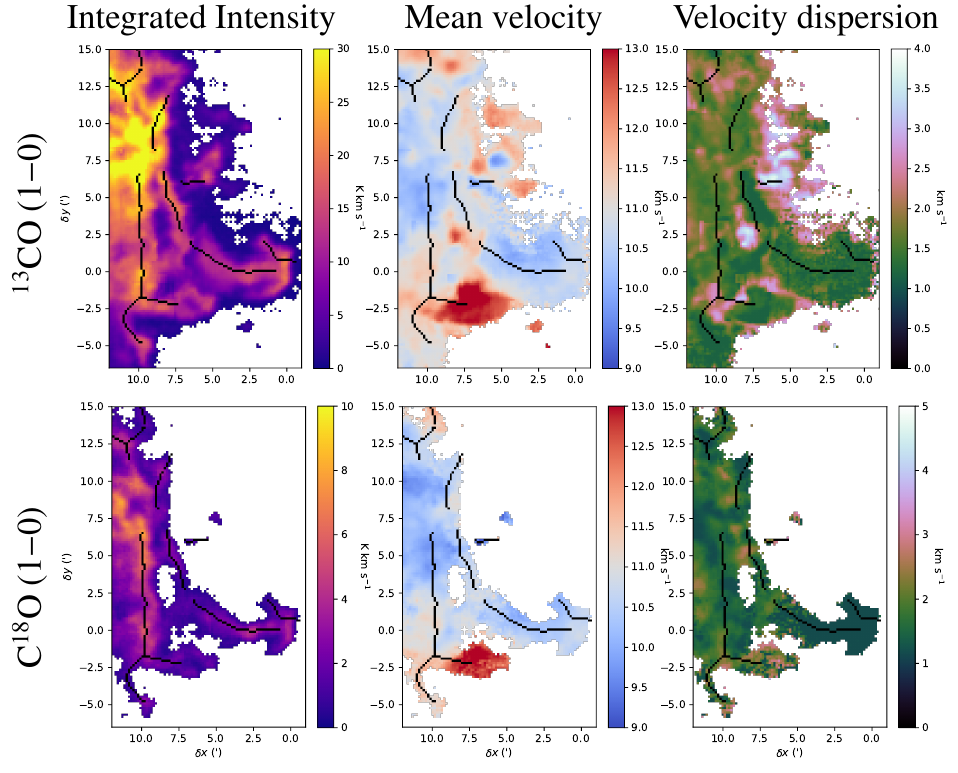}
    \caption{Moment maps for the main-velocity layer of the Horsehead
      Nebula area. The figure layout is identical to
      Fig.~\ref{fig:moment-maps:cloak}.}
    \label{fig:moment-maps:horsehead}
  \end{figure}}

\newcommand{\FigAngGradientsHorsehead}{%
  \begin{figure}
    \centering %
    \includegraphics[width=\linewidth]{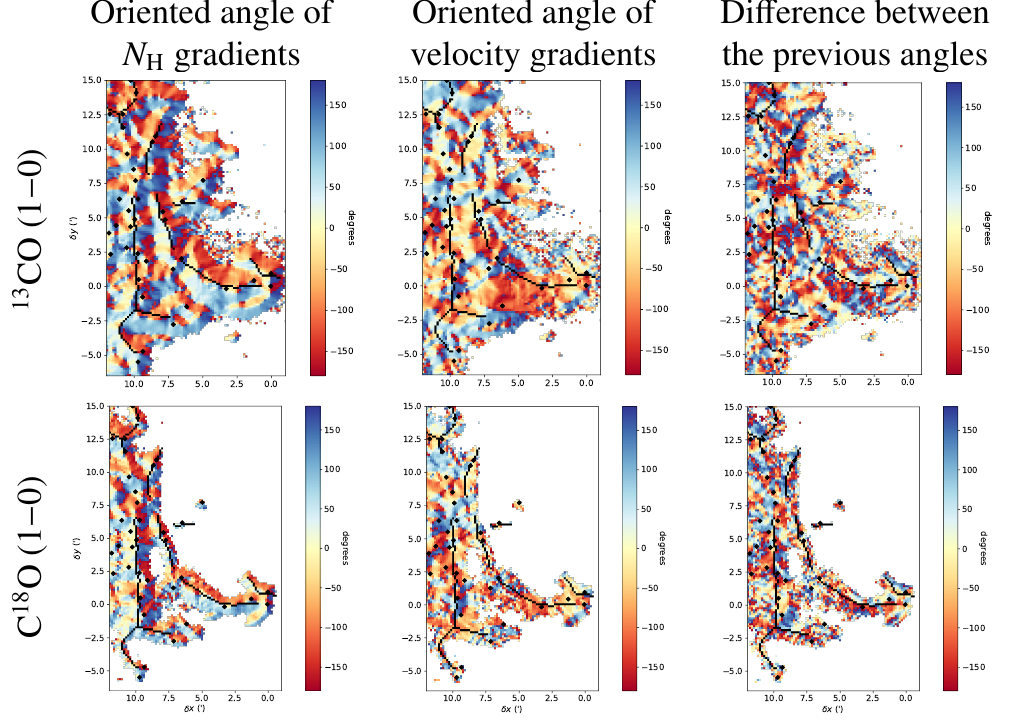}
    \caption{Maps of oriented angles of the gradients of the column density
      and centroid velocity gradients for the main-velocity layer of the
      Horsehead Nebula area.  The figure layout is identical to
      Fig.~\ref{fig:oriented-angle-maps:cloak}}
    \label{fig:oriented-angle-maps:tete-cheval}
  \end{figure}}

\newcommand{\FigDivMapsHorsehead}{%
  \begin{figure}
    \centering %
    \includegraphics[width=\linewidth]{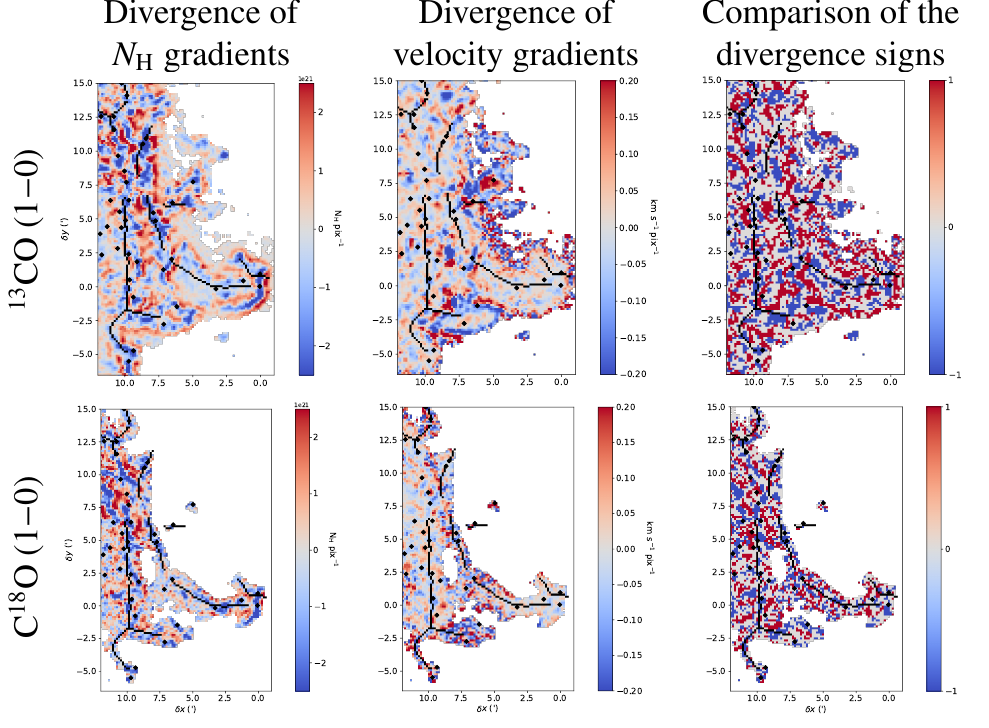}
    \caption{Divergence maps for the main-velocity layer of the Horsehead
      Nebula area.  The figure layout is identical to
      Fig.~\ref{fig:oriented-angle-maps:cloak}}
    \label{fig:divergence-maps:tete-cheval}
  \end{figure}}

\newcommand{\FigDecompROHSA}{%
  \begin{figure*}
    \centering %
    \includegraphics[width=\linewidth]{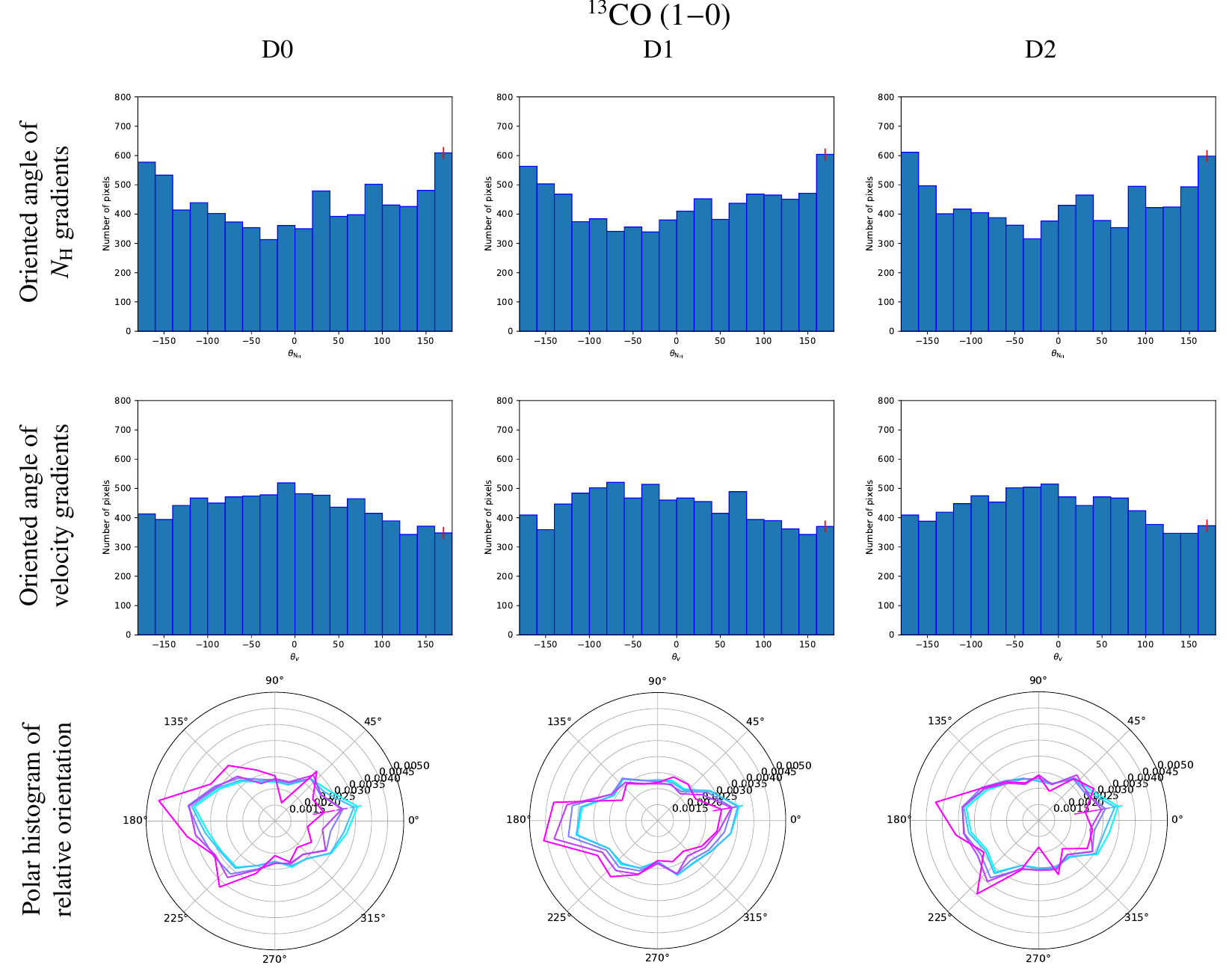}
    \caption{$^{13}$CO distribution histogram of the oriented angles of the
      column density (top row) and velocity gradients (middle row) in the
      Horsehead nebula from the reference decomposition $\mathrm{D0}$
      (left), decomposition $\mathrm{D1}$ (middle) and decomposition
      $\mathrm{D2}$ (right), respectively. Last row: histograms of the
      relative orientation between column density and velocity. Data with
      $\log(N_\mathrm{H}^{13}) > 20.00$ cm$^{-2}$ are shown in cyan,
      $\log(N_\mathrm{H}^{13}) > 21.00$ cm$^{-2}$ in blue,
      $\log(N_\mathrm{H}^{13}) > 21.55$ cm$^{-2}$ in blue-violet,
      $\log(N_\mathrm{H}^{13}) > 22.00$ cm$^{-2}$ in purple, and
      $\log(N_\mathrm{H}^{13}) > 22.40$ cm$^{-2}$ in fuchsia.  The colored
      bars at 10$^{\circ}$ show the error bars of the histograms.}
    \label{fig:comparison-results-decomposition-ROHSA-13CO}
  \end{figure*}}

\newcommand{\FigDecompROHSAbis}{%
  \begin{figure*}
    \centering %
    \includegraphics[width=\linewidth]{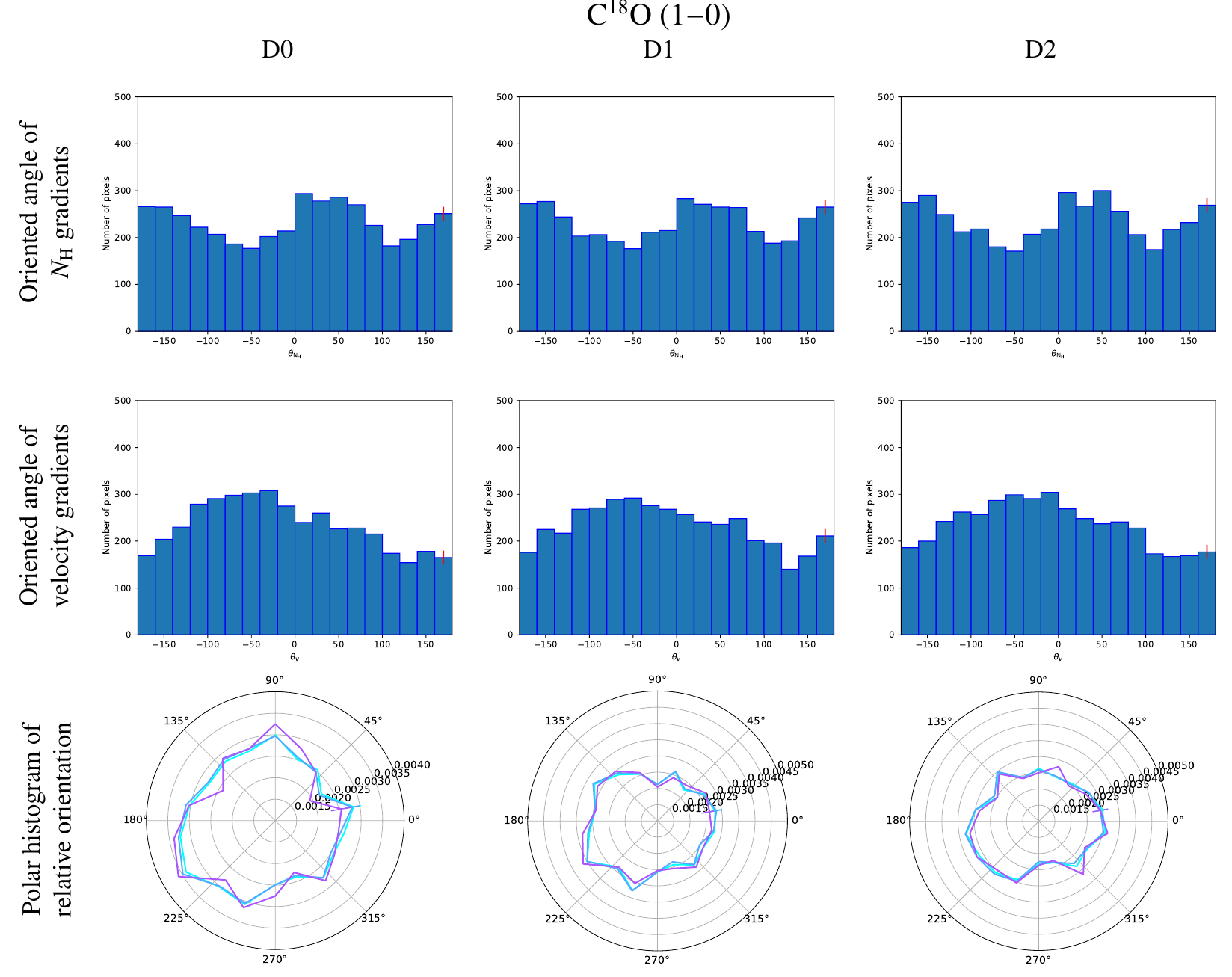}
    \caption{C$^{18}$O distribution histogram of the oriented angles of the
      column density (top row) and velocity gradients (middle row) in the
      Horsehead nebula from the reference decomposition $\mathrm{D0}$
      (left), decomposition $\mathrm{D1}$ (middle) and decomposition
      $\mathrm{D2}$ (right), respectively. Last row: histograms of the
      relative orientation between column density and velocity. Data with
      $\log(N_\mathrm{H}^{18}) > 21.00$ cm$^{-2}$ are shown in cyan,
      $\log(N_\mathrm{H}^{18}) > 21.55$ cm$^{-2}$ in blue-violet,
      $\log(N_\mathrm{H}^{18}) > 22.00$ cm$^{-2}$ in purple, and
      $\log(N_\mathrm{H}^{18}) > 22.40$ cm$^{-2}$ in fuchsia.  The colored
      bars at 10$^{\circ}$ show the error bars of the histograms.}
    \label{fig:comparison-results-decomposition-ROHSA-C18O}
  \end{figure*}}

\newcommand{\FigCompDecompROHSA}{%
  \begin{figure*}
    \centering %
    \includegraphics[width=0.75\linewidth]{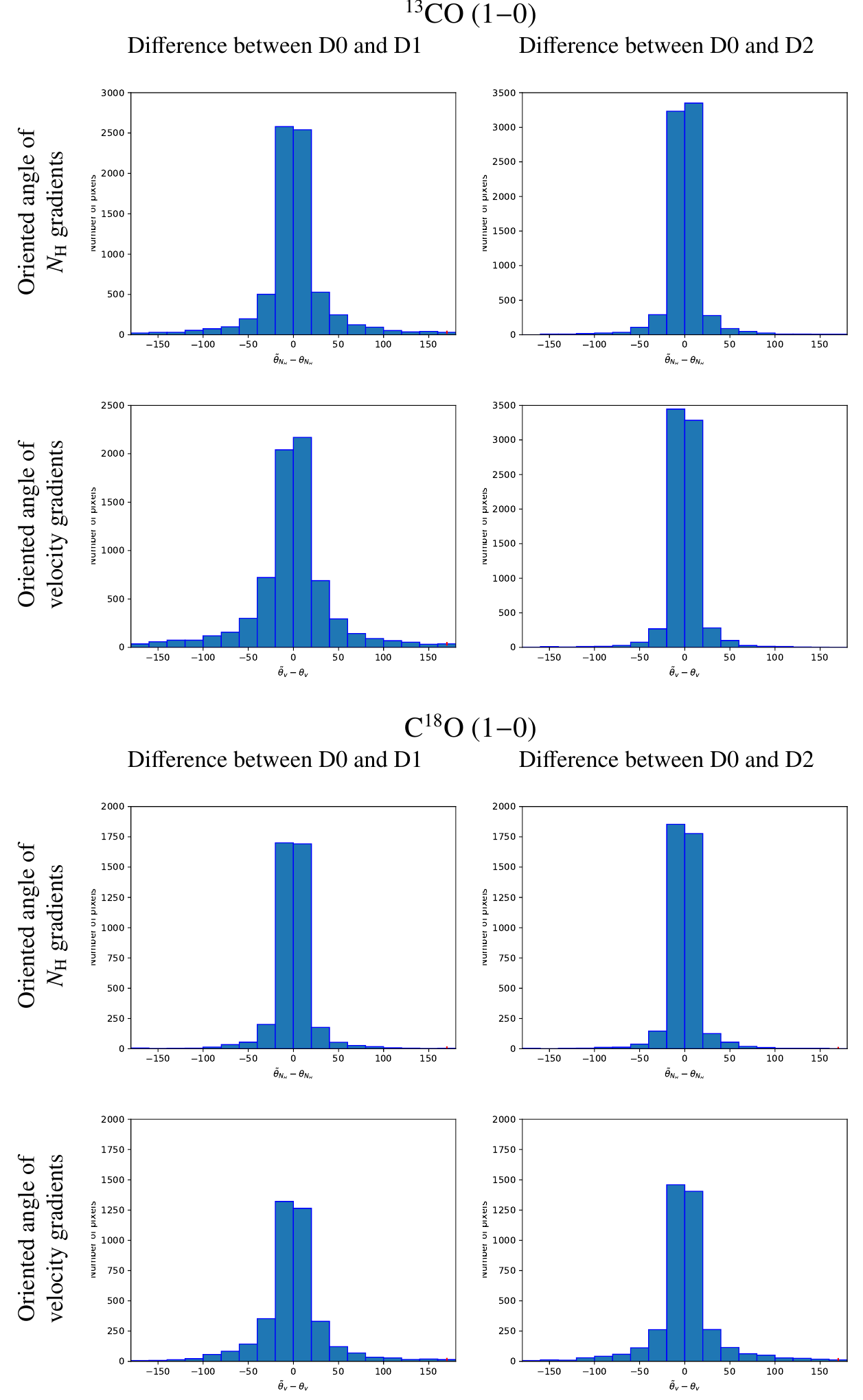}
    \caption{Difference between the oriented angles of the gradients
      obtained in the Horsehead nebula from $\mathrm{D0}$ and from
      $\mathrm{D1}$ (left), and from $\mathrm{D0}$ and from $\mathrm{D2}$
      (right), respectively.}
    \label{fig:residus-oriented-angle-gradients-tete-cheval}
  \end{figure*}}

\newcommand{\FigModelCuts}{%
  \begin{figure*}
    \centering %
    \includegraphics[width=\linewidth]{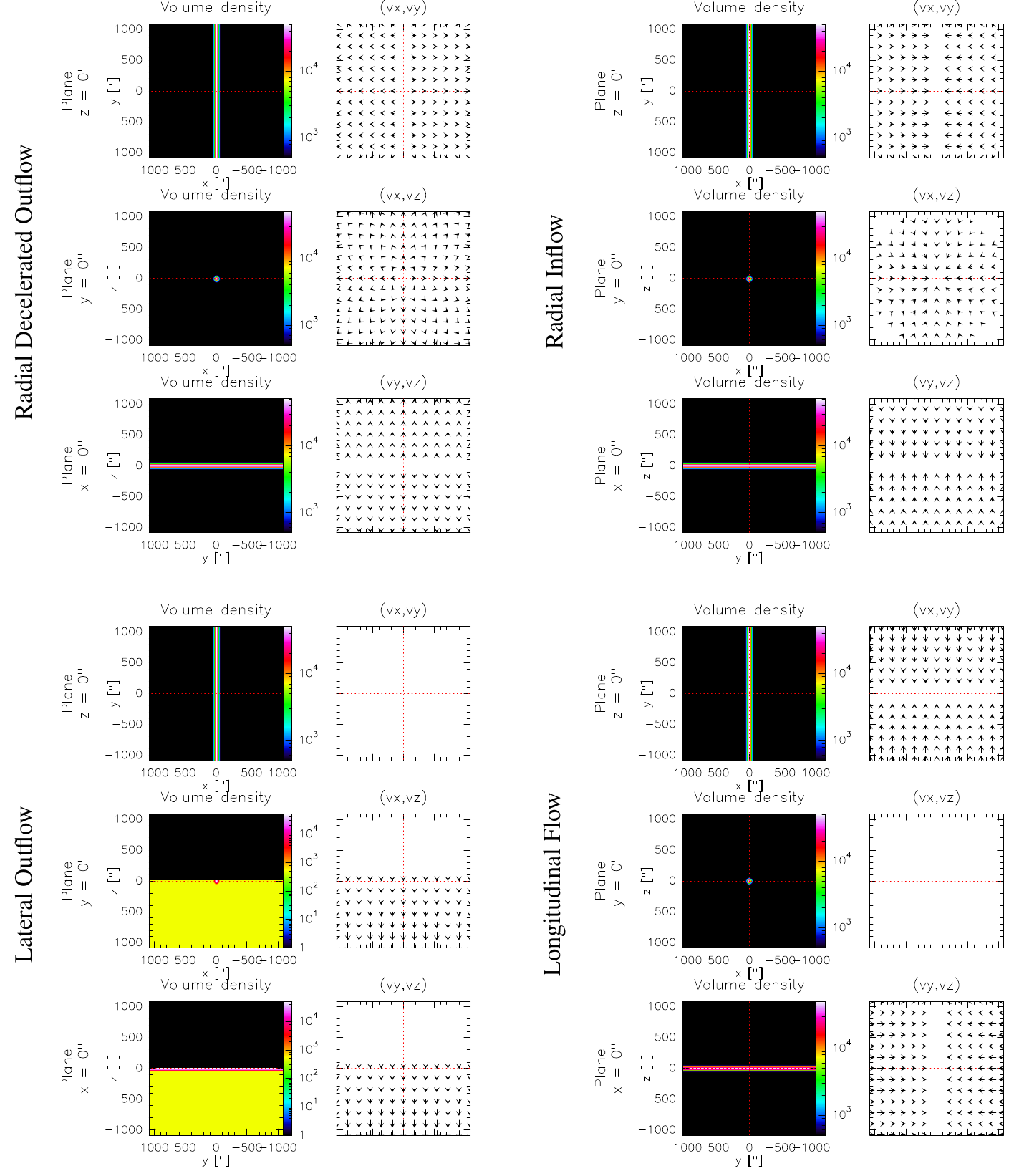}
    \caption{2D cuts along different plane of the 3D density (left) and
      velocity (right) fields for the four different models. The radial
      decelerated outflow is shown in the top left corner, the radial
      inflow in the top right corner, the lateral outflow in the bottom
      left corner and the longitudinal flow in the bottom right corner.}
    \label{fig:model-3D}
  \end{figure*}}
  
\newcommand{\FigModelProj}{%
  \begin{figure*}
    \centering %
    \includegraphics[width=\linewidth]{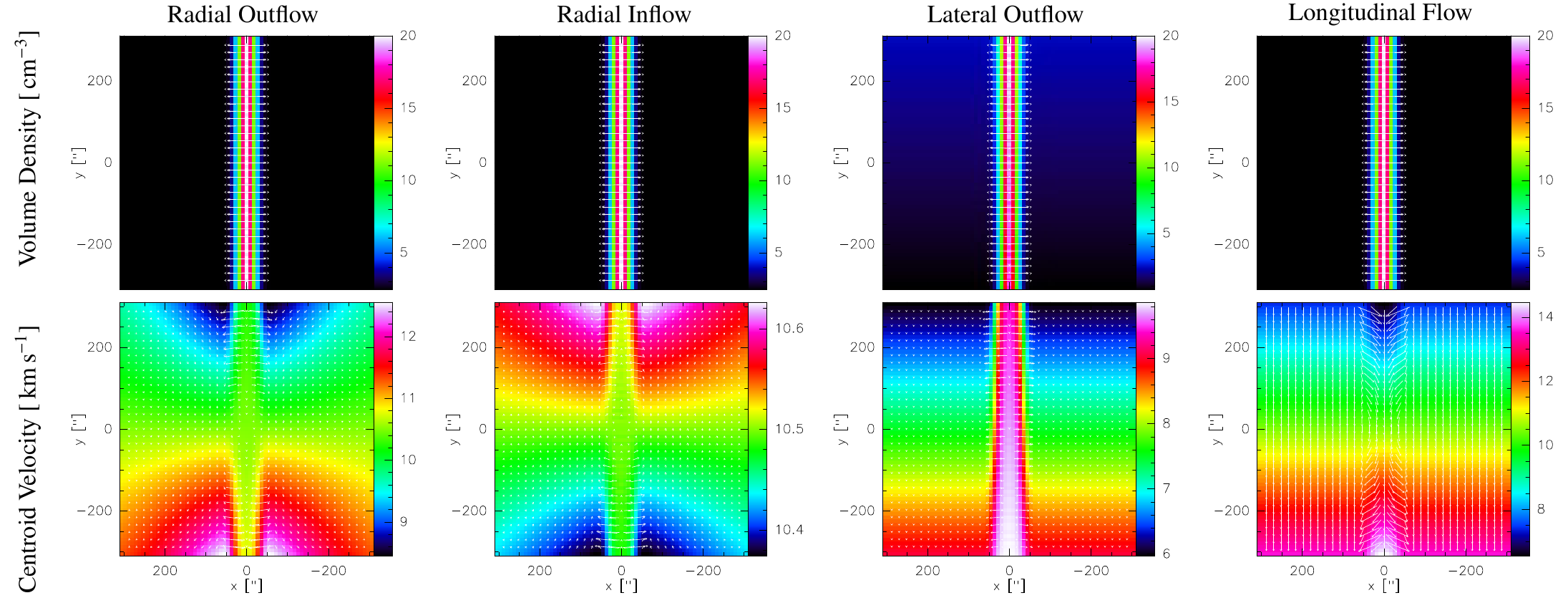}
    \caption{Spatial distribution of the projected column density (top) and
      centroid velocity (bottom) for the different models. The gradient
      vectors are overlaid.}
    \label{fig:model-2D}
  \end{figure*}}

\begin{document} 

\title{Gas kinematics around filamentary structures in the Orion B cloud}

\author{Mathilde Gaudel\inst{1} %
  \and Jan H. Orkisz\inst{2} %
  \and Maryvonne Gerin\inst{1} %
  \and Jérôme Pety\inst{3,1} %
  \and Antoine Roueff\inst{4} %
  \and Antoine Marchal\inst{5} %
  \and François Levrier\inst{6} %
  \and Marc-Antoine Miville-Deschênes\inst{7} %
  \and Javier R. Goicoechea\inst{8} %
  \and Evelyne Roueff\inst{9} %
  \and Franck Le Petit\inst{9} %
  \and Victor de Souza Magalhaes\inst{3} %
  \and Pierre Palud\inst{9} %
  \and Miriam G. Santa-Maria\inst{8} %
  \and Maxime Vono\inst{10} %
  \and Sébastien Bardeau\inst{3} %
  \and Emeric Bron\inst{9} %
  \and Pierre Chainais\inst{11} %
  \and Jocelyn Chanussot\inst{12} %
  \and Pierre Gratier\inst{13} %
  \and Viviana Guzman\inst{14} %
  \and Annie Hughes\inst{15} %
  \and Jouni Kainulainen\inst{2} %
  \and David Languignon\inst{9} %
  \and Jacques Le Bourlot\inst{1} %
  \and Harvey Liszt\inst{16} %
  \and Karin Öberg\inst{17} %
  \and Nicolas Peretto\inst{18} %
  \and Albrecht Sievers\inst{3} %
  \and Pascal Tremblin\inst{19} %
}
       
\institute{LERMA, Observatoire de Paris, PSL Research University, CNRS,
  Sorbonne Université, 75014 Paris, France %
  \and Chalmers University of Technology, Department of Space, Earth and
  Environment, 412 93 Gothenburg, Sweden %
  \and IRAM, 300 rue de la Piscine, 38406 Saint Martin d’Hères, France %
  \and Aix Marseille Université, CNRS, Centrale Marseille, Institut
  Fresnel, Marseille, France %
  \and Canadian Institute for Theoretical Astrophysics, University of
  Toronto, 60 Saint George Street, 14th floor, Toronto, ON,M5S 3H8,
  Canada %
  \and Laboratoire de Physique de l’Ecole normale supérieure, ENS,
  Université PSL, CNRS, Sorbonne Université, Université de Paris Cité, Paris, France %
  \and Laboratoire AIM, CEA Saclay/DRF/IRFU, CNRS, Université Paris-Saclay,
  Université de Paris Cité, F-91191 Gif-sur-Yvette,
  France %
  \and Instituto de Física Fundamental (CSIC). Calle Serrano 121, 28006,
  Madrid, Spain %
  \and LERMA, Observatoire de Paris, PSL Research University, CNRS,
  Sorbonne Universités, 92190, Meudon, France %
  \and University of Toulouse, IRIT/INP-ENSEEIHT, CNRS, 2 rue Charles
  Camichel, BP 7122, 31071, Toulouse cedex 7, France %
  \and Univ. Lille, CNRS, Centrale Lille, UMR 9189 - CRIStAL, 59651,
  Villeneuve d'Ascq, France %
  \and Univ. Grenoble Alpes, Inria, CNRS, Grenoble INP, GIPSA-Lab,
  Grenoble, 38000, France %
  \and Laboratoire d'Astrophysique de Bordeaux, Univ. Bordeaux, CNRS, B18N,
  Allee Geoffroy Saint-Hilaire, 33615, Pessac, France %
  \and Instituto de Astrofísica, Pontificia Universidad Católica de Chile,
  Av. Vicuña Mackenna 4860, 7820436, Macul, Santiago, Chile %
  \and Institut de Recherche en Astrophysique et Planétologie (IRAP),
  Université Paul Sabatier, Toulouse cedex 4, France %
  \and National Radio Astronomy Observatory, 520 Edgemont Road,
  Charlottesville, VA, 22903, USA %
  \and Harvard-Smithsonian Center for Astrophysics, 60 Garden Street,
  Cambridge, MA, 02138, USA %
  \and School of Physics and Astronomy, Cardiff University, Queen's
  buildings, Cardiff, CF24 3AA, UK %
  \and Université Paris-Saclay, UVSQ, CNRS, CEA, Maison de la Simulation,
  F-91191 Gif-sur-Yvette, France %
}

\date{Received 30 August 2021; accepted 14 November  2022}

 
\abstract
{
Understanding the initial properties of star-forming material and how they
  affect the star formation process is key. From an
  observational point of view, the feedback from young high-mass stars on
  future star formation properties is still poorly constrained.}
{In the framework of the IRAM 30m ORION-B large program, we obtained
  observations of the translucent (2 $\leq$ A$_\mathrm{V} < $ 6 mag) and
  moderately dense gas (6 $\leq$ A$_\mathrm{V} < $ 15 mag), which we used to
  analyze the kinematics over a field of 5 deg$^{2}$ around the filamentary
  structures.}
{We used the Regularized Optimization for Hyper-Spectral Analysis (ROHSA) algorithm 
to decompose and de-noise the C$^{18}$O(1$-$0)
  and $^{13}$CO(1$-$0) signals by taking the spatial coherence
  of the emission  into account. We produced gas column density and mean velocity maps to
  estimate the relative orientation of their spatial gradients.}
{We identified three cloud velocity layers at different systemic velocities
  and extracted the filaments in each velocity layer. The filaments are
  preferentially located in regions of low centroid velocity gradients.  By
  comparing the relative orientation between the column density and
  velocity gradients of each layer from the ORION-B observations and
  synthetic observations from 3D kinematic toy models, we distinguish two
  types of behavior in the dynamics around filaments: (i) radial flows
  perpendicular to the filament axis that can be either inflows (increasing
  the filament mass) or outflows and (ii) longitudinal flows along the filament
  axis. The former case is seen in the Orion B data, while the latter is
  not identified. We have also identified asymmetrical flow patterns,
  usually associated with filaments located at the edge of an \HII{} region.}
{This is the first observational study to highlight feedback from \HII{}
  regions on filament formation and, thus, on star formation in the Orion B
  cloud. This simple statistical method can be used for any molecular cloud
  to obtain coherent information on the kinematics.}


\maketitle{} %

\section{Introduction}

\FigRGB{} %

Observations reveal the ubiquitous presence of filamentary structures in
molecular clouds, which play a fundamental role in the star formation
process \citep{Andre10, Molinari10, Palmeirim13, Konyves15}. Indeed,
pre-stellar cores and protostars mainly form in supercritical filaments
(linear mass $M_\emr{line} > M_\emr{line, crit} = 2 c_S^2/G$), which are
expected to be gravitationally unstable and, thus, the site of
self-gravitational fragmentation \citep{Andre10, Tafalla15}.

The angular momentum of a core is expected to be inherited from the initial
core formation conditions.
Historical studies
\citep{Goodman93,Caselli02} showed that the observed mean angular momentum
within pre-stellar cores depends on their spatial scale. The gradients
observed from single-dish mapping at $r>$3000~au have been interpreted as
rotation and used to quantify the amplitude of the angular momentum problem
for star formation and the expected disk radii around protostellar
embryos. However, recent studies suggest that this trend could be due to
pure turbulence motions or gravitationally driven turbulence
\citep{Tatematsu16, Pineda19, Gaudel20}. This in turn suggests that the connection
between cores and the molecular cloud may be still present in the inner
envelopes of young Class 0 protostars ($\sim$1600~au;
\citealt{Gaudel20}). Hence, the analysis of the gas dynamics in filaments
could shed light on the origin of the angular momentum of the cores within
them.

From molecular line observations, we know that filamentary structures exhibit complex
motions with longitudinal collapse along the main filament \citep{Kirk13,
  Fernandez-Lopez2014, Tackenberg2014, Gong2018, Dutta2018, Lu2018,
  Chen2020} and radial contractions \citep{Kirk13, Fernandez-Lopez2014,
  Dhabal18}. Several studies have also highlighted velocity gradients
perpendicular to the main filament, which  suggests that the material may be
accreting along perpendicular, low-density striations into the main
filaments \citep{Palmeirim13,Dhabal18,Arzoumanian18, Shimajiri19}.  While
the filaments seem to accrete radially from their surroundings, the
longitudinal collapse feeds central hubs or clumps, namely sites where the
filaments converge \citep{Myers2009, Schneider2010, Sugitani2011,
  Peretto2014, Rayner2017, Baug2018, Trevino-Morales2019}.

Moreover, young massive stars are expected to have a strong impact on the
surrounding environment, in particular by generating expanding \HII{}
regions driven by UV ionizing radiation and stellar winds
\citep{Pabst2019,Pabst2020}.  From a theoretical point of view, the
ionization feedback from \HII{} regions would modify the properties of the
molecular cloud, such as turbulence \citep{Menon2020}. From an
observational point of view, the impact of young high-mass stars on
future star formation properties is still poorly constrained. Several
observational studies identified direct signatures of star formation,
namely jets and outflows, at the edge of the \HII{} regions, suggesting
that star formation could be triggered by the global effects of ionizing
radiation and stellar winds \citep{Sugitani2002, Billot2010, Smith2010,
  Chauhan2011, Reiter2013, Roccatagliata2013}. Using high-angular-resolution ALMA (Atacama Large Millimeter/submillimeter Array) 
  observations of the 1.3 mm continuum emission in eight
candidate high-mass starless clumps, \citet{Zhang2021} suggest that \HII{}
regions modify the dense material distribution around them, which tends to
assist the fragmentation and, thus, the formation of future stars. However,
there is no direct study of the kinematics around filamentary structures
close to \HII{} regions to quantify their feedback effect on the filament
formation and, thus, on star formation.

Understanding the initial properties of star-forming material and how they
affect the star formation process is key. It is therefore
crucial to robustly identify the physical mechanisms active in molecular clouds  (turbulence,
gravitational collapse, and magnetic fields) by
analyzing the kinematics at different scales around the filamentary
structures.

\section{The ORION-B large program}

The Outstanding Radio-Imaging of OrioN B (ORION-B) IRAM (Institut de Radioastronomie Millim\'etrique) 
large program
(co-PIs: J. Pety and M. Gerin) is a survey of the southern part (five
square degrees) of the Orion B molecular cloud, carried out with the IRAM
30-meter telescope (30m) at a wavelength of 3 mm.

Orion B, also known as the L1630 cloud, is located at a distance of
$\sim$400 pc \citep{Menten2007, Zucker2019} in the Orion Giant Molecular
Cloud complex \citep{Kramer1996, Ripple2013}, east of the Orion Belt. Close
to us, Orion B is a perfect object of study to fully understand stellar
formation.  This region is a active star-forming region: \cite{Konyves2020}
identified thousand of dense cores closely associated with highly
filamentary structures of matter from \textit{Herschel} Gould Belt Survey
(HGBS) observations. The Orion B cloud hosts several OB stars and well studied
areas such as the Horsehead pillar \citep{Zhou1993, Bally2018} and the
associated photodissociation region \citep{Goicoechea2009,Guzman2013}, the
NGC\,2023 and NGC\,2024 star-forming regions, and the Flame and the
Hummingbird filaments. Thus, it is a perfect target to study the kinematics
around filamentary structures and the influence of feedback from massive OB
stars on it.

The ORION-B program allows us to study in detail the properties of the
molecular gas in the Orion B cloud such as the relationships between line
intensities and the total gas column density \citep{Pety2017, Gratier2017,
  Gratier2021, Roueff2021}, the structure of the different
physical and chemical 
environments \citep{Bron2018, Bron2021} and the
characterization of turbulence modes \citep{Orkisz2017}. The main goals of
this large observing program are to establish Orion B cloud as a local
template for interpreting Galactic and extra-galactic molecular line
observations \citep{Pety2017}.
The Orion B cloud also hosts a large statistical sample of star-forming
cores, which allows us to better understand star formation. In a previous
work, \cite{Orkisz2019} developed a tool to identify and characterize the
filament population in the southwestern part the Orion B cloud including
NGC\,2023 and NGC\,2024. This paper presents an analysis of the kinematics
around filaments in the ORION-B field of view, which contains the NGC\,2023
- NGC\,2024 complexes, the Orion B9 and the Cloak areas (see
Fig. \ref{fig:regions-orionb-cloud}).

\FigData{} %
\TabData{} %

\section{Observations and data set reduction}

The observations were obtained at the IRAM 30-meter telescope (hereafter
30m) using the heterodyne Eight MIxer Receiver (EMIR) in one atmospheric
windows E090 band at 3~mm \citep{Carter2012} from 84.5 to 115.5 GHz. The
Fast Fourier Transform Spectrometer (FTS) was connected to the EMIR
receiver. The FTS200 backend provided total bandwidth of $\sim$32~GHz with
a spectral resolution of 195~kHz (0.5~km s$^{-1}$). The full field of view
of 5 square degrees was covered using rectangular tiles with a position
angle of 14$^{\circ}$ in the Equatorial J2000 frame to follow the global
morphology of the Orion B cloud. Details of the observation strategy can be
found in \cite{Pety2017}.

Data reduction was carried out using CLASS, which is part of the
GILDAS\footnote{See \url{http://www.iram.fr/IRAMFR/GILDAS/} for more
  information about the GILDAS software \citep{Pety2005}.} software,
following the standard steps: removal of atmospheric emission, gain
calibration, baseline subtraction, and gridding of individual spectra to
produce regularly sampled maps with pixels of 9$\arcsec$ size. Details of
each step can be found in \cite{Pety2017}. Each datacube covers a velocity
range of 120~km s$^{-1}$ (240 channels spaced by 0.5~km s$^{-1}$) centered
at the rest line frequency (see Table
\ref{tab:details-resulting-maps}). The spatial coordinates of the data
cubes are centered onto the photon-dominated region (PDR) of the Horsehead
Nebula at the coordinates (05h40m54.270s,
-02$^{\circ}$28$\arcmin$00.00$\arcsec$) and rotated by an angle of
14$^{\circ}$ counterclockwise with respect to the Equatorial J2000 frame. The
properties of the resulting maps, including the spatial resolution of the
molecular line emission maps and the root mean square (rms) noise levels from the mean spectra
are reported in Table \ref{tab:details-resulting-maps}.

%

\section{Gaussian decomposition}

From the mean spectra of the C$^{18}$O and $^{13}$CO (1$-$0) molecular
lines reported in Fig. \ref{fig:moment0-maps:initial-datasets}, we
identified several velocity components that seemed to belong to different
layers of the Orion B cloud along the line of sight. To robustly
investigate the kinematics, we need first to disentangle the different
layers of the cloud. To do this, we used the Gaussian decomposition
algorithm Regularized Optimization for Hyper-Spectral
Analysis (ROHSA\footnote{See
  \url{https://antoinemarchal.github.io/ROHSA/index.html} for more
  information about ROHSA}) developed by \cite{Marchal2019}. This algorithm, based on a
multi-resolution process from coarse to fine grid, uses a regularized
non-linear least-square criterion to take into account the spatial
coherence of the emission. This allows us to follow each Gaussian velocity
component in the field of view, even at the transitions between spatial
areas where the number of velocity components changes. In ROHSA, the model
used to fit the data is a sum of Gaussians, where each Gaussian is
parametrized by an amplitude $A_i$, a centroid velocity $\mu_i$ and a
velocity dispersion $\sigma_i$. ROHSA takes as free parameters the number
$N$ of Gaussian components, three hyper-parameters ($\lambda_\mathrm{A}$,
$\lambda_\mathrm{\mu}$, $\lambda_\mathrm{\sigma}$) ensuring the spatial
correlation of the model parameter maps for neighboring pixels, and
$\lambda'_\mathrm{\sigma}$ the variance of the velocity dispersion maps
\citep{Marchal2019}. Using the parameter maps, we can reconstruct a
de-noised signal with either the sum of all components defined as
$\Sigma_i A_i \exp({-(\mathrm{v}-\mu_i)^{2}/(2 \sigma_i^{2})})$, or with a
single component.

A visual inspection allows us to identify at least three velocity
components in the C$^{18}$O and $^{13}$CO mean spectra (see
Fig. \ref{fig:moment0-maps:initial-datasets}). However, non-Gaussian
effects, such as line saturation, line wings, or velocity component
asymmetry, must be taken into account by adding more Gaussian components
than can be seen by a visual inspection to best reproduce the spectrum. We
note that each emission line may trace different regions of the cloud (see
Fig. \ref{fig:moment0-maps:initial-datasets}), and hence the C$^{18}$O and
$^{13}$CO signals may not require the same number of Gaussian components to
be reconstructed by ROHSA. To deduce the best total number of Gaussian
components for each tracer, we initialized the algorithm to decompose the
signal with several numbers $N = [3, 4, 5,6]$ of Gaussian components using
an initial value $\sigma_\mathrm{init}= 1.5\kms$ for C$^{18}$O and
$^{13}$CO. The details of the different decompositions (residuals,
$\chi^2$, velocity and velocity dispersion histograms) are given in
Appendix \ref{sec:ROHSA-decomposition}. Based on a trade-off between
$\chi^2$ minimization and minimizing the number of components, the best
total number of Gaussian components $N_\mathrm{best}$ is four and five to
reconstruct the C$^{18}$O and the $^{13}$CO (1$-$0) signals, respectively.

To find the best hyper-parameters, we also tried several values
($\lambda_\mathrm{i}= [10, 100, 1000],$ $\lambda'_\mathrm{\sigma}= [0, 1]$)
for a given number $N$ of Gaussian components (see Appendix
\ref{sec:ROHSA-decomposition}). We selected $\lambda_\mathrm{i}$=100 for
each hyper-parameter. The variance of the velocity dispersion maps
$\lambda'_\mathrm{\sigma}$ is set to zero to let free the variation of the
velocity dispersion differences throughout the full field of view.

\section{Identification and characteristics of the different velocity
  layers of the Orion B cloud}
\label{sec:moment-maps}

\subsection{Moment maps}

\FigLayerSpectra{} %
\FigLayerMoments{} %

We built the zero-order (integrated intensity), the first-order (mean
velocity) and the second-order (velocity dispersion) moment maps of each
Gaussian component fit by ROHSA for the C$^{18}$O and $^{13}$CO (1$-$0)
lines (see Appendix~\ref{sec:moment-maps:Gaussian-components}). We
integrated each de-noised spectrum rebuilt by ROHSA on the velocity range
$[-5,+20\kms]$. Because of the spatial coherence, ROHSA can detect the
Gaussian components up to a low signal-to-noise ratio (S/N
$\lesssim$1$\sigma$). When including low S/N pixels, the amplitude and the
width of the average spectrum inferred with ROHSA are larger than those of
the average spectrum from the original data set. To avoid this effect, we
only kept the pixels initially robustly detected, namely a spectrum
detected with a signal-to-noise ratio higher than 3 in the initial
data sets, to build the moment maps inferred with
ROHSA. Figures~\ref{fig:mean-spectrum-ROHSA-gaussian-fit}-\ref{fig:moment-maps:ROHSA-gaussian-fit-13co}
show the mean spectrum and the moment maps for each Gaussian components
estimated for the two tracers.

On these moment maps, we observe that some Gaussian components trace the
same spatial structures in the same velocity range and with a similar
velocity dispersion. Given the close velocity ranges with respect to the
velocity dispersion, the similarities in the structure of the integrated
intensities from the C$^{18}$O (1$-$0) moment maps (see
Fig. \ref{fig:moment-maps:ROHSA-gaussian-fit-c18o}), we identified the two
last Gaussian components as coming from the same layer of the Orion B
cloud. We added them together to recover the total signal of this cloud
layer, which is not purely Gaussian. Three layers stood out (ordered
by increasing systemic local standard of rest velocity):
(i) the low-velocity layer with a systemic velocity around 2.5~kms$^{-1}$,
(ii) the intermediate-velocity layer with a systemic velocity around 6~km
  s$^{-1}$,
and (iii) the main-velocity layer of the Orion B cloud with a systemic velocity
  around 10~km s$^{-1}$.

In the following, we call these three layers the "cloud" layers to
differentiate them from the "ROHSA" Gaussian components.  From the
$^{13}$CO (1$-$0) moment maps (see
Fig. \ref{fig:moment-maps:ROHSA-gaussian-fit-13co}), the last two Gaussian
components trace the main-velocity layer: as previously with the C$^{18}$O
data set, we added them together to recover the total signal of the
layer. The first and the third Gaussian components trace the low and
intermediate-velocity layers, respectively.

\FigLayerNH{} %

Figures~\ref{fig:mean-spectrum-ROHSA-cloud-layers}
and~\ref{fig:moment-maps:ROHSA-layers-cloud} show the mean spectrum and the
moment maps from the C$^{18}$O and $^{13}$CO emission rebuilt by ROHSA for
each identified layer of the cloud. From the zeroth-order moment maps, we
observe that $^{13}$CO (1$-$0) traces more diffuse and more extended gas
than C$^{18}$O (1$-$0), which traces moderately dense gas (n$_\mathrm{H}$
$\gtrsim$ 10$^4$ cm$^{-3}$). The cloud layers we identified above are
velocity layers, but they may be located at different distances along the
line of sight. In the \cite{Gaia18} catalog, we identified young stellar
objects (YSOs) with ages of less than 4 Myr located in the ORION-B field of view and
with mean radial velocities corresponding to the velocity range of each
cloud layer we identified. Young stellar objects are of interest because
they are drifting slowly from their parent cloud and are still located
within the boundaries of their parent cloud \citep{Gaia18}. The selected
YSOs are all located at the same distance of $\sim$300-500
pc as the Orion B cloud showing that these velocity layers are associated
with the Orion Complex. In this paper, we study the translucent medium (2
$\leq$ A$_\mathrm{V} <$ 6 mag) and the moderately dense gas around the
filamentary structures (6 $\leq$ A$_\mathrm{V} <$ 15 mag) in the Orion B
cloud, and thus we did not consider the low density turbulent medium (1 $\leq$
A$_\mathrm{V} <$ 2 mag), which may connect these cloud layers. For example,
we note that we need a fifth Gaussian component to recover the $^{13}$CO
(1$-$0) signal, which traces more diffuse gas compared to C$^{18}$O
(1$-$0). This Gaussian component seems to trace the transition between the
low-velocity and the intermediate-velocity layers with a systemic velocity
around 4\kms{} (see component 2 in
Figs.~\ref{fig:mean-spectrum-ROHSA-gaussian-fit}
and~\ref{fig:moment-maps:ROHSA-gaussian-fit-13co}). In the following, we
did not use this Gaussian component to analyze the kinematics of the
cloud. This component represents only $\lesssim$25\% of the total flux; thus, removing it has little impact on the analysis on the kinematics of
the translucent and moderately dense gas.

In the low-velocity layer first-order moment maps in $^{13}$CO (1$-$0) (see
Fig.~\ref{fig:moment-maps:ROHSA-layers-cloud}), we notice a structure in
the center-south. With a velocity higher than the rest of the layer, it
seems not to be connected to this layer. Even if the nature of this
structure remains unclear, it may correspond to a cloud located along the
line of sight but at a different distance from Orion B. With a systemic
velocity close to that of the low-velocity layer, the structure could be
responsible for the line wing observed in the $^{13}$CO (1$-$0) mean
spectrum of the layer (see
Fig. \ref{fig:mean-spectrum-ROHSA-cloud-layers}). The weak intensity of the
structure does not allow ROHSA to separate it from the low-velocity layer.

The spatial coherence of the velocity components extracted in the ROHSA
decomposition in a way that allows continuous variations of the line
parameters, as described in \cite{Marchal2019}. The maps of the integrated
intensity, mean velocity and velocity dispersion of each velocity component
shown in Fig. \ref{fig:moment-maps:ROHSA-layers-cloud} indeed exhibit
spatial variations of these parameters, tracing the internal structure and
dynamics of the three layers. For each velocity layer, the ROHSA
decomposition has clearly captured the existing systematic intensity and
velocity gradients.

\subsection{Column density maps}

\FigSubRegions{} %

From the C$^{18}$O and $^{13}$CO (1$-$0) zero-order moment maps (see
Fig.~\ref{fig:moment-maps:ROHSA-layers-cloud}), we derived the column
density $N_\mathrm{H}$ maps of each cloud layer by using the following
equations \citep{Orkisz2019, Roueff2021}:
\begin{equation}
  \begin{gathered}
    N_\mathrm{C^{18}O}/N_\mathrm{H}= \frac{5.6 \times 10^{-7}}{2}, \\
    N_\mathrm{^{13}CO}/N_\mathrm{H}= \frac{2.5 \times 10^{-6}}{2}.
  \end{gathered}
\end{equation}
In these equations, $N_\mathrm{H}$ represents the total column of hydrogen,
including the contributions of atomic and molecular hydrogen,
$N_\mathrm{H} = N(\mathrm{HI})+2N(\mathrm{H_2})$.  We assumed local
thermodynamic equilibrium and an optically thin emission. We used the
standard equations described in \cite{Mangum2015}, the parameters values
from the \href{http://www.astro.uni-koeln.de/cdms/catalog}{Cologne Database
  for Molecular Spectroscopy} (CDMS; \citealt{Endres2016}), and the dust
temperature $T_\mathrm{dust}$ map from \cite{Lombardi2014}. We assumed that
the dust temperature is a good estimate for the excitation temperature of
$^{13}$CO (1$-$0) and C$^{18}$O (1$-$0). The Orion B cloud layers we
identified are not distinguished by \cite{Lombardi2014}: the
$T_\mathrm{dust}$ map is therefore the same for all cloud layers.  More
details on the assumptions made to compute the column density can be found
in Sect. 2.4 of~\cite{Orkisz2019}.
Figure~\ref{fig:NH-maps:ROHSA-layers-cloud} shows the column density maps
of each layer of the cloud deduced from the $^{13}$CO (1$-$0) and C$^{18}$O
(1$-$0) data sets. In the following, we distinguish the column density
deduced from the $^{13}$CO (1$-$0) and from the C$^{18}$O (1$-$0) data sets
with the notation $N_\mathrm{H}^{13}$ and $N_\mathrm{H}^{18}$,
respectively.  For the C$^{18}$O (1$-$0) emission, we only kept the pixels
with a column density $N_\mathrm{H}^{18}$ higher than 10$^{21}$ cm$^{-2}$
because at lower values, the signal to noise ratio becomes marginal. For
the $^{13}$CO (1$-$0) emission, we only kept the pixels with a column
density $N_\mathrm{H}^{13}$ higher than 10$^{20}$ cm$^{-2}$. The highest
contour of $N_\mathrm{H}^{13}$, 2.5 $\times$ 10$^{22}$ cm$^{-2}$, spatially
encompasses regions with even higher hydrogen column densities determined
using the C$^{18}$O (1-0) line. Thus, the $^{13}$CO and C$^{18}$O molecules
are complementary tracers of the kinematics. $^{13}$CO (1$-$0) becomes
partially optically thick at high $N_\mathrm{H}$ \citep{Roueff2021}, so in
the following we used the C$^{18}$O data set to probe the kinematics
robustly at high $N_\mathrm{H}$.

\subsection{Error bars}

The ROHSA algorithm does not provide the error bars of the output
parameters (amplitude, centroid velocity and velocity dispersion) of the
Gaussian components. Thus, we used the Monte Carlo method to estimate the
error bars on the $N_\mathrm{H}^{18}$, on the $N_\mathrm{H}^{13}$ and on
the centroid velocity maps of each cloud layer. We added a random Gaussian
noise on each noiseless spectrum (i.e., each pixel) of each Gaussian
component fit by ROHSA. The standard deviation $\sigma$ of the noise was
chosen so that each individual spectrum (i.e., at each pixel) of the total
signal rebuilt by ROHSA, namely the sum of all the Gaussian components, has
a noise consistent with the one observed in the individual spectrum from
the original data set.

From the randomly noised Gaussian components, we rebuilt the signal of each
cloud layer and estimated the $N_\mathrm{H}^{18}$, the $N_\mathrm{H}^{13}$
and the first-order moment maps by selecting only the velocity channels
with a signal-to-noise ratio higher than 3. We generated 1000 realizations
to estimate the mean and the standard deviation. The mean maps and the
standard deviation maps are shown in Appendix
\ref{sec:monte-carlo-results}. The mean maps of $N_\mathrm{H}^{18}$,
$N_\mathrm{H}^{13}$ and first-order moment of each cloud layer are
consistent with the maps inferred by ROHSA. In order to study the
kinematics in a robust way, we chose to keep the pixels in the velocity
maps with a standard deviation smaller than a half of the velocity channel
width ($< 0.25\kms$). We note that the main structure of each layer has
small velocity error bars, that are always lower than the selected
criterion of 0.25\kms. This criterion allows us to exclude the pixels with
faint and noisy emission.


\section{Extraction of the filamentary network}
\label{sec:detection-filaments}

\TabHIIregions{} %

To identify and locate the filamentary structures precisely in the
different layers of the Orion B cloud, we applied on the
$N_\mathrm{H}^{18}$ map estimated from the C$^{18}$O (1$-$0) emission an
algorithm developed and described in details in \cite{Orkisz2019}. The
algorithm is performed in a multi-scale fashion by re-scaling the
$N_\mathrm{H}^{18}$ map by an $arcsinh$ function before computing the Hessian
matrix. Then, the local aspect ratio of the structures and the column
density gradient is used to refine the ridge detection.  The raw skeleton
obtained by this algorithm was cleaned by following several steps (see
\citealt{Orkisz2019} for more details). 

First, the skeleton geometry was analyzed to remove the very short
  filaments, namely with an aspect ratio lower than two. Thus, we only kept
  the filaments with a length longer than twice the typical filament width
  (i.e., $\sim 0.2\unit{pc}$, \citealt{Arzoumanian2011, Orkisz2019}).

Second, we took the widths of the filaments into account. We removed the
  filaments with an unresolved width, namely lower than the resolution of
  the data sets (i.e., $\sim 0.02\unit{pc}$). Excessively large filaments (i.e.,
  $\gtrsim 0.28\unit{pc}$) are also removed because they are unstable.

Third, we analyzed the shape of the filaments by computing their mean
  curvature radius and comparing it to the width of the filament in order
  to remove the very curvy filaments (i.e., ratio between curvature radius
  and width $<1.5$) but keeping the large-scale loops.

Finally, we measured the relative contrast to determine how many times denser
  the filament is compared to its surrounding medium. We put a lower limit
  on the relative contrast at $\sim 0.5$.

Filaments are removed if they are rejected by more than one
criterion. Figure \ref{fig:zoom-regions-NH-maps:layers-cloud} shows the
cleaned skeleton overlaid on the $N_\mathrm{H}^{13}$ maps of each layer of
the Orion B cloud. Some filamentary structures within the main-velocity
layer, as in the Horsehead Nebula or in the Intermediate area, are at the
edge of \HII{} regions created by the intense UV ionizing from massive
stars at large scales (see Fig. \ref{fig:zoom-regions-NH-maps:layers-cloud}
and Table \ref{tab:regions-HII}).  The radii of the circles drawn in
Fig. \ref{fig:zoom-regions-NH-maps:layers-cloud} are determined according
to the size of the \HII{} region emission either in H$\alpha$ or in radio
continuum (\citealt{Pety2017}; e.g., \citealt{Martin-Hernandez2005} for
NGC\,2023 and NGC\,2024 for example).  Using the Heracles code as described
in \cite{Tremblin2014} and knowing the parameters of the exciting stars,
the age of the \HII{} region and the expansion velocity of the neutral
shell beyond the ionization front can be computed. The NGC\,2024 \HII{}
region is predicted to exert the strongest radiative feedback effect
(P. Tremblin, priv. comm.).

\section{Kinematics}

\FigAngGradientsExample{} %
\FigDivMapsExample{} %

\subsection{Computation of velocity and column density gradients}

From the first-order moment maps (see
Fig.~\ref{fig:moment-maps:ROHSA-layers-cloud}), we observe clear velocity
patterns in the different layers of the Orion B cloud. In order to estimate
the velocity gradients, we used the unbiased filter $[-1\,0\,1]/2$ in the
horizontal and vertical directions, which means that, for a given pixel, we
computed the difference between the velocity value at this pixel and the
values at the pixels located forward and backward along the x- and the
y-axis, and we divided by twice the pixel size. The pixels without forward
or backward neighbors along the x- and y-axis are ignored. We determined
the velocity gradients at each pixel for the different layers of the Orion
B cloud. Using the same method, we determined the $N_\mathrm{H}^{13}$ and
$N_\mathrm{H}^{18}$ gradients at each pixel. To facilitate the
visualization of the gradients, we zoomed into different subregions of the
cloud layers.  Figure~\ref{fig:zoom-regions-NH-maps:layers-cloud} shows the
subregions of interest on the cloud layer $N_\mathrm{H}^{13}$
maps. Figure~\ref{fig:oriented-angle-maps:example} shows as an example the
maps of the oriented angle of the velocity and the column density gradients
obtained for the main-velocity layer of the Horsehead Nebula.  The oriented
angle maps of the other subregions are provided in
Appendix~\ref{sec:comments-indiv-subregions}.

As expected, we observe that in all subregions most of the filaments are
located at the convergence of both $N_\mathrm{H}^{13}$ and
$N_\mathrm{H}^{18}$ gradients, namely the gradients have opposite
directions on both sides of the filaments. This is due to the filament
extraction algorithm, which robustly detects the $N_\mathrm{H}^{18}$
ridges. Maps of $N_\mathrm{H}^{13}$ and $N_\mathrm{H}^{18}$ gradient
oriented angles may be used as a simple way to detect the filamentary
structures. We investigate the possible convergence of gradients in the
following Sect. \ref{sec:convergence-velocity-density-gradients}. Maps of
the velocity gradient oriented angles show more complex patterns that are
not easy to interpret from the maps. We investigate these results further
in Sect. \ref{sec:comparison-velocity-density-gradients}.

\subsection{Convergence of the velocity and column density gradients}
\label{sec:convergence-velocity-density-gradients}

To quantify the possible convergence of gradients on localized areas, we
calculated their 2D divergence using the following equation:
\begin{equation}
  \emr{div}~\overrightarrow{\emr{grad}}=\frac{\partial \emr{grad}_{x}}{\partial x}+\frac{\partial \emr{grad}_{y}}{\partial y}.
\end{equation}
When the value of the divergence is positive at a position, the gradients
diverge from this same point, whereas the gradients converge to that point
when the value is negative. Figure~\ref{fig:divergence-maps:example} shows
as an example the divergence maps obtained for the main-velocity layer of
the Horsehead Nebula. The divergence maps of the other subregions are
provided in Appendix~\ref{sec:comments-indiv-subregions}. On these maps, we
display the cores identified in the HGBS
\citep{Konyves2020}. To select the self-gravitating objects, we only kept
the objects identified as starless and pre-stellar cores with a
Bonnor-Ebert mass ratio $<$2, combined with the cores that hosts confirmed
YSOs. As \citet{Konyves2020} had no access to the cloud layers identified
in this paper, the core selection is the same for all cloud layers.

A pattern of blue and red striations corresponding to convergence and
divergence, respectively, of velocity and column density gradients, is
visible. As expected from the oriented angle maps of the column density
gradients, the filamentary structures in most subregions are located in
convergence areas of both $N_\mathrm{H}^{13}$ and $N_\mathrm{H}^{18}$
gradients. The absolute values of the divergence of the gradient of
$N_\mathrm{H}$ are larger along and near the filamentary structures in most
subregions. In contrast, the filaments are mostly located in regions of
low absolute value of the divergence of the centroid velocity
gradient. This behavior is present in most of the analyzed subregions:
the Horsehead Nebula, the intermediate area, the flame filament, NGC\,2024,
and the Hummingbird filament (see Fig.~\ref{fig:divergence-maps:example}
and Appendix~\ref{sec:comments-indiv-subregions}). It suggests that
velocity gradients become more coherent in the vicinity of coherent
structures (i.e.,  filaments). In optically thin conditions, centroid
velocity increments, built from molecular line profiles, are related to the
average along the line of sight of some of the shear
components~\citep{Lis96}. A high absolute value of the divergence of
centroid velocity gradients can thus be interpreted as a signature of
shearing motions along the line of sight associated with the current
pixel. The relatively low absolute value of the divergence of centroid
velocity gradients near filaments could indicate less active shearing
motions in these regions. Such an effect was previously discussed by, for example,
\citet{Pety2003}, in the case of the L1512 dense core.
The much wider field
of view allows us to increase the number of probed regions.

Figure~\ref{fig:divergence-maps:example} compares the sign of the
divergence from the velocity and the column density gradients obtained in
the Horsehead Nebula in order to highlight whether gradients in column
velocity and density converge and diverge at the same place. The comparison
between both gradient divergence signs of the other subregions are
provided in Appendix~\ref{sec:comments-indiv-subregions}.  The convergence
and the divergence areas do not systematically correspond between the
column density and velocity gradient. In most subregions, the gradients
have opposite behaviors: when one converges, the other gradient diverges.

In the different subregions, cores are statistically located mostly at the
convergence of the $N_\mathrm{H}^{13}$ and $N_\mathrm{H}^{18}$ gradients
but are generally equally distributed in the convergence and divergence
areas from the velocity gradients.

\subsection{Comparison between velocity and column density gradients}
\label{sec:comparison-velocity-density-gradients}

In order to identify patterns and obtain coherent information on the
kinematics in such a wide field of view, we developed a statistical and
reproducible method that we applied in different subregions of the
cloud. We compared the orientation between the velocity gradients and the
column density gradients that characterize the density structures of the
cloud. We computed the relative orientation between the oriented angles of
the two gradients at each pixel without taking into account their
magnitude. The result of this operation is characterized by a polar
histogram (see
Fig.~\ref{fig:histogram-difference-oriented-angle-gradients}). We were
inspired by the method of the histogram of relative orientations 
developed by \cite{Soler2013} and used in \cite{Planck2016} that compares
the relative orientation of the magnetic field with respect to the density
structures.  To identify any preferential orientation of the velocity field
with respect to the density structures, we built this histogram for
different column density thresholds. We kept only the histograms with
enough pixels to build them, namely with small error bars ($<$0.0005).  We
chose five values of the threshold based on the visual extinction values
estimated by \cite{Pety2017}, representing diffuse gas ($10^{20}\pscm$ and
$10^{21}\pscm$), translucent gas $(3.5 \times 10^{21}\pscm)$, the
environment of filaments $(10^{22}\pscm)$, and the dense cores
$(2.5 \times 10^{22}\pscm)$. In the different subregions, the histograms
peak around 0$^{\circ}$ and/or 180$^{\circ}$ for the $^{13}$CO and
C$^{18}$O data set. The peaks are wider and noisier for the maps using the
C$^{18}$O (1$-$0) data set because the histograms are built on a smaller
number of pixels than those using the $^{13}$CO (1$-$0) data set.

To illustrate the impact of an error on the choice of the ROHSA parameters,
we show in Appendix \ref{sec:influence-gaussian-decomposition} some results
of the polar histograms obtained from another decomposition, such as
$\lambda_\mathrm{i}=10$. We see that there is very little difference with
the results for the chosen decomposition. By selecting the pixels with a
signal-to-noise ratio higher than 3 and with velocity error bars lower than
$0.25\kms$, we only kept the pixels with a robust kinematic signal. We
obtained a robust estimation of the gradients by using forward or backward
neighbors along the x- and y-axis for each pixel.  The kinematic patterns
highlighted in the histograms are therefore not due to the ROHSA
decomposition or to random noise in the data sets but reflect the complexity
of the kinematic motions of the gas in the Orion-B cloud.

In the histograms of NGC\,2024, the Cloak, and the low-velocity layer
associated with Orion B9, a change of pattern is visible between the
histogram angles built from the $^{13}$CO(1$-$0) data set, and those using
the C$^{18}$O(1$-$0) one.  In NGC\,2024, the histogram peaks at
180$^{\circ}$ in $^{13}$CO whereas in C$^{18}$O, the histogram peaks at
0$^{\circ}$. The $^{13}$CO and C$^{18}$O emissions do not trace the gas
from the same spatial area in NGC\,2024: a more extended region is detected
in $^{13}$CO (1$-$0) compared with the restricted area showing C$^{18}$O
(1$-$0) emission. Thus, this result suggests a change in the mechanisms
responsible for the kinematics depending on the traced volume of gas.  For
Orion B9 from the low-velocity layer and the Cloak, the difference between
the two lines can be explained by the low number of pixels available in the
C$^{18}$O data set, resulting in large error bars on the histograms. In the
following, only the histograms from the $^{13}$CO emission for these two
regions will be considered.

In all regions except the intermediate area, the shape of the polar
histograms is relatively stable when increasing the column density
threshold.  In the intermediate area, the histograms peak at 180$^{\circ}$
both in $^{13}$CO and in C$^{18}$O at high column density, namely the dense
core column density ($N_\mathrm{H} >$ 2.5 $\times$ 10$^{22}$ cm$^{-2}$),
whereas at lower column density thresholds, the histograms peak only at
0$^{\circ}$. This suggests a change in the mechanisms responsible for the
high column density gas kinematics in $^{13}$CO and in C$^{18}$O as
compared with the gas at a lower column density.

\FigCircularHistoOne{} %
\FigCircularHistoTwo{} %


\section{Discussion}

\subsection{Physical processes that dominate the velocity field}

To identify the physical mechanisms responsible for the patterns observed
in the histograms, we created 3D toy models of the velocity and volume
density fields around an isolated filament. After processing these models
in the same way as the observed data, we produced polar histograms
from the models which can be compared with those obtained from the
observations. This allows us to distinguish two main behaviors of the
kinematics around filaments: (i) velocity fields that mainly depend on the
radial distance to the filament, such as a radial inflow or a radial
outflow, and (ii) velocity fields along the main filament axis (i.e., longitudinal flows), where the velocity varies along the filament axis
to promote the formation of a core \citep{Hacar2011}.

\subsubsection{Results from toy models}

\FigCircularHistoSim{} %

The models are here meant to represent simple geometric
configurations. Hence, they only implement regular density and velocity
fields, and do not include velocity fluctuations nor random motions due to
turbulence for example. In the ORION-B observations, the filaments are
located at the convergence of the column density gradients. We thus modelled
the filament volume density with an infinite cylinder whose section follows
a 2D Gaussian function.  In order to explore the variety of kinematic
behaviors, we considered four different patterns for the velocity field:
three flows perpendicular to the filament axis, and a flow along the
filament axis. In all cases, we used simple analytical formulae to describe
the modulus of the velocity. The three flows perpendicular to the filament
axis are (i) a gravitational inflow motion, (ii) an accelerated outflow
motion, and (iii) a one sided flow. The latter flow is devised to model the
filaments that are located at the edges of \HII{} regions
(\citealt{Pety2017}; see Fig.
\ref{fig:zoom-regions-NH-maps:layers-cloud}). In this case, one side of the
filament is ionized, and the kinematics information of the molecular gas is
accessible only from the other side of the filament. The last model is
intended to represent a core forming flow, along the main filament
axis. Details of each model are given in Appendix~\ref{sec:3D-models}.

The cubes of volume density and velocity are then inclined by 30$^{\circ}$
with respect to the plane of sky and integrated along the line of sight to
compute the maps of column density and centroid velocity. We then computed
their gradients, the associated oriented angles, and their differences.
Figure~\ref{fig:histogram-difference-oriented-angle-gradients-modeles}
shows the histograms of the relative orientation between the gradients of
column density and centroid velocity. Models where the flow is
perpendicular to the filament axis show a preferred orientation of the
polar histograms at 0$^{\circ}$ and 180$^{\circ}$. In contrast, the model
with a longitudinal flow exhibits a pattern with peaks at 90$^{\circ}$ and
270$^{\circ}$. The difference between the patterns predicted for the
studied inflow and outflow models is moderate. 
The lateral flow model
shows the same 0$^{\circ}$ -- 180$^{\circ}$ orientation but with a pronounced
asymmetry with low column density contours on one side and high column
density contours on the other side.

More generally, the symmetry between the histogram peaks located at
$180^\circ$ from each other comes from the symmetry of the toy models. When
the axis of inclination of the filament on the plane of sky is moved upward
or downward along the filament axis, the histogram
becomes asymmetrical : the amplitude of one of the peaks (for example the peak at
0$^{\circ}$) becomes higher and that at of the other peak  becomes lower.


\subsubsection{Comparison between models and observations}

A comparison of the modeled
(Fig.~\ref{fig:histogram-difference-oriented-angle-gradients-modeles}) and
observed (Fig.~\ref{fig:histogram-difference-oriented-angle-gradients})
histograms indicates that the models with a velocity field perpendicular to
the main filament axis clearly reproduce the observations better. Indeed,
the histograms built from the observed data show clear peaks at $0^{\circ}$
and/or $180^{\circ}$ rather than at other angles. In addition peaks at
angles $90^{\circ}$ or $270^{\circ}$ that are expected in the case of a
longitudinal velocity field are not detected in any of the studied regions.

By exhibiting a main peak at 0$^{\circ}$ in their $^{13}$CO and C$^{18}$O
histograms, and a weaker secondary peak at 180$^{\circ}$, the velocity
fields of the Orion B9 area in the main-velocity layer and the Flame
filament are consistent with a radial inflow on the filaments in the case
when the projection center is not localized in the center of the studied
field.  The C$^{18}$O histogram corresponding to the Hummingbird filament
does not show any clear orientation due to the low number of pixels having
a high S/N ratio C$^{18}$O emission.  The $^{13}$CO histogram shows two
peaks at 0$^{\circ}$ and $135 - 190^{\circ}$.
 We also notice that at high
column density $(> 10^{22}\pscm)$, the second peak around 135$^{\circ}$ is
less pronounced than at lower column densities while the main peak
increases. This histogram is not consistent with the radial inflow or outflow models
when the center of symmetry is located in the middle of the filament
because we expect two equivalent peaks at 0$^{\circ}$ and
180$^{\circ}$. With a clear main peak at 0$^{\circ}$, the mechanism
dominating the kinematics in the Hummingbird filament is radial inflow with
a displacement of the center of projection. Depending on the orientation, a
lateral flow would also be possible, reinforcing the diagnostic of a
significant asymmetry.

By showing two equivalent peaks at 0$^{\circ}$ and 180$^{\circ}$ in their
$^{13}$CO histograms, the velocity fields of the Horsehead Nebula, the
Cloak, and the Orion B9 area from the low-velocity layer are consistent
with the velocity field resulting from radial inflow or outflow
motions. The Horsehead Nebula is at the edge of the \HII{} region, created
by the O9 star $\sigma$Ori, whose expansion can exert a lateral pressure on
the Horsehead Nebula, and inflowing and/or outflowing 
motions explaining the
histogram with two peaks at 0$^{\circ}$ and 180$^{\circ}$ (see
Fig.~\ref{fig:zoom-regions-NH-maps:layers-cloud} and
Table~\ref{tab:regions-HII}).  Orion B9 (both layers) and the Cloak are
located in a quiet region of the Orion B cloud without \HII{}
region. However, their $^{13}$CO histograms from the low and
intermediate-velocity layers also indicate the presence of an asymmetry and
velocity fields perpendicular to the filament axes.

In NGC\,2024, the different patterns of the $^{13}$CO and C$^{18}$O
histograms, which are both aligned along the 0$^{\circ}$ -- 180$^{\circ}$ line
but on opposite directions, indicate a strong asymmetry. This is consistent
with the lateral flow model in the ionized case, a likely scenario because
the IRS2b star in the center of the NGC\,2024 region generates a dense
\HII{} region, the expansion of which can push the molecular gas outward. 
The
NGC\,2024 region is also located at the edge of the \HII{} region created
by the Alnitak star which can cause lateral pressure (see Fig.
\ref{fig:zoom-regions-NH-maps:layers-cloud} and Table
\ref{tab:regions-HII}). As the $^{13}$CO emission traces the translucent
gas in the cloud, it is differently affected by \HII{} regions than the gas
in the filamentary structure traced by C$^{18}$O. This explains the
structure of the C$^{18}$O histogram which shows a main peak at 0$^{\circ}$
consistent with the effect of the expansion of the \HII{} region.  In the
intermediate area, the $^{13}$CO and C$^{18}$O histograms peak at
180$^{\circ}$ at the high column density threshold whereas they peak at
0$^{\circ}$ at lower column density thresholds, a pattern associated with
lateral flows.  Indeed, at high column density, the gas traced by the
$^{13}$CO and C$^{18}$O emission corresponds to the filamentary structures
located in the south of the intermediate area at the edge of the \HII{}
region created by the IRS2b star in the center of the NGC\,2024 region (see
Fig. \ref{fig:zoom-regions-NH-maps:layers-cloud} and Table
\ref{tab:regions-HII}). Thus, the kinematics at high column density can be
due to a lateral pressure on partially ionized filaments.

\subsection{Star formation induced by \HII{} regions}

Our study of selected regions in the ORION-B field of view shows the effect
of \HII{} regions in driving the kinematics of the molecular gas and the
formation of filamentary structures.  A previous study done with the
$^{13}$CO (1$-$0) emission by the ORION-B team \citep{Orkisz2017}
identified compressive modes of turbulence around the active star-forming
regions NGC\,2024 and NGC\,2023, while the quieter regions, the Hummingbird
and Flame filaments, exhibit mostly solenoidal motions. The southern part
of the Intermediate region is also a highly compressive region while the
northern part seems to be dominated by solenoidal motions. These two
observational ORION-B studies put together are consistent with recent
numerical simulations that suggest that \HII{} regions drive
compressive motions in pillar-like structures, thus triggering star
formation \citep{Menon2020}.  This is also consistent with the conclusion
of \cite{Bally2018}. They studied the kinematics of the IC434 \HII{} region
ionized by $\sigma$Ori, and the nearby Horsehead Nebula using SOFIA-GREAT
observations of the [CII] 158$\mu$m line. They conclude that the ionization
front induced a pressure shock triggering a compression of the western edge
of the cloud. They also suggest that the compression would not only be due
to $\sigma$Ori but also to older generations of massive stars.

The Horsehead Nebula is an active star-forming region. Two dense
condensations were identified as dense cores in approximate gravitational
equilibrium by \citet{Ward-Thompson2006}. \citet{Bowler2009} found five
candidate young stars, a class II YSO, and two protostars
in this region. For the NGC\,2024 region, maps of dust emission at 1 mm by
\citet{Mezger1988,Mezger1992} exhibit high-density condensations aligned in
an elongated ridge running north to south, namely the filamentary structure we
identified at the edge of the \HII{} region generated by Alnitak (see
Fig. \ref{fig:zoom-regions-NH-maps:layers-cloud}). Submillimeter and
millimeter observations reveal outflows associated with some of the dense
condensations in NGC\,2024, which would therefore be young protostars
\citep{Moore1995,Richer1990,Richer1992}. \citet{Choi2012} also detected
H$_2$O and CH$_3$OH maser lines, which suggests that the NGC\,2024 region is
actively forming stars. The methanol masers suggest the existence of shocks
that may be driven by YSOs outflows \citep{Urquhart2015} or the expanding
\HII{} regions.

Therefore, the NGC\,2024 and NGC\,2023 \HII{} regions could be good
examples of the triggering effect of future star formation by exerting a
lateral pressure that causes compressive motions, favoring the formation of
filamentary structures at their edges and, thus, the formation of new stars.
\cite{Hosokawa2006} studied the expansion of \HII{} regions and
dissociation fronts due to a massive star performing numerical calculations
of radiation-hydrodynamics. They conclude that the \HII{} region expands,
accumulating molecular gas at the edge where gravitational fragmentation is
then expected.

The compression phenomenon could probably also be reinforced by the
photo-evaporation flow at the illuminated edges of molecular clouds. Indeed,
the UV-illuminated gas at the ionization and dissociation fronts expands
and exerts a compression on the molecular part of the cloud
\citep{Bertoldi1989, Bertoldi1996}. \cite{Bron2018} and \cite{Joblin2018}
recently model photo-evaporation flows from the illuminated edges of
molecular clouds and find that they can induce high pressures that could
explain the origin of the dense structures found at the edges of \HII{}
regions. The presence of dust in the IC\,434 \HII{} region and the
kinematics of ionized carbon \citep{Bally2018} indeed indicates the
presence of an evaporation flow at the edge of the Horsehead Nebula.

\subsection{Inflow due to gravitationally driven turbulence}

In the absence of \HII{} regions, radial flows seem to be the main
mechanism dominating the kinematics around filaments as the observed
pattern are consistent with radial inflows or outflows toward the filament
ridges. Radial inflows are more likely as they contribute to the accretion
of material onto the filaments to increase their linear mass and bring them
closer to the critical state for collapse. We have found a decrease of the
amplitude of centroid velocity gradients divergence in the filaments,
indicating a more coherent behavior of the velocity field in the filaments
as compared with their surroundings, an expected behavior in the case
accretion on filaments \citep{Heigl2020}. Such a velocity pattern could be
due to different mechanisms, including pure infall induced by gravitational
collapse, pure externally driven turbulence, gravitationally driven
turbulence, or accretion flows channeled by magnetic field. Below, we
discuss these four possibilities.

Low-density striations, which may be unresolved in our single-dish
observations of the Orion B cloud, are usually observed perpendicular to
filaments, suggesting that the cloud material is accreting along these
striations onto the main
filaments~\citep{Palmeirim13,Dhabal18,Arzoumanian18,Shimajiri19}.
\citet{Shimajiri19} produce a toy accretion model and compare the modeled
velocity patterns to those observed by \citet{Palmeirim13} around the
B211/B213 filament. They suggest that the gas is located in a thin shell at
the edge of a super-bubble and that the radial accretion flow in this shell
onto the filament is controlled by the gravitational potential of the
filament. In this scenario, the inflow we observed in the Orion B cloud
would be due to pure infall induced by gravitational collapse.  From
estimates of the effective gravitational instability criterion,
\citet{Orkisz2019} identify super- or trans-critical filaments in the
NGC\,2024 region. In this region, the flow we observed in the moderately
dense gas traced by the C$^{18}$O (1$-$0) emission could be consistent with
gravitational collapse motions. However, the other filaments would be
subcritical structures, suggesting that they are not collapsing to form
stars yet. Thus, the radial flows we identified in the Flame and
Hummingbird filaments for example, would contribute to the accretion of
matter on these young filaments to increase their linear mass and bring
them closer to the critical state. In this case, the radial velocity flow
might not be due to pure gravitational infall but could correspond to a
combination of infall, compressive motions or accretion channels along the
magnetic field.

Several filament formation scenarios support the fact that filaments are
formed by an external turbulent compression \citep{Smith2014, Smith2016} or
the collision of two supersonic turbulent gas flows \citep{Pety2000,
  Padoan2001,Tafalla15,Clarke2017}. In this picture, the turbulent cascade
generated initially could be responsible for the inflow observed around
some of the filamentary structures in the Orion B cloud. The Orion B
molecular cloud is highly supersonic with a mean Mach number between six
and eight \citep{Schneider2013, Orkisz2017}. This high Mach number is
usually interpreted as due to the turbulent cascade from the scale of
molecular cloud (tens of pc) down to the scale of a filament (0.1 pc)
\citep{Kritsuk2013,Federrath2016,Padoan2016}. However, some regions, as
NGC\,2024, the Flame and Hummingbird filaments, are moderately supersonic,
with a Mach number between 3 and 5 \citep{Orkisz2017}. For the actively
star-forming region NGC\,2024, we could expect a lower Mach number if
collapse motions were contributing significantly to the velocity
field. Indeed, it is expected that the gravitational instability will
convert potential energy into kinetic energy, and the associated shocks
will contribute to the dissipation of a fraction of the turbulence.  The
decrease of the Mach number in dense filaments is also often interpreted as
a sign of dissipation of turbulence \citep[e.g.,][]{Orkisz2019} as this
reduction is associated with a smaller velocity dispersion. For the Flame
and Hummingbird filaments, the moderately supersonic Mach number as
compared with the bulk of the cloud, also suggests that externally driven
turbulence is not responsible for the radial flow motion. Further work is
needed to determine the degrees of turbulent, thermal and magnetic support
of the filaments and how they change with their evolutionary stage to
confirm whether filaments with lower Mach number than the bulk of the gas
indeed show infall motions.

The gravitational potential of the filaments could lead to radially
dominated accretion and could generate turbulent motions at the same
time. Numerical studies on gravitationally driven turbulence associated
with accretion on filaments (i.e., neglecting the self-gravity of
filaments) succeed in reproducing the linear relation between velocity
dispersion, size, and subsonic inflow velocity
\citep{Ibanez-Mejia2016,Heigl2018}, which is usually expected to be due to
pure turbulent motions. From these results, they argue that
gravitationally driven turbulence due to the gravitational potential of the
filaments is a significant source of turbulent motions.  The velocity and
velocity dispersion pattern predicted for self gravitating filaments by
\citet{Heigl2020} (radial accretion, constant turbulent pressure) is
consistent with the properties of the Flame filament. The inflow we
observed in the Flame filament, and by extension in the Hummingbird
filament and in the Orion B9 region, should be further studied as a test
case of gravitationally driven turbulence around filaments.

Polarization observations revealed a change of the magnetic field direction
with density: the magnetic field is parallel to the diffuse gas structures
\citep{Chapman2011,Palmeirim13,Planck2016} and becomes perpendicular to
dense gas structures, especially the filamentary structures
\citep{Sugitani2011, Planck2016}. This change has been also observed in 3D
and 2D models of magnetized molecular cloud formation \citep{Seifried20}
and interpreted as due to compressive motions, which can be the result of
gravitational collapse or converging flows along filamentary structures
\citep{Soler17}. Shocks located along the filament edge is another
possibility \citep{Hu2019,Hu2020}. Thus, the global magnetic field would be
distorted by the gas becoming self-gravitating after flowing along the B
field lines in low density diffuse regions
\citep{Heyer2016,Tritsis2016,Chen2017}.  Low-density and thermally
subcritical filaments, such as the perpendicular striations, are expected
to be parallel to the magnetic field, while high-density and
self-gravitating filaments should be preferentially oriented perpendicular
to the field lines \citep{Nagai1998}. This may explain the distribution of
filament orientations with two orthogonal groups found in several clouds
\citep{Peretto2012, Palmeirim13, Cox2016,Ladjelate2020}. The relative
orientation of filaments with respect to the magnetic field is therefore an
important element to fully understand their dynamical state. However, the
\textit{Planck} polarization data set does not have a sufficient angular
resolution (2.5$\arcmin$) to estimate robustly the relative orientation of
magnetic field with respect to the filaments we studied in the ORION-B
field \citep{Orkisz2019}. Polarization measurements at higher angular
resolution, reaching 0.1\, pc or better, are thus required.

\subsection{Longitudinal flows}

Among the studied fields we have not found evidence for longitudinal flows
as discussed by, for example, \cite{Hacar2011} in the context of core formation
within a filament. Because of the limited spatial resolution of the ORION-B
data ($\sim 0.05$\, pc, these phenomena may not contribute much to the
statistics of the centroid velocity gradients as compared to the other
types of flows that are more extended. Nevertheless, the toy models clearly
indicate that the specific geometry of core forming flows can be clearly
identified with the proposed method. Further studies at a higher linear
resolution may therefore help in identifying these flows.

\section{Conclusions}

In the framework of the ORION-B program, we analyzed the kinematics around
filamentary structures in the Orion B cloud. The main results of our study
are listed below.
\begin{enumerate}
\item Using the ROHSA algorithm, we distinguished three cloud velocity
  layers at different systemic velocities (2.5, 6, and 10 km s$^{-1}$) in
  the translucent and moderately dense gas.
\item We identified the filamentary network of each velocity layer. The
  filaments are preferentially located in regions of low amplitude centroid
  velocity gradients.
\item By comparing the orientation of density column gradients and the
  orientation of velocity gradients from the ORION-B observations and
  synthetic observations from 3D toy models, we were able to identify the
  physical mechanisms at work around the filamentary structures. This
  simple statistical method can be reproduced in any molecular cloud to
  obtain coherent information on the kinematics. This allows us to
  distinguish two types of behavior in the kinematics around filaments:
  (i) a radial flow of the molecular matter toward the filament ridges and
  (ii) a longitudinal flow associated with the formation of a core within
  the filament.
\item \HII{} regions tend to generate compressive motions via an
  outflow or a lateral pressure on the translucent gas, favoring the
  formation of filamentary structures at their edges, in, for example, NGC\,2024.
  This is the first observational study to highlight feedback from \HII{}
  regions on filament formation and, thus, on star formation in the Orion B
  cloud.
\item Without nearby \HII{} regions, radial inflow seems to be the main
  mechanism dominating the kinematics around the filaments (Hummingbird,
  Flame, and Orion B9 areas in the main-velocity layer) and may reveal the
  accretion processes generating gravitationally driven turbulence.
\end{enumerate}

{These results were obtained by observing molecular line data at high S/N, high spectral resolution,  over a wide area encompassing more than two decades of spatial scales. The methods we developed  could be applied to other molecular clouds to investigate the relationships of filaments with their parent molecular cloud in different environments and at different evolutionary stages.}

\begin{acknowledgements}
  We thank the referee for thoughtful comments that helped to improve the
  paper significantly. This work was supported in part by the French Agence
  Nationale de la Recherche through the DAOISM grant ANR-21-CE31-0010 and
  by the Programme National ``Physique et Chimie du Milieu Interstellaire''
  (PCMI) of CNRS/INSU with INC/INP, co-funded by CEA and CNES. We thank
  ``le centre Jules Jensen'' from Observatoire de Paris for its hospitality
  during the workshops devoted to this project. This research has made use
  of data from the Herschel Gould Belt Survey (HGBS) project
  (\url{http://gouldbelt-herschel.cea.fr}). The HGBS is a Herschel Key
  Programme jointly carried out by SPIRE Specialist Astronomy Group 3 (SAG
  3), scientists of several institutes in the PACS Consortium (CEA Saclay,
  INAF-IFSI Rome and INAF-Arcetri, KU Leuven, MPIA Heidelberg), and
  scientists of the Herschel Science Center (HSC). JRG and MGSM thank the
  Spanish MCIYU for funding support under grant PID2019-106110GB-I00. JO
  acknowledges funding from the Swedish Research Council, grant
  No. 2017-03864.
\end{acknowledgements}

\bibliographystyle{aa} 
\bibliography{bibliography} 

\appendix{} %
\section{Comparison between the Gaussian decomposition
  parameters} \label{sec:ROHSA-decomposition}

\subsection{Best total number of Gaussian components}

\Figchidist{} %

To identify the best total number of Gaussian components for C$^{18}$O and
$^{13}$CO, we initiated the ROHSA algorithm to decompose the signal with a
series of Gaussian components with an initial velocity dispersion of
$\sigma_\mathrm{init}=$1.5~km s$^{-1}$. We computed a decomposition with
$N$ = [3, 4, 5, 6] Gaussian components for a given value of
hyper-parameters ($\lambda_\mathrm{i}$=100,
$\lambda'_\mathrm{\sigma}$=0). For the different numbers of Gaussian
components, we estimated the residuals between the original and the ROHSA
reconstructed cubes, and computed the $\chi^2$ of the fit on each spectrum
(i.e., at each pixel). We computed the $\chi^2$ on the velocity range -5
$-$ 20~km s$^{-1}$ following the equation
\begin{equation}
  \chi^2 = \frac{1}{n} \sum_{i=1}^n \frac{\mathrm{residuals^2}}{\mathrm{rms^2}},
\end{equation}
where $rms$ is the rms noise level of each individual spectrum. The
Gaussian decompositions have similar residual and $\chi^2$
distributions. This indicates that the ROHSA algorithm reproduces the
signal well regardless of the number of Gaussian components (see
Fig.~\ref{fig:chi2-residuals-distribution-ngauss-ROHSA}).  Then, we built
the distributions in velocity and in velocity dispersion of each Gaussian
component from the different decompositions (see
Fig.~\ref{fig:distribution-position-sigma-ngauss-c18o-ROHSA} for C$^{18}$O,
and Fig.~\ref{fig:distribution-position-sigma-ngauss-13co-ROHSA} for
$^{13}$CO). From these distributions, we observe that the more Gaussian
components are added, the more they overlap in velocity and their velocity
dispersions become small. To avoid over-fitting, we chose the decomposition
with the maximum number of Gaussian components with the constraint that the
differences between the mean values of the velocity distribution of each
Gaussian component are higher than the spectral resolution (0.5~km
s$^{-1}$). This criteria indicates that adding more Gaussian components as
an initial parameter in ROHSA does not allow us to recover more signal. If
more Gaussian components were added, they would only subdivide a previous
single Gaussian component. From this criteria, the best total number of
Gaussian components $N_\mathrm{best}$ is four and five to reconstruct the
C$^{18}$O (1$-$0) and the $^{13}$CO (1$-$0) signals, respectively.

\FigDistCompOne{} %
\FigDistCompTwo{} %

\subsection{Best hyper-parameters}

\FigchidistTwo{} %

To find the best hyper-parameters, we also set up the algorithm with several
values ($\lambda_\mathrm{i}$= [10, 100, 1000], $\lambda'_\mathrm{\sigma}$=
[0, 1]) for the best total number of Gaussian components
$N_\mathrm{best}$. After building the associated $\chi^2$ and residual
distributions, we observe that the Gaussian decompositions do not show
distributions very different from each other to identify robustly the best
hyper-parameters (see
Fig. \ref{fig:chi2-residuals-distribution-param-ROHSA}). We selected
$\lambda_\mathrm{i}$=100 for each hyper-parameter.  The hyper-parameter
$\lambda'_\mathrm{\sigma}$ have been set to $0$ to let free the variation
of the velocity dispersion across the map throughout the full field of
view.

\clearpage{} %

\section{Moment maps of the Gaussian components for each tracer}
\label{sec:moment-maps:Gaussian-components}

\FigGaussSpectra{} %
\FigGaussMomentsCeO{} %
\FigGaussMomentstCO{} %
\clearpage{} %

\section{Monte Carlo simulations}
\label{sec:monte-carlo-results}

\FigMonteCarloNH{} %
\FigMonteCarloVelo{} %
\clearpage{}

\section{Comments on individual subregions}
\label{sec:comments-indiv-subregions}

This section shows the detailed results for all subregions sorted from
east to west, and then from north to south.

\subsection{Intermediate-velocity layer, Cloak}

\FigMomentsCloak{} %
\FigAngGradientsCloak{} %
\FigDivMapsCloak{} %

According to Greek mythology, the constellation of Orion represents a
hunter. The CO emission from the main-velocity layer looks like a human
skeleton with the head at the Hummingbird filament and the two feet at the
Flame filament and the Horsehead Nebula (see
Fig. \ref{fig:moment0-maps:initial-datasets}). The Cloak area was named
as such because the CO emission from this layer is filamentary and located
to the northeast of the NGC\,2023 and NGC\,2024 star-forming regions, like
a knight's cloak.

The HGBS mapped this area and reveals
that it is actually composed of a long filamentary structure that follows
the Cloak shape from east to west in which several starless and pre-stellar
cores appear to be embedded \citep{Konyves2020}. This is quite consistent
with the filamentary network we extracted from the C$^{18}$O emission of
the intermediate-velocity layer. However, we identified a network of small
filaments that follow the Cloak shape from east to west instead of a long
single filament. This is due to the structure of the C$^{18}$O emission,
which is faint and not continuous (see Fig. \ref{fig:moment-maps:cloak}).

There are very few or no studies of the kinematics of this region. The
velocity map from the C$^{18}$O emission (see
Fig. \ref{fig:moment-maps:cloak}) shows an oscillation in blue- and
red-shifted velocities along the entire filamentary structure. Other
filaments, as G350.5-N filament \citep{Liu2019}, exhibit this kind of
velocity pattern. It may suggest either an oscillation of the filament
toward and away from the observer or motions due to core formation along
the filamentary structure. However, at larger scales, from the $^{13}$CO
emission, the velocity patterns are more complex than an oscillation along
the filamentary network.

\clearpage{} %

\subsection{Low- and intermediate-velocity layer, Orion B9}

Orion B9 is an active star-forming region of the Orion B molecular
cloud. \citet{Miettinen2009} mapped the Orion B9 region at 870 $\mu$m and
identified 12 dense cores. Six of them are associated with Spitzer 24
$\mu$m sources, and are therefore identified as protostellar objects. The
remaining cores have no mid-infrared counterparts but are gravitationally
bound \citep{Miettinen2010}, and thus they can be considered as pre-stellar
cores. The HGBS images revealed that
Orion B9 is actually composed of supercritical filaments in which pre- and
protostellar cores appear to be embedded \citep{Konyves2020}. However, the
HGBS data set does not distinguish the different components in velocity we
identified.

From multiple line observations, \citet{Miettinen2012,Miettinen2020} find
two separated velocity components at about 1.5$-$5 and 8.5$-$9.5 km
s$^{-1}$ in several protostellar objects. These velocity components are
consistent with the low-velocity layer at 0$-$4 km s$^{-1}$ and the
main-velocity layer at 8$-$10.5 km s$^{-1}$ we identified in this study
from both $^{13}$CO and C$^{18}$O (1$-$0) observations (see
Figs. \ref{fig:moment-maps:orion-b9-low} and
\ref{fig:moment-maps:orion-b9-main}).  \citet{Miettinen2012} detect a
velocity gradient along the northwestern-southeastern direction in the two
different velocity components from C$^{18}$O (2$-$1) observations. The
direction of this gradient is consistent with the one we observe in the
Orion B9 area from $^{13}$CO and C$^{18}$O (1$-$0) observations, from both
the low-velocity and the main-velocity layers (see
Figs. \ref{fig:moment-maps:orion-b9-low} and
\ref{fig:moment-maps:orion-b9-main}).  \citet{Miettinen2012,Miettinen2020}
interpret this gradient as due to a collision between two clouds. The
collision may be triggered by the expansion of the \HII{} region of
NGC\,2024 that lies about 4 pc to the southwest of Orion B9 region. This
collision would be responsible for filament and dense core formation in
Orion B9 area.

\clearpage{} %

\FigMomentsBnineLow{} %
\FigAngGradientsBnineLow{} %
\FigDivMapsBnineLow{} %

\newpage{} %

\FigMomentsBnineMain{} %
\FigAngGradientsBnineMain{} %
\FigDivMapsBnineMain{} %

\clearpage{} %

\subsection{Main-velocity layer, Hummingbird filament}

The velocity map from the C$^{18}$O emission (see
Fig. \ref{fig:moment-maps:hummingbird}) shows an oscillation in blue- and
red-shifted velocities along the main filamentary structure. Other
filaments, as G350.5-N filament \citep{Liu2019} and the Cloak filament (see
Fig. \ref{fig:moment-maps:cloak}), exhibit this kind of velocity
pattern. It may suggest either an oscillation of the filament toward and
away from the observer or motions due to core formation along the
filamentary structure. \citet{Orkisz2019} also identify hints of periodic
longitudinal fragmentation along the Hummingbird filament. However, they
estimated that this filament is gravitationally subcritical and very few
known YSOs are embedded in it. Thus, this velocity oscillation is unlikely
to be due to  core formation.

\newpage{} %

\FigMomentsHummingbird{} %
\FigAngGradientsHummingbird{} %
\FigDivMapsHummingbird{} %

\clearpage{} %

\subsection{Main-velocity layer, intermediate area}

The Intermediate area was named this way because it is located between two
well-studied areas: NGC\,2024 and the Hummingbird filament. There are very
few or no studies of the kinematics of this region but its kinematics is
interesting to study because it encompasses the northern edge of the
NGC\,2024 \HII{} region.

\newpage{} %

\FigMomentsIntermediate{} %
\FigAngGradientsIntermediate{} %
\FigDivMapsIntermediate{} %

\clearpage{} %

\subsection{Main-velocity layer, NGC\,2024}

NGC\,2024, also known as the Flame Nebula, is a well-known massive star-forming region.  From $^{12}$CO and $^{13}$CO observations,
\citet{Emprechtinger2009} found that most of this region consists of a hot
(75 K) and dense (9 $\times$ 10$^5$ cm$^{-2}$) gas. From $^{13}$CO (2$-$1)
observations, \citet{Enokiya2021} find two velocity components with a
velocity separation of $\sim$2 km s$^{-1}$ (8.5$-$9.5 km s$^{-1}$ and
10.5$-$11.5 km s$^{-1}$) in the region including NGC\,2023, NGC\,2024 and
the Horsehead Nebula. Both components are associated with the main-velocity
layer. This suggests that a major cloud-cloud collision, with a timescale
of 2 $\times$ 10$^5$ yr, triggered the formation of massive stars in
NGC\,2024, including the late O or early B massive star IRS2b. IRS2b has
been identified as the main ionizing source of NGC\,2024
\citep{Bik2003}. The associated \HII{} region is expected to expand into
the molecular cloud and to trigger star formation.

Maps of dust emission at 1 mm by \citet{Mezger1988,Mezger1992} exhibit
high-density condensations aligned in an elongated ridge running north to south,
consistent with the filamentary structure we identified at the center of
the region (see Fig. \ref{fig:moment-maps:ngc2024}). Submillimeter and
millimeter observations revealed outflows associated with some of the dense
condensations, which would therefore be young protostars
\citep{Moore1995,Richer1992, Richer1990}. \citet{Choi2012} also detected
H$_2$O and CH$_3$OH maser lines in NGC\,2024, suggesting that this region
is actively forming stars.

\newpage{} %

\FigMomentsNGC{} %
\FigAngGradientsNGC{} %
\FigDivMapsNGC{} %

\clearpage{} %

\subsection{Main-velocity layer, Flame filament}

\newpage{} %

\FigMomentsFlame{} %
\FigAngGradientsFlame{} %
\FigDivMapsFlame{} %

\clearpage{} %

\subsection{Main-velocity layer, Horsehead Nebula}

The Horsehead Nebula, also known as Barnard 33, is one of the most
easily recognizable shape in the sky.  At visible wavelengths, it appears
as a 5-arcminute dark patch against the bright H$\alpha$
emission. \citet{Philipp2006} determined that the UV radiation primarily
comes from $\sigma$Ori.  High density gas and UV stellar light interact at
the edge of the western condensation, forming a photon-dominated region. At 400pc, the
Horsehead Nebula is the closest known pillar \citep{Pound2003}.
\citet{Reipurth1984} suggested that the Horsehead Nebula formed through the
photo-evaporation of low density material around the neck, shaping the
molecular cloud into the famous Horsehead.  \citet{Hily-Blant2005} measured
velocity gradients from the C$^{18}$O (2$-$1) emission and suggested that
the gas is rotating around the neck axis with a rotation period of $\sim$4
Myr. They also found two dense condensations that were identified as dense
cores by \citet{Ward-Thompson2006}. The Horsehead Nebula is an active
star-forming region since \citet{Bowler2009} found also five candidate
young stars, a class II YSO, and two protostars.

While the majority of the emission of the Horsehead Nebula is associated with
the main-velocity layer, we can see that a spatial component corresponds to
NGC\,2023 and the Horsehead Nebula in the C$^{18}$O and $^{13}$CO (1$-$0)
emissions of the Cloak layer (see
Fig. \ref{fig:moment-maps:ROHSA-layers-cloud}). Thus, NGC\,2023 and the
Horsehead Nebula have two velocity components, a first one at 6$-$9.5 km
s$^{-1}$ (associated with the intermediate-velocity layer) and a main one at
10$-$12 km s$^{-1}$ (associated with the main-velocity layer).

\newpage{} %

\FigMomentsHorsehead{} %
\FigAngGradientsHorsehead{} %
\FigDivMapsHorsehead{} %

\clearpage{} %

\section{Influence of the Gaussian decomposition on the kinematic results}
\label{sec:influence-gaussian-decomposition}

\FigDecompROHSA{} %
\FigDecompROHSAbis{} %
\FigCompDecompROHSA{} %

To characterize the influence of the choice of hyper-parameters of the
ROHSA Gaussian decomposition on the moment maps and, thus, on velocity and
column density gradients, we set up the algorithm to decompose the $^{13}$CO
and C$^{18}$O signals with different hyper-parameters for the best total
number of Gaussian components determined above. As example, we chose the
two following sets of hyper-parameters:
($\lambda_\mathrm{A} = \lambda_\mathrm{\mu} = \lambda_\mathrm{\sigma} =
10$, $\lambda'_\mathrm{\sigma} = 0$) and
($\lambda_\mathrm{A} = \lambda_\mathrm{\mu} = \lambda_\mathrm{\sigma} =
100$, $\lambda'_\mathrm{\sigma} = 1$). In the following, these two
decompositions are denoted as $\mathrm{D1}$ and
$\mathrm{D2}$. $\mathrm{D1}$ forces the spatial coherence of
the Gaussian components in the field of view a little less than the decomposition chosen
and presented in the main text, namely the decomposition $\mathrm{D0}$ with
the set of hyper-parameters:
$\lambda_\mathrm{A} = \lambda_\mathrm{\mu} = \lambda_\mathrm{\sigma} =
100$, $\lambda'_\mathrm{\sigma} = 0$. $\mathrm{D2}$ puts a constrain on the
$\sigma$ variation of Gaussian components along the field of view compared
to $\mathrm{D0}$.

From these decompositions, we followed the same methodology as described
above, namely, the construction of moment maps to identify the different
layers in the Orion B cloud and the computation of the errors on the
velocity and the column density via a Monte Carlo process to select only
the robust pixels (i.e., $N_\mathrm{H}>10^{20}$ cm$^{-2}$ for $^{13}$CO,
$N_\mathrm{H}>10^{21}$ cm$^{-2}$ for C$^{18}$O, and error on the velocity
$<$0.25 km s$^{-1}$).

In the following, we present the kinematic results obtained from
$\mathrm{D1}$ and $\mathrm{D2}$ for the $^{13}$CO and C$^{18}$O emission of
the Horsehead Nebula and we compare them with the $\mathrm{D0}$ results
showed in the main text.  Figures
\ref{fig:comparison-results-decomposition-ROHSA-13CO} and
\ref{fig:comparison-results-decomposition-ROHSA-C18O} show the distribution
histogram of the oriented angles of the column density and velocity
gradient in the Horsehead Nebula and the histograms of the relative
orientation between both gradients to compare the results from the
different decompositions. The trend of the histograms is the same and the
small visible variations between the different decompositions are
consistent with the error bars of the histograms.
Figure~\ref{fig:residus-oriented-angle-gradients-tete-cheval} shows the
variation of the column density and velocity gradient orientation in the
Horsehead Nebula from D1 and D2 decompositions with respect to D0
one. There is very little variation in the orientation of the
gradients. For velocity gradients, the variation in orientation is mostly
between -20$^{\circ}$ and 20$^{\circ}$, which is consistent with the error
distribution estimated from the Monte Carlo process.

\clearpage{} %

\section{Simple 3D model of a filament and its associated velocity field}
\label{sec:3D-models}

\FigModelCuts{} %
\FigModelProj{} %

\subsection{Volume density}

\subsubsection{Typical case}

In a 3D cube of dimensions $241 \times 241 \times 241$ pixels with a
central position at the coordinates $(\mu_x,\mu_y,\mu_z)=(121,121,121),$ we
modeled a simple isolated filament using an infinite cylinder along the $y$-axis, whose section in the $xz$ plane follows a 2D Gaussian function. We
used $n_\mathrm{H_0}$ as an amplitude factor for the density, and
$\sigma_x = \sigma_z = \sigma_{xz}$ the standard deviation of the density
distribution along the $x$- and $z$-axes.  We define the radius
perpendicular to the filament as $r = \sqrt{(x-\mu_x)^2+(z-\mu_z)^2}$. With
these definitions, the density distribution of the filament
$n_\mathrm{fil}$ is
\begin{equation}
  n_\mathrm{fil}(x,y,z) = n_\mathrm{H_0}
  \exp{\left\{ - \frac{r^2}{2 \sigma_{xz}^2} \right\} }.
\end{equation}
  
We modeled the filament at the central position of the 3D cube with a
Gaussian full width at half maximum of 4\,pixels for the $x$- and $z$-axes.  The pixel size is
$9''$ or $17.5\unit{mpc}$ at a distance of 400\,pc. The filament is
immersed in a constant density background $n_\mathrm{bg}$. The total
density of the cube is therefore represented by
$n_\mathrm{tot}(x,y,z) = n_\mathrm{fil}(x,y,z)+n_\mathrm{bg}(x,y,z)$ with
$n_\mathrm{H_0} = 5\times10^4\pccm$, and
$n_\mathrm{bg}(x,y,z) = 5\times10^2\pccm$. Figure~\ref{fig:model-3D} shows
the three 2D cuts of the volume density.

\subsubsection{Ionized-like case}

Some of the filaments in the ORION-B field of view are localized at the
interface between the molecular cloud and an \HII{} region. We thus
developed a model where only the molecular half of the space emits. To do
this, we kept the same typical volume density field, except that we
set the volume density to zero for half of the cube along the $z$-axis.

\subsection{Velocity field}

We developed four models of velocity fields representative of the different
physical processes: 1) a radial inflow corresponding to accretion onto the
filament, 2) a radial decelerated outflow in case the filament would host a
star, 3) a lateral flow from the filament to the diffuse environment
representing a filament at the edge of an \HII{} region, and 4) a
longitudinal flow along the filament axis. Figure~\ref{fig:model-3D} shows
the associated velocity vectors in three 2D planes.

\subsubsection{Radial inflow and outflow}

For the radial flows, we introduced $\beta(x,z)$, the velocity angle in the
$(x,z)$ plane so that
\begin{equation}
  \cos \beta(x,z)= \frac{x-\mu_x}{r}, %
  \quad \mbox{and} \quad %
  \sin \beta(x,z)= \frac{z-\mu_z}{r},
\end{equation}
where $\mu_x$ and $\mu_z$ are the positions of the center of the
filament. The components of the radial velocity gradient are then defined
as
\begin{eqnarray}
  V_x(x,y,z) &=& V(r)\,\cos{\beta(x,z)},\\
  V_y(x,y,z) &=& 0,\\
  V_z(x,y,z) &=& V(r)\,\sin{\beta(x,z)},
\end{eqnarray}
where $V(r)$ is the amplitude of the velocity field at the distance $r$
from the filament center. We considered two possibilities for the velocity
amplitude, namely,
\begin{description}
\item[a decelerated outflow,] \mbox{}
  \begin{equation}
    V(r) = \frac{50\kms}{\sqrt{1+\paren{\frac{r}{r_0}}^2}\,\bracket{1+\frac{n_\mathrm{H_0}}{n_\mathrm{bg}}\exp\cbrace{-\paren{\frac{r}{r_0}}^2}}};
  \end{equation}
\item[and a gravitational inflow,] \mbox{}
  \begin{equation}
    V(r) = - 2 \sqrt{G\, m_\emr{lin}(r) \,\log\paren{\frac{r_\emr{max}}{r}}},
  \end{equation}

\noindent where $m_\emr{lin}$ is the linear mass inside the cylinder of radius $r$,
  given by
  \begin{equation}
    m_\emr{lin}(r) = \pi  \, \mu \, n_\emr{H_0} \, r_0^2 \bracket{1-\exp\paren{-\frac{r^2}{r_0^2}}}.
  \end{equation}
  This follows the models by \citet{heitsch13a,heitsch13b}. It can be
  rewritten as
  \begin{equation}
    V(r) = - V_0
    \sqrt{\bracket{1-\exp\paren{-\frac{r^2}{r_0^2}}}\,\log\paren{\frac{r_\emr{max}}{r}}}
  \end{equation}
  \begin{equation}
    \mbox{with} \quad V_0 = 2\sqrt{\pi\,G\,\mu\,n_\emr{H_0} \, r_0^2}
    \sim 0.23\kms.
  \end{equation}
\end{description}
For these two possibilities, $r_0$ is a scaling parameter that defines the
scale of the gradient. We chose the same value as the full width at half maximum of the
filament, 4 pixels. For the gravitational inflow, $r_\emr{max}$ is the
starting point of the inflow and $\mu$ is the mean molecular mass of the
gas (i.e., $\mu \simeq 2.5 m_\emr{H}$). We chose $r_\emr{max} = L/2$
with $L$ the length of the cube.

\subsubsection{Lateral flow}

The gas is accelerating from one side to another side of the 3D cube
through the filament. This simple scenario could be expected by the
expansion of a \HII{} region pushing matter outward, the other part of the
gas being ionized, or to a lateral pressure exerted on the high density
filament from a lower density expanding structure, a bubble of hot gas for
instance.  In this model, the gas is decelerating from the outside region
to the filament.

\subsubsection{Longitudinal flow}

In this case, the flow is parallel to the filament axis (i.e., along the $y$
direction) and converges toward the plane $y=0''$:
\begin{eqnarray}
  V_x(x,y,z) &=& 0,\\
  V_y(x,y,z) &=& -V_0\,y,\\
  V_z(x,y,z) &=& 0,
\end{eqnarray}
where $V_0$ is a constant that we set to $0.5\kms$. This case belongs
to the family of scenarios where the gas flow converges parallel to the
filament to form a dense core.

\subsection{Projection onto the plane of the sky}
\label{sec:3D-model-projection}

The models are computed in three dimensions in a coordinate frame that is
adapted to the geometry of the filament. However, the filament is probably
inclined on the plane of the sky and the observer will only have access to
values projected onto the plane of the sky. In order to deal with the
inclination on the plane of the sky, we rotated the position and velocity
vectors around an axis parallel to the $x$-axis and we then interpolated
the values of the density and velocity coordinates into the inclined
coordinate system with new axes $(x',y',z')$. This generates edge effects
because the initial cubes has a limited size. To avoid these ``artifacts,''
we extracted an inner cube of $69^3$ pixels. The column density was then
computed by integrating the volume density along the line of sight with
\begin{equation}
  N(x',y')= \frac{L}{N_\emr{pix}} \sum_{z'} n_\mathrm{tot}(x',y',z').
\end{equation}
We computed the centroid velocity and its gradient in the optically thin limit, 
\begin{equation}
  C_V(x',y')= \frac{\sum_{z'} n_\mathrm{tot}(x',y',z') V_{z'}(x',y',z')}{\sum_{z'} n_\mathrm{tot}(x',y',z')} 
.\end{equation}
All models were computed for different inclinations of the filament along
the line of sight to investigate the effect of inclination and we selected
the models for a typical inclination of $30^{\circ}$.
Figure~\ref{fig:model-2D} shows the resulting maps of column density and
centroid velocity. The vectors of their gradient are overlaid. All the
models shown in this appendix are inclined on the plane of sky around an
axis located in the middle of the field of view but we discuss in the main
text the effect of varying the position of the inclination axis.
  
\end{document}